\pdfoutput=1
\documentclass[12pt]{article}
\usepackage{graphicx}

\usepackage[colorlinks=true, linkcolor=blue, citecolor=black, urlcolor=black]{hyperref}
\usepackage{pdflscape}
\usepackage[margin=1in,left=1in, top=1in]{geometry} 
\usepackage{booktabs}
\usepackage{tikz}
\usepackage{colortbl}
\usetikzlibrary{positioning, arrows.meta, calc}
\usepackage{authblk}
\usepackage{amsmath}
\usepackage{amssymb}
\usepackage{dsfont}
\usepackage[utf8]{inputenc}
\usepackage[ruled, vlined]{algorithm2e}
\usepackage{subcaption}
\usepackage[section]{placeins}
\usepackage{multirow}
\usepackage{caption}
\usepackage[backend=biber,style=authoryear, maxcitenames=2, doi=false, url=false, bibencoding=utf8, hyperref=true,dashed=true, uniquelist=true, natbib=true, date=year, urldate=long]{biblatex}
\usepackage{setspace} 
\usepackage{eurosym}
\usepackage{siunitx}
\usepackage{tabularx}
\newcolumntype{L}{>{\raggedright\arraybackslash\hangindent=1em\hangafter=1}X}
\newcommand{\tablenote}[1]{\par\smallskip\begin{minipage}{\linewidth}\footnotesize #1\end{minipage}}
\newcommand{\LTV}{\mathrm{LTV}}
\usetikzlibrary{positioning, arrows.meta}

\DeclareUnicodeCharacter{20AC}{\EUR}

\DontPrintSemicolon %
\SetNlSty{textbf}{}{.} %
\SetKwInput{KwInput}{Input}
\SetCommentSty{texttt}

\SetKwBlock{Begin}{}{end}
\SetKwFor{For}{for}{}{end}
\SetKwFor{ForEach}{for}{}{end}
\SetKwIF{If}{ElseIf}{Else}{if}{}{else if}{else}{end}

\DefineBibliographyStrings{english}{%
  urlseen = {accessed},
  january = {January}, february = {February}, march = {March},
  april = {April}, may = {May}, june = {June}, july = {July},
  august = {August}, september = {September}, october = {October},
  november = {November}, december = {December}}
\DeclareSourcemap{
  \maps[datatype=bibtex]{
    \map{
      \pernottype{online}
      \pernottype{dataset}
      \pernottype{report}
      \pernottype{techreport}
      \pernottype{thesis}
      \step[fieldset=url, null]
      \step[fieldset=urldate, null]
      \step[fieldset=issn, null] %
      \step[fieldset=doi, fieldsource=doi, final]
    }
  }
}

\renewbibmacro*{doi+eprint+url}{%
  \iftoggle{bbx:doi}
    {\printfield{doi}}
    {}%
  \newunit\newblock
  \iftoggle{bbx:url}
    {\usebibmacro{url+urldate}}
    {}}

\makeatletter
\newcommand*{\addFileDependency}[1]{%
  \typeout{(#1)}%
  \@addtofilelist{#1}%
  \IfFileExists{#1}{}{\typeout{No file #1.}}%
}
\makeatother

\newcommand{\SIfig}[1]{SI Fig.~\ref{#1}}
\newcommand{\SItab}[1]{SI Table~\ref{#1}}
\newcommand{\SIsec}[1]{SI Section~\ref{#1}}
\newcommand{\SIeq}[1]{SI Equation~(\ref{#1})}
\newcommand{\SIalg}[1]{SI Algorithm~\ref{#1}}

\author[1,2]{Matteo Valle\thanks{Corresponding author: \href{mailto:m.valle@uva.nl}{m.valle@uva.nl}}}

\author[1,3]{Drona Kandhai}
\author[2]{Cees Diks}
\author[1]{Debraj Roy}

\affil[1]{Computational Science Lab, University of Amsterdam}
\affil[2]{Amsterdam School of Economics, University of Amsterdam}

\affil[3]{Korteweg de Vries Institute, University of Amsterdam}

\title{Neighbourhood flood damage amplifies mortgage default cascades}

\renewcommand{\SIfig}[1]{Fig.~\ref{#1}}
\renewcommand{\SItab}[1]{Table~\ref{#1}}
\renewcommand{\SIsec}[1]{Section~\ref{#1}}
\renewcommand{\SIeq}[1]{Eq.~(\ref{#1})}
\renewcommand{\SIalg}[1]{Algorithm~\ref{#1}}

\begin{document}

\maketitle

\begin{abstract}
Floods threaten mortgage repayment by damaging homes, disrupting incomes, and worsening conditions in surrounding neighbourhoods. However, it remains unclear how these neighbourhood effects turn local damage into systemic mortgage distress, as Europe’s housing crisis leaves households that lose their homes with few affordable alternatives. Using an agent-based model calibrated to Limburg, the Netherlands, we show that damage to surrounding properties can trigger mortgage defaults among households that can still afford to pay, including those whose own homes remain undamaged. In the most severe flood scenario, the annual foreclosure rate more than doubles. As each foreclosure lowers the value of nearby homes, it triggers a cascade. Floods make such large foreclosure cascades more frequent, with the share of foreclosures that belong to these cascades rising from 31\% to 45\%. They also shift foreclosures towards younger mortgages, when more debt is still outstanding. These findings show how neighbourhood damage extends mortgage risk beyond directly flooded homes and causes foreclosures to arrive in concentrated waves earlier in repayment.

\end{abstract}


\section{Introduction}
More frequent and severe floods cause substantial physical damage and economic losses, placing financial pressure on households and lenders. Climate-risk research increasingly examines how physical shocks cascade through social and economic systems \citep{Simpson_2021,IPCC2022WGII,Ebi_2024}. Residential mortgages sit at the centre of such cascades, because their performance depends on the condition of the property and of the neighbourhood around it. A flood can propagate from damaged homes through mortgage defaults into the housing market, where forced sales and falling prices can sort low-income households into high-risk areas, a process known as climate gentrification, and from there into poverty traps that leave them with little capacity to adapt to the next shock \citep{dupont2025emergent, garibay2026pricing}. These risks are compounded by Europe’s housing crisis: house prices in the European Union have risen by about 65\% since 2015 and rents by more than 20\% \citep{Eurostat2026HousePrices}, prompting the first European Affordable Housing Plan \citep{EC2025AffordableHousingPlan}. Yet the credit-risk models used in climate stress tests typically treat defaults as independent once regional shocks are accounted for \citep{Alogoskoufis_etal_2021}, and so cannot represent interactions in which one household’s default changes the risk faced by its neighbours.

Repeated floods impose a dual financial shock on households. First, they reduce property values: evidence from the U.S., Italy and the Netherlands shows that properties in flood-prone areas sell at a discount relative to comparable properties outside high-risk zones \citep{Harrison2001,Zhang2019,Daniel2009,Bin_Polasky_2004, Bellaver_Costantini_Fosch_Monticelli_Scala_Pangallo_2025}.
Second, floods induce a liquidity shock by disrupting local labour markets \citep{Groen_Kutzbach_Polivka_2020}. These combined shocks often exceed households' financial capacity, leading to mortgage default \citep{Kousky_Palim_Pan_2020}. Existing research on mortgages distinguishes borrowers who cannot pay from those who can pay but choose to stop \citep{Foote_Willen_2018,Ganong_Noel_2023}. The latter are commonly described as \textit{strategic defaults}: borrowers stop paying because continuing the mortgage has become less attractive, rather than because they face an immediate cash shortfall. Negative equity is one reason for such a decision, although empirical work indicates that income shocks and their interaction with indebtedness explain much observed default \citep{Ganong_Noel_2023}. Repeated floods may shift this balance, because damage affects not only the resources available for repayment but also the value households attach to remaining in their homes.

Understanding the drivers of defaults remains a central challenge  \citep{Foote_Willen_2018}. \textcite{Ganong_Noel_2023} offer a classification of defaults into three distinct categories, based on the cause: the ``Equity'' channel, the ``Liquidity'' channel (cash flow default) and the ``Double trigger'' default, which combines both channels. Understanding strategic versus liquidity-driven defaults remains contested: earlier studies estimated strategic defaults to be $>30\%$ of the total \citep{Gerardi_Herkenhoff_Ohanian_Willen_2018,Guiso_sapienza_zingales2017}, while more recent findings using income trajectories show that only $\sim 6\%$ of underwater defaults are caused by negative equity alone, with the majority being due to liquidity or double-trigger mechanisms \citep{Ganong_Noel_2023}.

Flood damage may introduce a neighbourhood influence: a household that can still afford its mortgage may become less willing to continue paying if the surrounding area is extensively damaged. Increasing evidence has been documented of ``Foreclosure Contagion'', a phenomenon where a homeowner's default probability increases as a reaction to their neighbours. \textcite{Guiso_sapienza_zingales2017} document using survey data how an individual who observes someone in foreclosure is more likely to state they will default, and \textcite{Gupta_2019} provides causal evidence for the same mechanism, finding that each foreclosure within 0.10 miles generates 0.3--0.6 additional foreclosures through price depreciation and borrower reassessment of the stigma cost of default. \textcite{Towe_Lawley_2013} find strong evidence that social interactions influence the decision to default by homeowners who observe a nearby neighbour in foreclosure, while \textcite{Goodstein2017} find the same pattern: the delinquency rate at the zip-code level is a significant factor that drives an increase in strategic defaults.

These two neighbourhood channels have so far been studied separately. Flood damage lowers neighbourhood conditions for every household in the affected area at once, while foreclosure contagion carries distress from one household to the next. Together, flood damage could strengthen the contagion loop until foreclosures become self-sustaining, a tipping point beyond which distress spreads well past the directly damaged properties. Whether this can happen, and how much of flood-related mortgage risk lies outside directly damaged homes, remains unknown. Empirical data offer limited guidance: severe floods are rare, their effects are mixed with those of insurance and public compensation, and the counterfactual without neighbourhood spillovers is never observed.

We address these questions with a calibrated spatial agent-based model (ABM) that lets us observe what empirical data cannot: how flood damage, household finances and foreclosure spillovers combine, and what happens when the spillovers are switched off. We measure four outcomes: (i) the annual foreclosure rate, separated into financial foreclosures, caused by insufficient funds, and strategic foreclosures, driven by neighbourhood conditions and debt pressure; (ii) the amplification of this rate by foreclosure spillovers; (iii) the size distribution of the resulting foreclosure cascades; and (iv) the mortgage age at foreclosure. For lenders, these outcomes show where flood exposure lies beyond the flooded properties, how strongly defaults cluster, and which loans default first (see section~\ref{sec:method} for details). For society, they show where neighbourhood-level recovery could prevent households from losing their homes. The model combines heterogeneous household finances, damage to housing collateral, disruptions to employment, and a discount on the collateral of homes near recent foreclosures. We calibrate the model to Limburg, the Netherlands, using household and property distributions fitted to bank data and national flood-risk maps. We examine stylised flood stress scenarios with progressively heavier tails, identify the parameter region consistent with Dutch arrears and guarantee-fund data, and use global sensitivity analysis (GSA) \citep{bazyleva2024trajectory} to determine which parameters govern credit outcomes.

In the most severe flood scenario the model-implied annual foreclosure rate more than doubles from $0.42\%$ to $0.91\%$, through strategic foreclosures by households that retain the means to pay, while financial foreclosures fall slightly below their no-flood level. At this parametrisation, $37.6\%$ of strategically foreclosed households in the most severe scenario live in homes undamaged in the month of foreclosure, so the risk extends beyond the properties a lender would identify from flood maps alone. Foreclosure spillovers amplify the increase: nearby foreclosures weaken households' willingness to continue servicing their mortgages, potentially triggering further foreclosures. At a moderate spillover intensity, $45\%$ of foreclosures belong to large cascades under the most severe flood scenario, compared with $31\%$ without floods. For lenders, this clustering means that portfolio models assuming independent defaults understate flood-related credit risk. It also limits what conventional early-warning signals can detect. Strategic foreclosures occur at a median mortgage age of $15$ months, after a median of $6$ distress months, whereas financial foreclosures follow $27$ months of distress. Monitoring repeated distress would therefore miss most flood-driven foreclosures. Neighbourhood damage and nearby foreclosures, both observable to a lender, provide an earlier signal. These findings point to neighbourhood-level recovery as a key element of post-flood relief, and identify mortgage spillovers as an early link through which flood damage can propagate into housing markets and longer-term inequality.

\section{Results}
\label{sec:results}
\subsection{A spatial model of flood-exposed mortgage holders}
\label{sec:overview}
\begin{figure}[ht]
    \centering
    \includegraphics[width=\linewidth]{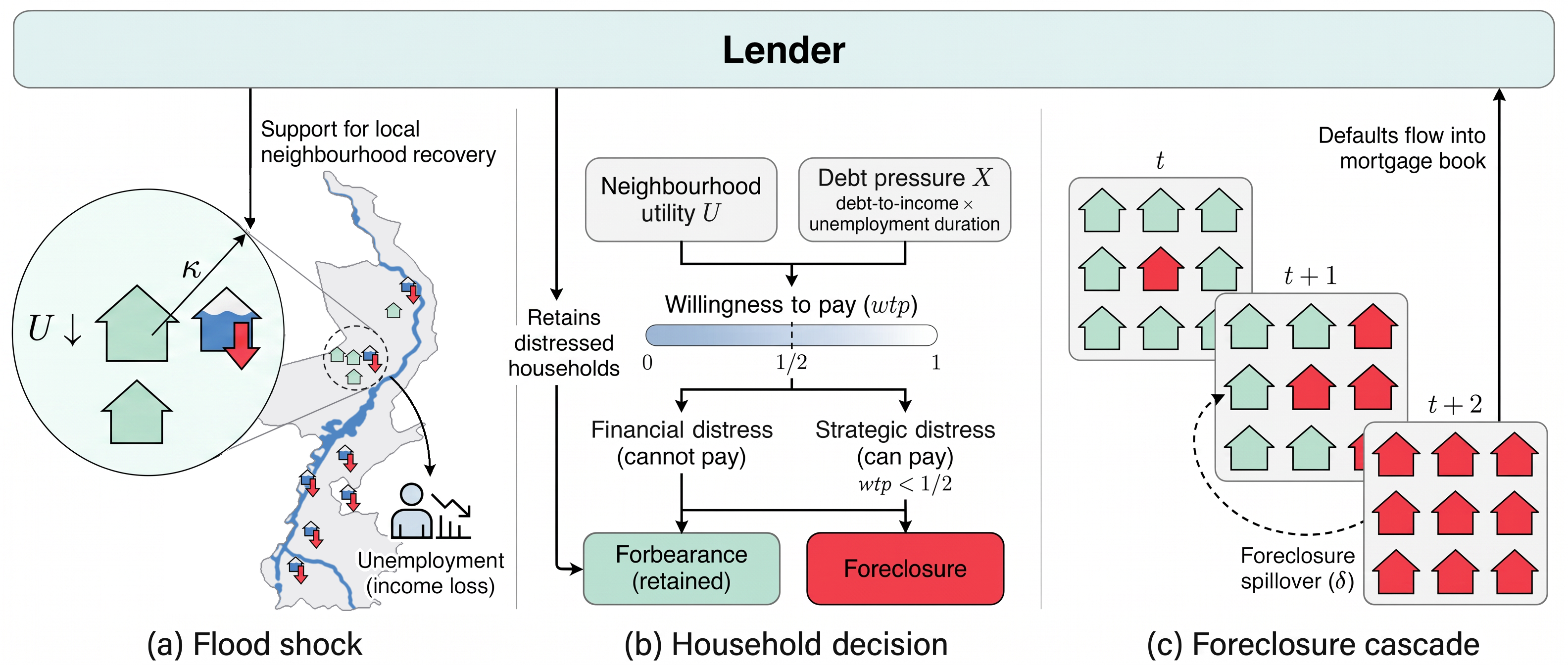}
    \caption{\footnotesize{Schematic model overview: (a) Flood shocks damage collateral values and neighbourhood utility, including for nearby households whose own properties remain undamaged. (b) Neighbourhood deterioration lowers the household's willingness to pay $\text{wtp}_{i,t}$. For financially able households the individual tipping point at $1/2$ separates foreclosure-possible and retention-only regimes. (c) The default cascades consist of increasing numbers of new defaults in one neighbourhood over consecutive time steps. The three vertical arrows link the lender and the households: the arrow in (a) highlights the policy of supporting localised neighbourhood recovery. The central arrow in (b) represents the lender retaining distressed households through forbearance, preventing distress from becoming default, and the arrow in (c) links household defaults to mortgage-book outcomes. Household locations and default counts are only a schematic illustration. The figure was generated with PaperBanana \parencite{Zhu2026PaperBanana}, an AI tool for generating academic illustrations.}}
    \label{fig:model_overview}
\end{figure}

We model how floods affect mortgage repayment through two mechanisms: unemployment and damage to housing collateral (Fig.~\ref{fig:model_overview}). Floods raise unemployment, therefore reducing affected households' income and their capacity to pay back the mortgage.
Second, floods lower the collateral value of exposed homes and worsen neighbourhood conditions that influence the households’ willingness to pay even when their own properties remain undamaged.
In our model a default can happen for two reasons. A financial default occurs when a household can no longer pay its mortgage. A strategic default occurs when a household can still pay but no longer wants to.

Whether a household wants to keep paying is summarised by its willingness to pay. The willingness to pay is decomposed into two measures. The first is the state of the neighbourhood; floods lower the value of homes used as collateral for mortgages, and we measure the condition of a neighbourhood by the combined value of nearby homes relative to their undamaged value. The second is debt pressure, the value of which reflects the outstanding debt relative to income and the length of any unemployment. Willingness to pay rises with neighbourhood utility and falls with debt pressure. A household that can still pay defaults strategically only when its willingness to pay falls below a threshold. 

Foreclosures can themselves weaken neighbouring households' willingness to continue paying, creating a feedback through which one foreclosure can trigger others. We call this feedback the foreclosure spillover. Its intensity, $\delta$, controls how strongly foreclosures in the previous month affect households' willingness to pay through neighbourhood utility. The model represents this effect through a reduction in nearby collateral values. Either financial or strategic foreclosures can initiate a cascade, which then spreads through strategic defaults and can reach households whose own homes are undamaged.

In our model households have different income, liquid funds, loan size and employment status. Each month they meet living expenses and mortgage payments. Floods also raise unemployment, reducing the income of affected households. Distress does not necessarily lead to default: a household can be retained through forbearance or lose its home through foreclosure, with the willingness to pay governing the choice.

We apply the model to the region of Limburg in the Netherlands, using Dutch flood-risk maps and household characteristics drawn from distributions fitted to banking and national statistics data. We test the model under a benchmark without floods, $\mathcal{S}_0$, and four stylised flood stress scenarios, $\mathcal{S}_1$ to $\mathcal{S}_4$, with progressively heavier tails in flood severity. Comparing simulations with and without foreclosure spillovers allows us to assess how this feedback contributes to defaults. We ran a Global Sensitivity Analysis (GSA) of the most important parameters of the model \citep{Saltelli_Aleksankina_Becker_Fennell_Ferretti_Holst_Li_Wu_2019}.
We report the full specification of the model in the Methods section and the parameter values in \SItab{tab:model_params} of the Supplementary Information (SI).

\FloatBarrier
\subsection{Model calibration and flood-driven strategic foreclosures}
\label{sec:empirical_validation}
\begin{figure}[htbp]
    \centering
    \includegraphics[width=\textwidth]{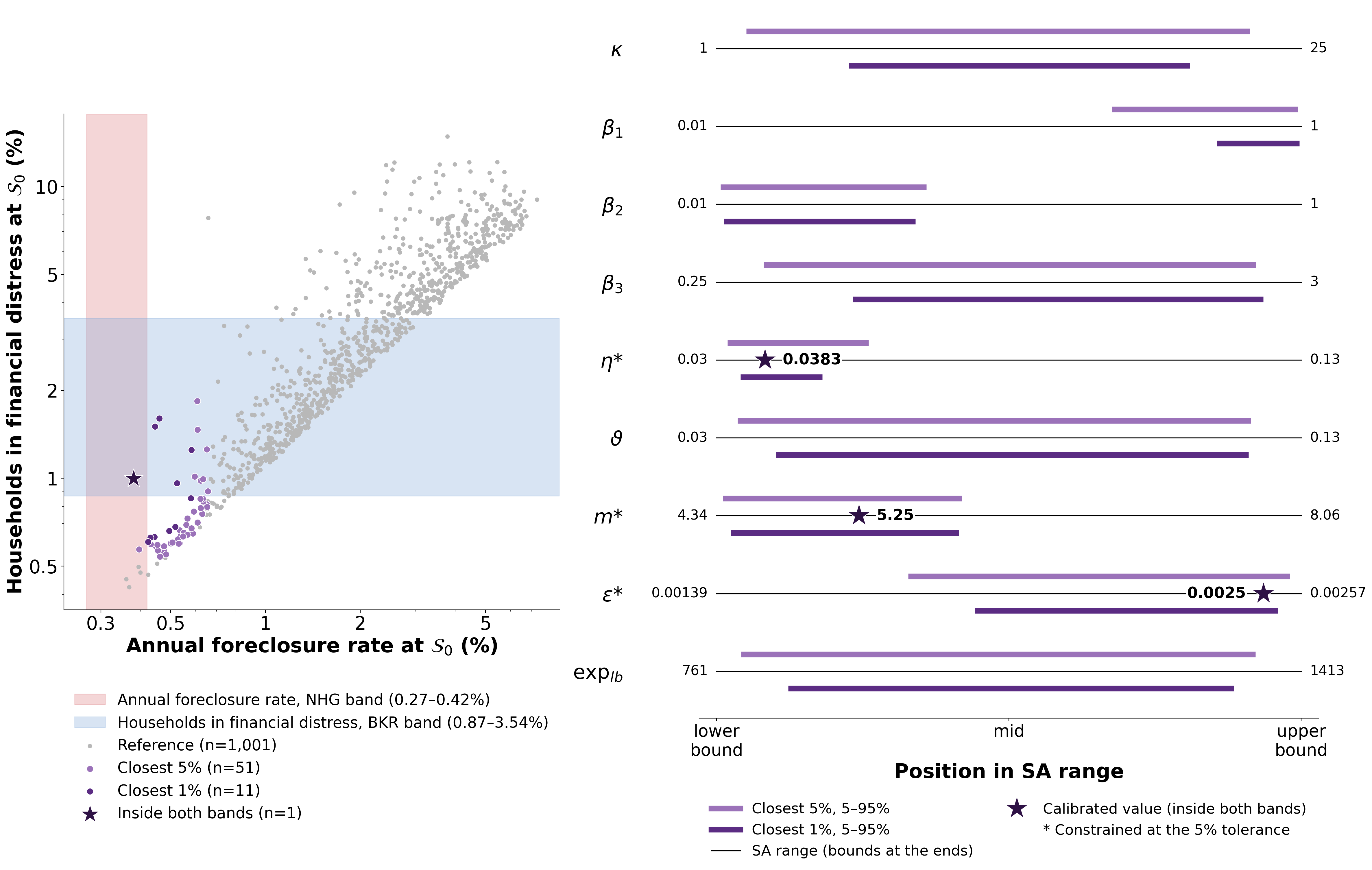}
    \caption{\small Calibration of the model to the empirical credit outcomes at $\mathcal{S}_0$ with $\delta=0$, by rejection approximate Bayesian computation. The reference is a sample of the prior: $1{,}001$ parameter combinations of a scrambled Sobol' sequence, uniform on the GSA ranges of the nine parameters active at $\mathcal{S}_0$ and restricted to $\beta_1-\beta_2\ge 1/2$, each averaged over $10$ seeds. The distance of a combination to the data is defined in \SIeq{eqn:abc_distance}; the $5\%$ and $1\%$ of combinations with the smallest distance are retained. Left: annual foreclosure rate and share of households in financial distress for every combination (grey), the retained combinations (purple) and the combination inside both bands (star), with the NHG band of the annual foreclosure rate (red) and the BKR band of the share of households in financial distress (blue); both axes are logarithmic. Right: position of each parameter in its sampled range (\SItab{tab:sa_ranges}). The black line spans the sampled range, with its lower and upper bounds printed at its ends, and the purple bars span the $5$th to $95$th percentile of the retained combinations at the $5\%$ tolerance (above the line) and the $1\%$ tolerance (below it). Parameters marked with $*$ are constrained by the calibration at the $5\%$ tolerance (\SIsec{sec:app:identification} and \SItab{tab:abc_ks}); for them the star marks the calibrated value, that of the combination inside both bands, used in \SItab{tab:model_params}.}
    \label{fig:abc_identification}
\end{figure}

Before analysing the effects of floods on the system, we compare the annual foreclosure rate and the percentage of households in financial distress under $\mathcal{S}_0$ with empirical data and use them to calibrate the parameters of the model (Figure~\ref{fig:abc_identification}). Every flood effect we report is a change relative to $\mathcal{S}_0$, so the comparison is only meaningful if the model without floods produces foreclosure and distress levels close to those observed. We evaluate $\mathcal{S}_0$ without the foreclosure spillover ($\delta=0$), so that the comparison isolates the financial distress mechanism and the initialisation from the foreclosure externality, whose magnitude we treat as a stress dimension for flooded markets (Section~\ref{sec:neighbourhood}).

The closest empirical counterpart to the foreclosure rate in our model is the loss claims received by the Dutch mortgage guarantee fund (NHG): $0.27$--$0.42\%$ of active guarantees per year in 2013--2016 \citep{NHG_2014_2016}. These cover forced sales with insufficient proceeds to repay the guaranteed mortgage, providing a lower-bound benchmark. For financial distress, Dutch credit registry data report mortgages in arrears for $0.87$--$3.54\%$ of households holding a mortgage \citep{BKR2025,CBS83834NED}.

\begin{figure}[htbp]
    \centering
    \captionsetup[subfigure]{font=footnotesize}
    \begin{subfigure}[t]{0.48\textwidth}
        \centering
        \includegraphics[width=\linewidth]{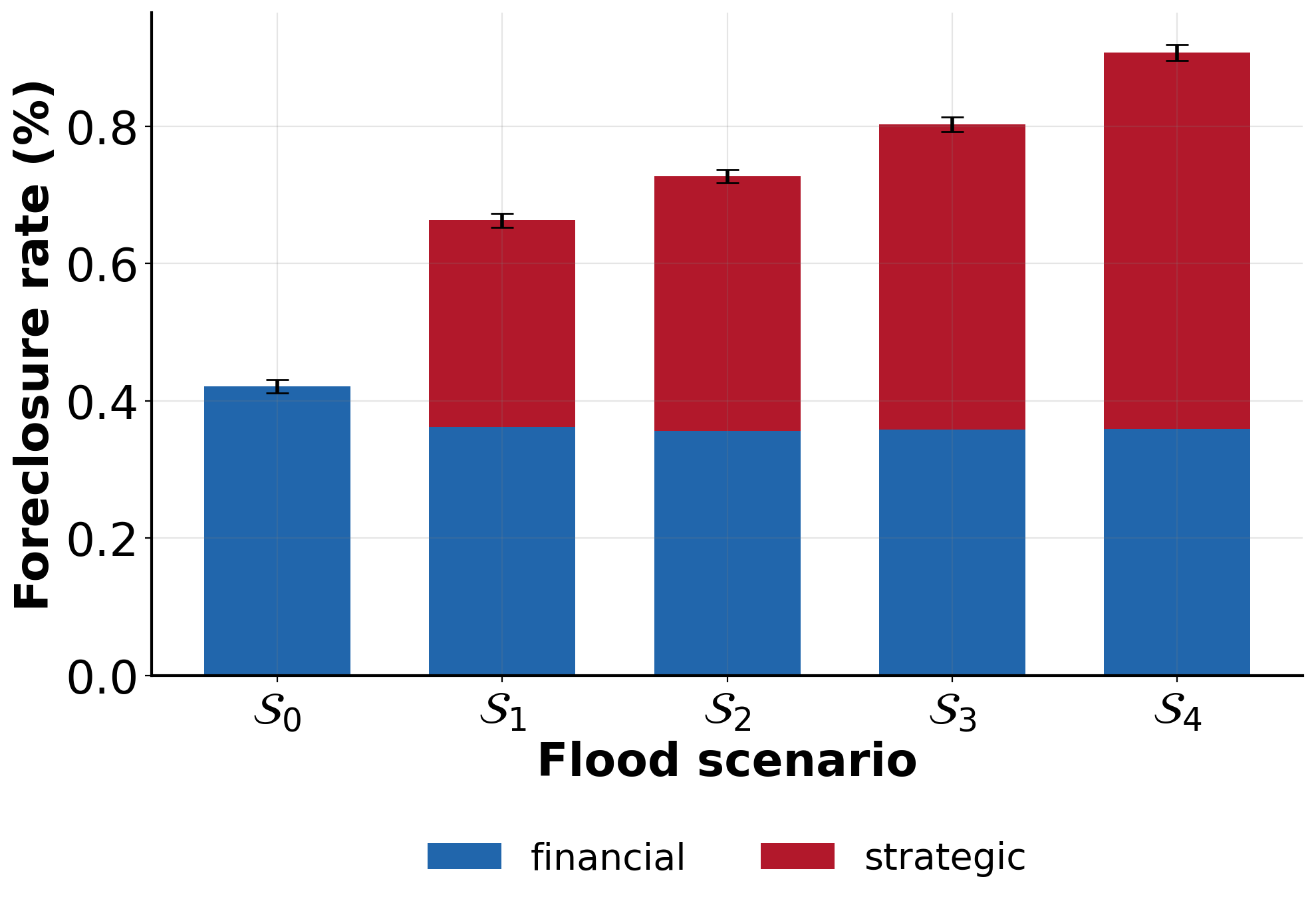}
        \caption{Annual foreclosure rate at the calibrated parametrisation.}
        \label{fig:annual_rate_channel}
    \end{subfigure}\hfill
    \begin{subfigure}[t]{0.50\textwidth}
        \centering
        \includegraphics[width=\linewidth]{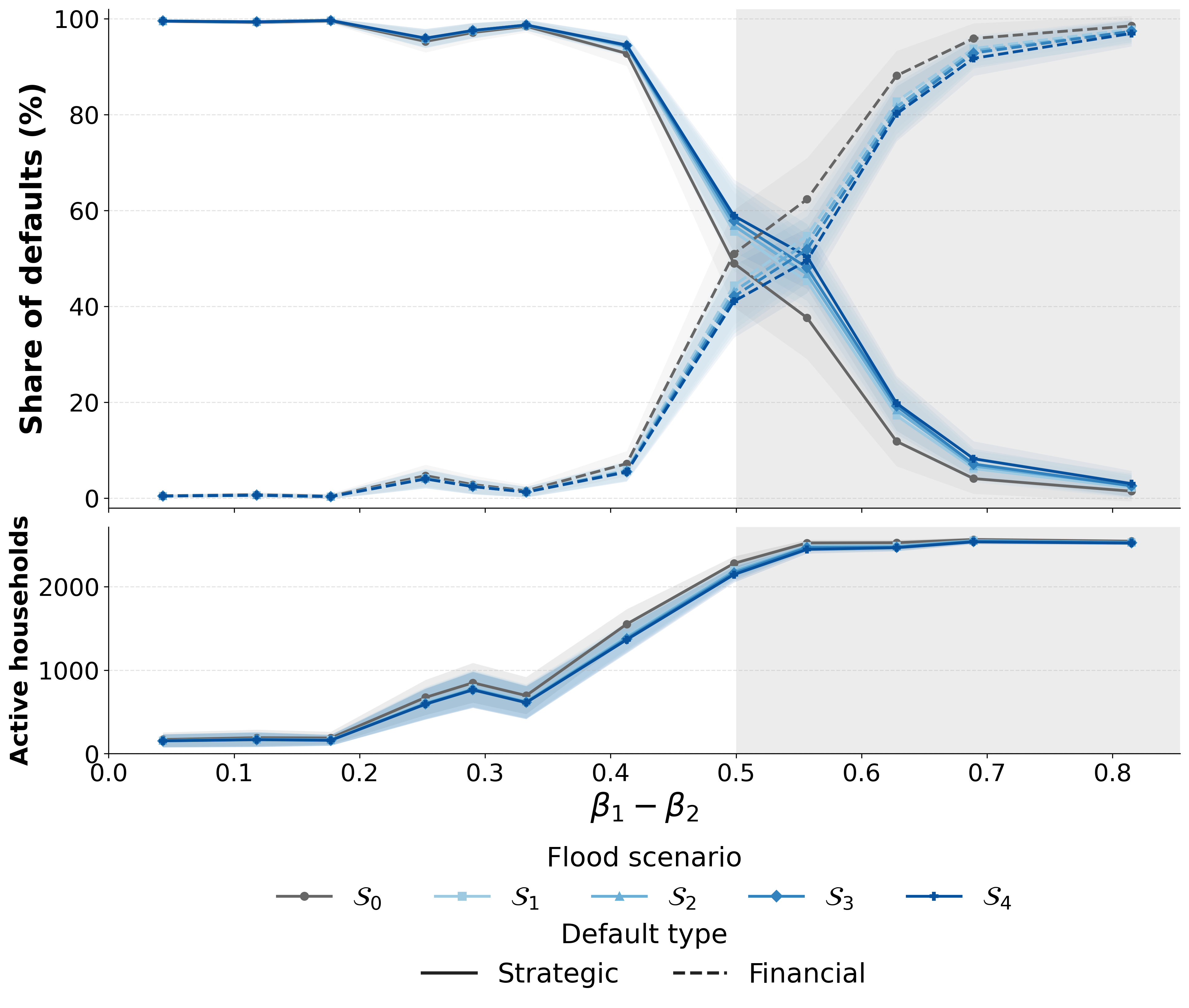}
        \caption{Default composition and active mortgages across the gap.}
        \label{fig:defaults_gap}
    \end{subfigure}
    \caption{\footnotesize Foreclosure rates, default composition and active mortgage population by flood scenario, over the $300$ months after a burn-in of $100$ months. (a) Annual foreclosures per $100$ active mortgage holders, split into financial and strategic channels, at the calibrated parametrisation ($\delta=0$, $\kappa=10$; \SItab{tab:model_params}). Each flood scenario pools twenty seeds for each $\lambda\in\{0.25,0.5,0.75\}$; $\mathcal{S}_0$, where $\lambda$ has no effect, uses twenty seeds. Error bars are $95\%$ Student-$t$ intervals for the mean total rate over the $20$ seeds, with $\lambda$ averaged within seed. Across $\lambda=0.25$, $0.5$ and $0.75$ the total rate is $0.49\%$, $0.65\%$ and $0.84\%$ under $\mathcal{S}_1$ and $0.57\%$, $0.88\%$ and $1.27\%$ under $\mathcal{S}_4$. (b) The $786$ GSA parameter vectors with $\delta\le0.15$, of which $291$ have $\beta_1-\beta_2\ge1/2$, divided into twelve equal-count bins of $\beta_1-\beta_2$ (\SIsec{sec:app:sa}). Top: strategic (solid) and financial (dashed) shares of all defaults in each bin, with $95\%$ confidence intervals for the ratio. Bottom: mean active mortgage population, averaged across seeds and then parameter vectors, with $95\%$ confidence intervals for the mean; the dashed line marks $1{,}000$ households. Grey shading marks $\beta_1-\beta_2\ge1/2$ (Section~\ref{sec:empirical_validation}).}
    \label{fig:foreclosure_composition}
\end{figure}

We restrict the analysis to $\beta_1-\beta_2\ge 1/2$ (see Methods, Equation~\ref{eqn:wtp}) because below this threshold strategic defaults dominate and the active mortgage population shrinks sharply even without floods (see Figure~\ref{fig:foreclosure_composition}, right panel). Without floods and without the spillover, neighbourhood utility remains at its baseline level.

The calibration therefore constrains the financial channel. The GSA identifies the expenditure parameters $m$ and $\varepsilon$ as influencing financial foreclosure counts, with little influence on strategic counts (\SIsec{sec:app:gsa_defaults}).
We calibrate the model by rejection Approximate Bayesian Computation (ABC), retaining the parameter combinations whose outcomes lie closest to the two empirical bands \parencite{Grazzini_Richiardi_Tsionas_2017}. The ABC calibration constrains these two expenditure parameters together with baseline unemployment $\eta$ (Figure~\ref{fig:abc_identification}). One sampled combination places both outcomes inside their bands; we use its values of these three parameters in the calibrated parametrisation (\SItab{tab:model_params}). The parameters governing strategic foreclosures remain unconstrained by these data. We vary them in the results, keeping other parameters at their baseline values unless stated otherwise.
At $\beta_1=1$, $\beta_2=0.5$, the values used in Figure~\ref{fig:annual_rate_channel}, the annual foreclosure rate without floods is $0.42\%$, and the share of households in financial distress is $0.51\%$. The calibration procedure and robustness checks are in \SIsec{sec:app:identification}, and the GSA results are in \SIsec{sec:app:gsa_defaults}.

Flooding raises the total annual foreclosure rate at almost every sampled parametrisation. At the calibrated parametrisation, even without foreclosure spillovers, the rate more than doubles, from $0.42\%$ without floods to $0.91\%$ in the most severe scenario. This increase is driven by strategic foreclosures, while financial foreclosures fall slightly (Figure~\ref{fig:annual_rate_channel}; detailed rates and the comparison across GSA parameter vectors are in \SIsec{sec:app:gsa_defaults}). The increase is concentrated in the floodplain of the river Meuse. Without floods, the local foreclosure rate is nearly uniform across the region; in the most severe scenario it rises about fivefold inside the flood extent and by less than half outside it (\SIfig{fig:combined_maps}, panels (c) and (d)).

The balance between financial and strategic defaults depends on the gap between the neighbourhood-utility loading and the maximum debt-pressure penalty, $\beta_1-\beta_2$ (Figure~\ref{fig:defaults_gap}). Strategic defaults dominate at narrow gaps, while financial defaults dominate at wider gaps. Floods increase the strategic share near $\beta_1-\beta_2=1/2$, where neighbourhood deterioration can push households across the strategic-default threshold. For the sampled combinations with $\beta_1-\beta_2\ge1/2$ and $\delta\le0.15$, the strategic share rises from $20\%$ without floods to $28\%$ under the most severe scenario. Detailed comparisons of default composition, active mortgage populations and spillover intensities are in \SIsec{sec:app:gsa_defaults}.

The GSA combines Sobol' \citep{Sobol_2001} and PAWN \citep{Pianosi_Wagener_2015} indices over $\delta\in[0.05,0.5]$, with active foreclosure feedback (Sections~\ref{sec:app:gsa_method} and~\ref{sec:app:gsa_defaults}, Figs.~\ref{fig:sa_perc_fin}--\ref{fig:sa_n_strategic}). Both measures identify $\beta_1$ as the leading parameter for the financial share and both counts in every flood scenario. Sobol' places $\beta_2$ second, while under PAWN it stays below the permutation threshold. The consumption intercept $m$ moves the financial count but not the strategic one; the spillover intensity $\delta$ and its radius $\kappa$ move the strategic count. The damage share $\lambda$ has no detectable influence on these outcomes over the sampled range of $\delta$.

\FloatBarrier
\subsection{Floods increase large foreclosure cascades}
\label{sec:cascades}

Foreclosure spillovers generate a distinct pattern of defaults: foreclosures are concentrated in isolated events and large cascades, with relatively few cascades of intermediate size. Floods increase the number of large cascades most at intermediate spillover intensities, without increasing their typical size. Foreclosure spillovers have empirical support: nearby foreclosures depress property prices, and foreclosure contagion can operate through both price declines and a lower perceived cost of default \parencite{Campbell_Giglio_Pathak_2011,Gupta_2019}.

We measure the amplification caused by foreclosure spillovers using the foreclosure-rate multiplier $M=\Phi(\delta)/\Phi(0)$, where $\Phi$ is the annual foreclosure rate per $100$ active mortgage holders. The numerator and denominator refer to simulations with and without spillovers under the same flood scenario. A multiplier of one indicates no net amplification, while a multiplier of two means that spillovers double the foreclosure rate. As in epidemic transmission, an initial event can trigger successive generations of events, amplifying the total impact. Here, $M$ measures the aggregate rate amplification; its relation to a reproduction number and the assumptions needed to interpret it as an expected cascade size are given in \SIsec{sec:app:cascade_method}.

\begin{figure}[htbp]
    \centering
    \includegraphics[width=\textwidth]{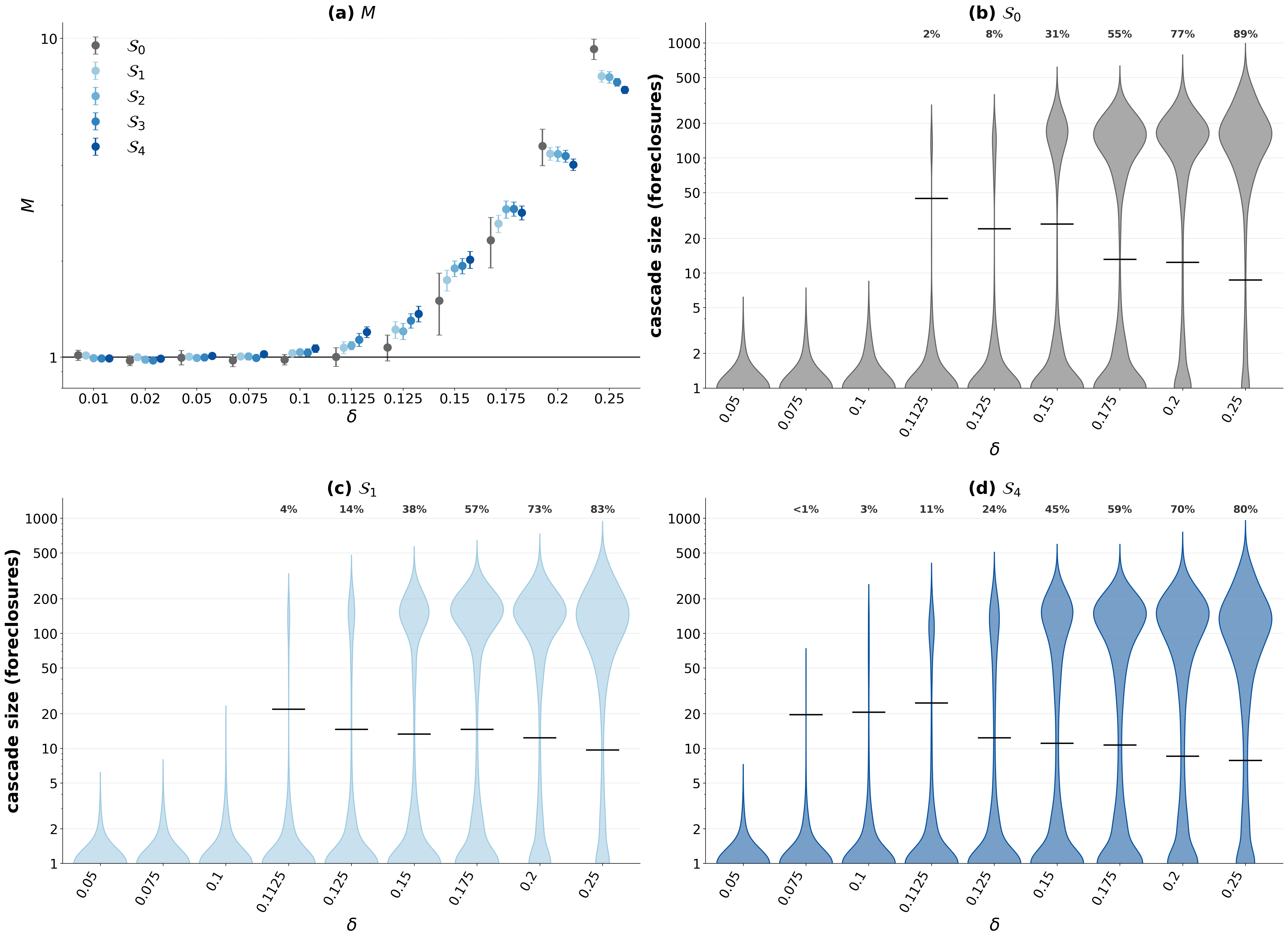}
    \caption{\footnotesize Foreclosure feedback and default cascades against the spillover intensity $\delta$. (a) Foreclosure-rate multiplier $M=\Phi(\delta)/\Phi(0)$ by flood scenario, on a logarithmic axis; $M=1$ indicates no net amplification. Points are means over $20$ seeds with $95\%$ Student-$t$ intervals, with the damage share $\lambda\in\{0.25,0.5,0.75\}$ averaged within seed; the percentage change in $M$ relative to the no-flood scenario is in \SIfig{fig:multiplier}. (b) to (d) Distribution of default cascade sizes without floods ($\mathcal{S}_0$), under $\mathcal{S}_1$ and under $\mathcal{S}_4$: each violin shows the share of foreclosures belonging to cascades of each size (logarithmic axis), scaled to its own maximum; the horizontal tick separates isolated foreclosures from large cascades, and the number above each violin is the share of foreclosures in large cascades. Cascade reconstruction in \SIsec{sec:app:cascade_size}. Calibrated parametrisation (\SItab{tab:model_params}) with $\kappa=10$, $\beta_1-\beta_2=1/2$ and $\vartheta=0.05$; rates and cascades over the $300$ months after a burn-in of $100$ months.}
    \label{fig:r_net}
\end{figure}

Without floods, the multiplier remains close to one at low spillover intensities (Figure~\ref{fig:r_net}, panel (a)). From $\delta=0.15$ it rises sharply, from $1.50$ at $\delta=0.15$ to $4.59$ at $\delta=0.2$ (\SItab{tab:cascade_gross_net}). Flooding increases this amplification most at intermediate spillover intensities: at $\delta=0.15$, the multiplier ranges from $1.74$ to $2.02$ across flood scenarios. The proportional change relative to the no-flood multiplier is small or negative where the feedback is already strong (\SIfig{fig:multiplier}); this does not imply a lower total foreclosure rate, because flooding also raises the rate without spillovers. At high spillover intensities, the lower multiplier is consistent with a saturation effect: foreclosure feedback already removes many susceptible households without floods, leaving fewer additional households for flood damage to push into foreclosure.

This multiplier does not show how the additional foreclosures arise. They could be spread evenly, or they could come from a few large default cascades. We reconstruct the shape of the cascades using the model's output (\SIsec{sec:app:cascade_size}). A foreclosure either stays isolated or belongs to a large default cascade, which in our model typically ranges between $100$ and $200$ foreclosures, with few cascades of intermediate size. Figure~\ref{fig:r_net} (b) to (d) shows the distribution of the cascades quantified by the number of defaulters that were part of each of them. The spillover intensity changes how many foreclosures belong to large cascades, not how large the cascades are: without floods, the share rises from $8\%$ at $\delta=0.125$ to $77\%$ at $\delta=0.2$. We check the robustness of this pattern using two attribution-based reconstructions, weighted trees and connected components, and an attribution-free method that groups foreclosures by proximity in space and time. All three recover similar large-cascade distributions, while the attribution-free method finds no large-cascade mode in the runs without spillovers (\SIsec{sec:app:cascade_size}).

Floods both initiate foreclosures and lower neighbouring households' willingness to pay, allowing the foreclosure feedback to propagate distress more readily. Each reconstructed cascade starts from a foreclosure that would also have occurred that month without the spillover, given the household's current state. Floods increase these starting points, while the feedback decides whether a starting point grows into a large cascade. Flood damage lowers neighbourhood utility, bringing households closer to the threshold for strategic foreclosure. The additional reduction caused by nearby foreclosures can then push them across that threshold. Floods can therefore facilitate propagation even though the spillover intensity $\delta$ remains unchanged.

Where the feedback is weak, the additional foreclosures that floods cause remain isolated. Where it is moderate, a larger share of foreclosures belongs to large cascades: at $\delta=0.15$, this share is $45\%$ under the most severe flood scenario, against $31\%$ without floods (Figure~\ref{fig:r_net}, panels (b) and (d)). Where the feedback is strong, large cascades occur with or without floods, and floods mainly add isolated foreclosures. Floods therefore increase the number of large foreclosure cascades without increasing their typical size.

The neighbourhood radius $\kappa$ and the gap $\beta_1-\beta_2$ determine where these regimes lie; a larger radius and a narrower gap bring the moderate regime to lower spillover intensities, and the same concentration of flood-induced amplification at intermediate spillover intensities holds across them, except at the smallest radius, where no large cascades form (\SIfig{fig:tipping_sensitivity}). For a lender, the additional foreclosures arrive as infrequent, concentrated events rather than as a uniform increase in the foreclosure rate. Floods matter most in markets where the feedback can sustain large cascades but rarely does, because there they turn occasional cascades into frequent ones.

\FloatBarrier
\subsection{Floods shift foreclosures towards younger mortgages}
\label{sec:forbearance_spells}
\begin{figure}[htbp]
    \centering
    \includegraphics[width=\textwidth, keepaspectratio]{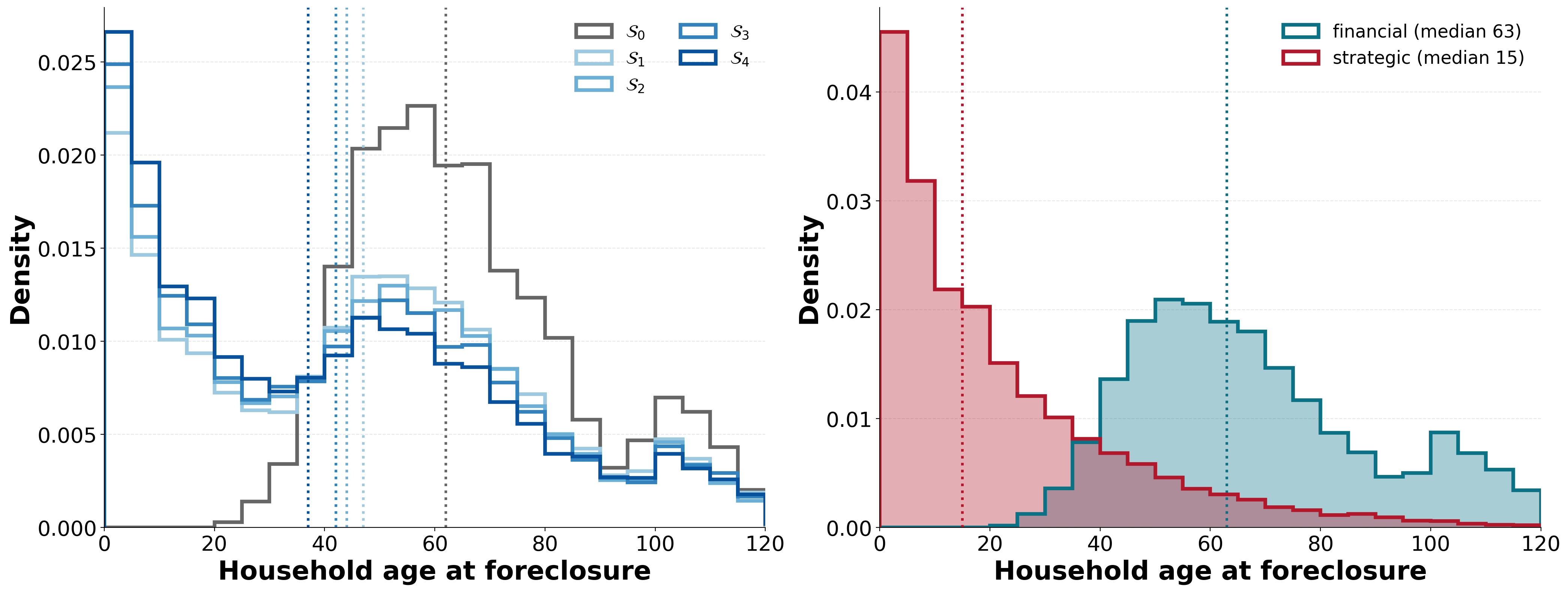}
    \caption{\small Household age at foreclosure, $a_{i,t}$, measured in months of the mortgage life. The model's standard mortgage term is $120$ months ($10$ years). Panel (a) pools default channels and separates flood scenarios; panel (b) separates financial and strategic foreclosures, pooling $\mathcal{S}_1$--$\mathcal{S}_4$. Histograms are normalised separately; dotted lines mark medians. The sample contains $68{,}927$ foreclosures recorded after warm-up at steps $100$--$399$, from ten seeds for each scenario and each value of the damage share $\lambda \in \{0.25, 0.5, 0.75\}$, with the three values of $\lambda$ pooled. Parameters are at the calibrated parametrisation: $\beta_1=1.0$, $\beta_2=0.5$, $\beta_3=1.0$, $\delta=0$, $\kappa=10$, $\eta=0.0383$, $\vartheta=0.05$, $m=5.25$, $\varepsilon=0.00250$ and $\mathit{exp}_{lb}=1087$ (\SItab{tab:model_params}). Mortgage age is measured from origination, including time before the observation window.}
    \label{fig:foreclosure_time}
\end{figure}

The impact of floods shifts foreclosures towards the early stages of the mortgage. At the calibrated parametrisation without foreclosure spillovers, the median mortgage age at foreclosure falls from $62$ months without floods to $37$ months under the most severe flood scenario (Figure~\ref{fig:foreclosure_time}, panel (a)). This shift mainly reflects an increasing share of strategic foreclosures, which occur earlier than financial foreclosures.

Strategic foreclosures occur at a median mortgage age of $15$ months, compared with $63$ months for financial foreclosures (Figure~\ref{fig:foreclosure_time}, panel (b)). Early in the mortgage, outstanding debt is high, increasing debt pressure and leaving households more susceptible to neighbourhood deterioration. Flood damage can therefore push willingness to pay below the strategic-default threshold while households still retain the means to pay.

Across flooded scenarios, the median age varies little; financial foreclosures occur at $62$--$64$ months and strategic foreclosures at $14$--$15$ months. Strategic defaults account for none of the foreclosures without floods, but for $60\%$ under the most severe flood scenario. Flooding therefore shifts the composition of foreclosures towards early strategic exits. Greater losses of collateral value strengthen this shift and also lower the age of strategic foreclosure (\SIfig{fig:foreclosure_time_lambda}). The distress months preceding foreclosure in each channel are reported in \SIsec{sec:app:fcltime}, and their sensitivity to the parameters in \SIsec{sec:appendix_sobol_distress}.
For lenders, the shift towards younger mortgages means that flood-related foreclosures occur earlier in repayment, when less principal has been repaid. The outstanding balance at foreclosure rises from $48\%$ of the original loan without floods to $68\%$ under the most severe flood scenario, and strategic foreclosures occur with $83\%$ of the loan still outstanding. Floods therefore affect both the number of foreclosures and the outstanding debt exposed at foreclosure.

\section{Discussion}
\label{sec:discussion}
Flood damage to surrounding properties can push households that can still afford their mortgage into foreclosure. In a simulated portfolio of Limburg mortgage holders, flooding more than doubles the annual foreclosure rate from 0.42\% to 0.91\% in the most severe scenario, even without foreclosure spillover. The increase comes from strategic foreclosures, which account for 60\% of foreclosures in that scenario and occur at a median mortgage age of 15 months, compared to 63 months for financial foreclosures.

Our results connect three strands of work. Empirical studies of floods and hurricanes link property-level damage to mortgage distress, with stronger effects where damage is uninsured or compensation is limited
\citep{gallagher2017household} %
. These studies identify the effect of a household's own damage. We add a channel that runs through damage to surrounding properties and also reaches households whose home was not flooded. Second, the foreclosure externality literature shows that foreclosures affect nearby house prices \citep{Campbell_Giglio_Pathak_2011, Gerardi_Rosenblatt_Willen_Yao_2015}, and
survey evidence suggests that knowing others who defaulted strategically lowers willingness to pay \citep{Guiso_sapienza_zingales2017}. We do not estimate such spillovers. We treat their intensity $\delta$ as a parameter and examine how a flood shock interacts with it. Third, climate stress tests and flood-scenario analyses of financial stability assess portfolio losses from physical risk \citep{battiston2017climate, lepore2024beyond, fitzmaurice2026flood}. These studies focus on own-property damage, collateral values and borrower finances rather than on neighbourhood-level damage and foreclosure spillovers. The new element in our model is the combination of these channels. Neighbourhood flood damage lowers utility for all nearby households, which generates strategic foreclosures among borrowers who can still pay, and these foreclosures can then propagate as cascades. The model's value lies in this mechanism and in its implications for how risk should be measured. 

For lenders, the first implication is that flood exposure extends beyond the flood footprint. In the model, $39\%$ of strategic foreclosures across the four flood scenarios occur in homes undamaged in the month of foreclosure: damage to nearby properties lowers neighbourhood utility for every household in the area. Screening a portfolio on each property’s own flood exposure therefore misses part of the risk. A neighbourhood-level measure, such as the share of collateral value damaged around each property, captures it.

The second implication concerns how defaults cluster. With foreclosure spillovers, a foreclosure either stays isolated or belongs to a cascade of $100$ to $200$ foreclosures, and flooding makes these cascades more frequent; at a moderate spillover intensity ($\delta = 0.15$), the share of foreclosures in large cascades rises from 31\% without floods to 45\% in the most severe scenario. Two portfolios with the same average foreclosure rate can therefore differ widely in their largest losses. Stress tests and capital plans that reproduce average default rates but assume independent defaults underestimate losses that are concentrated in a few months and a few neighbourhoods. Flood-driven foreclosures also occur earlier in the mortgage, when less principal has been repaid, so the outstanding balance at foreclosure is higher.

The third implication concerns early warning. Monitoring based on arrears is designed for borrowers who run out of funds, whereas strategic foreclosures involve households that retain the means to pay. The model points to three indicators that a lender can observe before such foreclosures occur: damage to surrounding properties after a flood; foreclosures nearby in the preceding month, from which cascades spread; and loan characteristics, since strategic foreclosures are concentrated among recently originated mortgages with high debt relative to income. We have not tested how well these indicators predict foreclosure; doing so against loan-level data is a natural next step.

For policymakers, the results support relief at the scale of the neighbourhood. In the model, damage to surrounding properties triggers strategic foreclosures, and repairing it restores neighbourhood utility for every nearby household, flooded or not. Coordinated support for the repair of neighbouring properties can therefore complement assistance to individual households, reducing both the foreclosures that start cascades and the conditions that allow them to spread. We do not simulate relief programmes here; comparing neighbourhood-level and household-level relief is a priority for future work.

These effects matter beyond lenders’ balance sheets. A foreclosure costs a household its home and its accumulated equity, at a time when affordable alternatives are scarce. The model holds house prices and the flow of new borrowers fixed, so it does not represent what follows. Extending it with endogenous house prices, household relocation and credit supply that responds to flood risk would allow the same framework to examine whether repeated floods sort lower-income households into higher-risk areas (climate gentrification) \citep{Keenan_Hill_Gumber_2018, deKoning_Filatova_2020}, whether lost housing wealth and restricted credit hold households in poverty traps \citep{dupont2025emergent, garibay2026pricing}, and how these processes change wealth inequality.

The contribution of the model is a transmission channel linking neighbourhood flood damage, repayment decisions and foreclosure spillovers. The quantitative results are conditional on the Limburg application and the stylised flood stress scenarios; their magnitude should not be assumed to carry over to other mortgage markets. Three limitations deserve emphasis. First, the spillover intensity $\delta$ is not constrained by the calibration data, so the cascade results describe how outcomes depend on $\delta$; they do not predict its value in Limburg. Second, the model does not represent flood insurance or public compensation, which would reduce the damage households bear. The empirical evidence suggests that insurance coverage is practically limited to about one third of the estimated loss (Section~\ref{sec:flood_impact}). Third, it does not represent alternative mortgage recourse rules governing borrowers' liability for debt remaining after foreclosure. It therefore cannot establish how differences in these rules affect willingness to pay, the strength of local spillovers or lenders' eventual recoveries. Applying the framework elsewhere would require local calibration and an explicit treatment of the relevant recourse arrangements to assess whether the same transmission channel remains quantitatively important.
\section{Methods}
\label{sec:method}
We build a spatial agent-based model, in which outcomes at the level of the mortgage market emerge from the decisions and interactions of heterogeneous households rather than from a representative agent. See \citet{axtell_farmer2025_jel} for a review on agent-based modelling in economics and finance.

\subsection{Model structure and time step}
\label{sec:update}
The model is simulated in discrete monthly steps. Each step begins with an update to the environment, where a flood severity value $\mathit{sev}_t$ is generated, and the global unemployment rate is adjusted accordingly (Equation~\ref{eqn:unemployment}). Following this, the model calculates the number of potential new households that join the simulation; we cap the total number of agents at $6{,}000$ active mortgages for computational constraints. If the active population is at the bound no new applicants are generated. We assume a stationary flow of mortgage applicants: a fixed number of candidates arrives each period,
\begin{equation}
\label{eqn:number_new_join}
N^{\text{created}} = \rho \cdot N_{nj} ,
\end{equation}
where $N_{nj}$ is the maximum number of new joiners per step and $\rho$ is a density scaling factor. Both are fixed parameters; entry demand is considered exogenous. 
At every step the available funds are updated through: $w_{i,t+1} = w_{i,t} + I_{i,t} - E_{i,t}$ (Section~\ref{sec:income_empl_dynamics}). 
The agents update the number of instalments $n_{i,t}$ decreasing it by one each month the instalment is paid. They also update the $\LTV$ ratio accordingly:
$\LTV_{i,t} =  \frac{P_i \cdot n_{i,t}}{c_{i,t}}$ (Equation~\ref{eqn:collateral}).
In the model updates each agent's financial and employment status are updated and default risk is assessed(Section~\ref{sec:distress}). After all agents have acted, those who have defaulted or whose mortgage term has matured are removed from the simulation.

\subsection{Flood scenarios}
\label{sec:environment}
\label{sec:flood_method}
We model unitless flood severity, $\mathit{sev}$, using a Generalised Extreme Value (GEV) distribution, a standard tool for characterising hydrological extremes \citep{Regional_parent_flood_frequency_distributions, Villarini_Smith_2010, morrison_smith_2002}. Each flood scenario is defined by a parameter vector $\mathcal{S}=(\mu,\sigma,\xi)^\top$, where $\mu$, $\sigma$, and $\xi$ denote location, scale, and shape of the distribution, respectively. We draw severity at each monthly time step $t$, matching the frequency of mortgage payments:
\begin{equation}
    \mathit{sev}_t \sim GEV(\mathcal{S}).
\end{equation}
The distribution's functional form is reported in \SIsec{sec:app:gev_construction}, \SIeq{eqn:cdf_gev}.

We consider five scenarios. Scenario $\mathcal{S}_0$ represents the absence of floods, while $\mathcal{S}_1$--$\mathcal{S}_4$ use Fréchet-type GEV distributions with shape parameters $\xi=0.1$, $0.5$, $0.8$, and $1.2$, respectively. Across the four flood scenarios, we adjust location and scale to approximately preserve the mode of the density peak while increasing the probability of high-severity floods in the tail (\SIfig{fig:gev_distrib}). The parameter choices and construction are detailed in \SIsec{sec:app:gev_construction}.

We treat $\mathcal{S}_1$ as the baseline and $\mathcal{S}_2$--$\mathcal{S}_4$ as stylised stress scenarios, used to examine mortgage vulnerability under progressively heavier-tailed flood regimes rather than as calibrated projections of future climate conditions. This approach follows the emphasis of the Basel Committee and the Network for Greening the Financial System on adverse climate scenarios as tools for exploring financial risk beyond central or most likely pathways \citep{BCBS2024,NGFS2025}.

The impact of floods in the model is twofold: the intensity $\mathit{sev}$ causes a wealth shock via the property value, and triggers unemployment for the agents. 
For the property value shock, the unitless draw of $\mathit{sev}$ selects one flood damage map for the entire model at each monthly step. The selected map determines the spatial extent of flooding and the water depth at each location, so the same flood realisation produces different damages to properties across households. We use four maps from the Dutch national flood risk maps (Overstromingsrisicokaarten, \citealp{IPO_2019_overstromingsrisicokaarten}; \SIfig{fig:combined_maps}, panel (b)), preprocessed to the model grid as described in \SIsec{sec:appendix_map_preprocessing}. The maps correspond to annual flood-probability categories ranging from frequent to extremely rare \citep{IPO_risicokaart_overstroming}. Lower values of $\mathit{sev}$ select the lower-damage maps, while higher values select the more destructive maps. The thresholds $\{30, 300, 3000\}$ that assign a draw of $\mathit{sev}$ to a damage tier take their numerical values from the return-period boundaries of the flood risk maps, whose scenarios are grouped by annual probability into the classes $>1/30$, $1/30$ to $1/300$, $1/300$ to $1/3000$, and exceptional events \citep{IPO_2019_overstromingsrisicokaarten,Slager2019handboek}. We adopt these values as fixed cutoffs on the unitless severity draw $\mathit{sev}$. The GEV distributions of $\mathcal{S}_1$--$\mathcal{S}_4$ have support below zero: a draw $\mathit{sev}_t \le 0$ selects no map, so that month carries no flood damage ($D=0$ for every agent) and unemployment stays at its baseline (Equation~\ref{eqn:unemployment}).
We do not identify $\mathit{sev}$ with a return period in years: the simulated frequency of selecting each map is determined by the scenario-specific GEV distribution, as reported in Table~\ref{tab:flood_severity}. At a given location, the damage factor is fixed for all severity draws assigned to the same map, while varying spatially within that map.

Moving from $\mathcal{S}_1$ to $\mathcal{S}_4$ lowers the probability of a month without flooding and increases the probability of selecting the severe and catastrophic tiers.

\begin{table}[ht]
    \centering
    \caption{Monthly selection probabilities (\%) for each damage tier under flood scenarios $\mathcal{S}_0$--$\mathcal{S}_4$. A month with $\mathit{sev}_t \le 0$ selects no map.}
    \label{tab:flood_severity}
    \begin{tabular}{ll S[table-format=3.1] S[table-format=2.1] S[table-format=2.1] S[table-format=2.1] S[table-format=2.1]}
        \toprule
        {Damage tier} & {Map} & {$\mathcal{S}_0$} & {$\mathcal{S}_1$} & {$\mathcal{S}_2$} & {$\mathcal{S}_3$} & {$\mathcal{S}_4$} \\
        \midrule
        \textbf{No flood}    & none               & 100.0 & 23.3 & 14.8 & 10.0 & 4.1 \\
        \textbf{Minor}       & $R_{grote}$        & 0.0 & 38.0 & 37.8 & 34.8 & 33.9 \\
        \textbf{Moderate}    & $R_{middel}$       & 0.0 & 38.6 & 44.6 & 47.3 & 46.0 \\
        \textbf{Severe}      & $R_{klein}$        & 0.0 &  0.1 &  2.7 &  7.5 & 13.4 \\
        \textbf{Catastrophic}& $R_{eklein}$       & 0.0 & {$<$0.1} & {$<$0.1} &  0.5 &  2.6 \\
        \bottomrule
    \end{tabular}
\end{table}

\subsection{Flood damage and collateral}
\label{sec:flood_impact}
Once a map is selected the location of the agents $(x_i, y_i)$ establishes the damage factor $D(x_i,y_i,\mathcal{S})_t \in [0,1]$, read from the flood risk map of the damage tier that is selected at the specific point in time. This factor expresses the expected fraction of building value lost given the water depth at that location. Within each damage tier, the flood risk maps encode spatially varying water depths. Locations outside the flood extent receive $D_t=0$, while inundated locations receive $D_t>0$ proportional to local water depth. The resulting monetary loss to property value is the share of the value of the house given by:
\begin{equation}
    \Delta p_{i,t} = \lambda \, v_i \, D(x_i,y_i,\mathcal{S})_{t}  \label{eqn:delta_p},
\end{equation}
where $\lambda \in [0,1]$ is the share of house value lost at the maximum damage factor, $D=1$. The damage factor is a fraction of the building value, while the value of a property also reflects its land and location, which a flood does not destroy; $\lambda$ converts the loss on the building into a loss on the value of the whole property. The loss therefore scales with the value of the house and never exceeds a share $\lambda$ of it. The collateral of agent $i$ at time $t$ is
\begin{equation}
\label{eqn:collateral}
c_{i,t} 
= v_i\left(1 - \lambda\, D(x_i,y_i,\mathcal{S})_{t}\right)(1-\delta)^{n^{\kappa}_{i,t-1}}.
\end{equation}
The first factor is the flood loss at time $t$. The second is the foreclosure spillover, a loss of share $\delta$ (the spillover intensity) for each of the $n^{\kappa}_{i,t-1}$ foreclosures within radius $\kappa$ of agent $i$ at time $t-1$, defined in Section~\ref{sec:neighbourhood}. Neither factor accumulates over time: the damage factor is read from the map selected at time $t$, and the count covers the foreclosures of time $t-1$ only. When no foreclosure occurred nearby, $n^{\kappa}_{i,t-1}=0$ and $c_{i,t} = v_i - \Delta p_{i,t}$.

The loss $\Delta p_{i,t}$ is borne in full by the household: the model does not include flood insurance or public compensation. The motivation is that empirical evidence suggests that flood losses are borne mostly by households themselves: about 70\% of expected flood losses to homes in the US are uninsured \citep{amornsiripanitch2025underinsurance}, and in Europe only about a quarter of losses from climate-related extremes are insured \citep{EIOPA2024dashboard}. In the Netherlands, flooding from the major rivers, such as the Meuse, and from the sea is largely uninsured: an estimated 99\% of households and businesses are not insured against it \citep{verbond2024floods}. Furthermore, public compensation under the Disaster Damage Compensation Act (Wts) is decided only after the event and covers the current value of damaged goods rather than their replacement cost \citep{ENW2023verzekerbaarheid}.

\subsection{Unemployment}
We model the unemployment response to flood severity using a hyperbolic tangent function (Eqn. \ref{eqn:unemployment}), which captures the empirically observed asymmetric and bounded nature of unemployment rate dynamics. This modelling choice is grounded in empirical studies that highlight the sudden nature of labour market disruptions during disasters \citep{Belasen_Polachek_2008} and a more general negative average effect of floods on employment \citep{Belasen_Polachek_2008, Sarmiento_2007}. A similar nonlinear response of economic variables to climate shocks has been used by \textcite{Burke_Hsiang_Miguel_2015}.
Unemployment increases with severity, but at a diminishing rate, approaching an upper bound for very large floods. We specify:
\begin{equation}
\label{eqn:unemployment}
u_t =
\begin{cases}
\eta + \vartheta  \tanh\left(\dfrac{\mathit{sev}_{t}}{100}\right) & \text{if } \mathit{sev}_t > 0, \\[6pt]
\eta & \text{if } \mathit{sev}_t \le 0.
\end{cases}
\end{equation}
Here, $\eta$ is baseline unemployment and $\vartheta$ is the maximum additional unemployment induced by flooding. The denominator of $100$ sets the severity scale over which unemployment approaches its upper bound, $\eta+\vartheta$. The ranges of $\eta$ and $\vartheta$ follow the historical range of Dutch unemployment \citep{Salverda2025,CBS2023arbeidsmarkt} (\SIsec{sec:baseline_parametrization}), and we explore them in a global sensitivity analysis (GSA) in \SIsec{sec:app:sa}. At $t=0$, $u_t=\eta$.

Employment depends on two quantities. The first is the unemployment rate $u_t$ of Equation~\ref{eqn:unemployment}, which is set at the model level and depends on the flood severity. The second is a layoff threshold $\psi^{e}_{i} \sim U[0,1]$, which each household draws when it enters the model and which stays fixed throughout the simulation. A household is employed when its threshold is at least the current rate:
\begin{equation}
\label{eqn:employment}
e_{i,t} = \mathbf{1}\left[\psi^{e}_{i} \geq u_t\right],
\end{equation}
where $e_{i,t}=1$ denotes employment. Since $P(\psi^{e}_{i} < u_t) = u_t$, the rate determines how many households are laid off and the thresholds determine which ones. A rise in $u_t$ lays off the households with the lowest thresholds first. A household that returns to work draws a new income from the initial income distribution, and $\tau_{i,t}$ counts its consecutive number of months out of work.

\subsection{Household initialisation and data}
\label{sec:agents}
\label{sec:agents_init}
\label{sec:application}
The agents in the simulation are households. We initialise them with several financial attributes relative to their property, coordinates on the map $(x_i,y_i)$ and value of the property ($v_i$). We add parameters relative to their loan with the lender, such as the monthly instalment ($P_i$), collateral ($c_{i,t}$), total mortgage amount ($\ell_{i,t}$) and loan-to-value ratio ($\LTV_{i,t}$). We also add parameters that reflect the working situation of each agent: income ($I_{i,t}$), working seniority ($\mathit{sen}_{i,t}$), disposable funds $w_{i,t}$, disposable funds over loan amount ratio at origination ($\zeta_{i,0}$) and number of instalments left to end the loan ($n_{i,t}$). We draw income and property value from the distributions fitted to real-world data by \textcite{Lin2024ESGClimate}, and the remaining attributes from the distributions specified in \SItab{tab:parameter_distributions} (\SIsec{sec:app:agent_initialization}). \SItab{tab:parameter_distributions} presents the distribution of each attribute for our case study, the Limburg region. The coordinates ($x_i,y_i$) of the agents in the model are initialised according to a population density map. The population density map we use covers the Limburg region, is based on data from the Dutch Association of Real Estate Agents (NVM) and is taken from \textcite{Lin2024ESGClimate}. The synthetic population at the initialisation closely mimics the characteristics of the real population, shown in \SIfig{fig:combined_maps}, panel (a). We detail the process of creating and updating the density of a synthetic population that mimics the real one in \SIsec{sec:app:agent_initialization} (\SIalg{alg:coordinate_agents}).

We initialise the system so that it is a snapshot of a system already in evolution. To achieve that, we initialise a heterogeneous population of agents at different stages of their mortgage life-cycle, which means different numbers of instalments $n_{i,t}$ left. We draw $n_{i,t}$ from a uniform distribution between 1 and 120 as specified in \SItab{tab:agent_attributes}, where 120 represents the longest time (in months, which is 10 years) in our system (we use 10-year loans for simplicity but this assumption can be relaxed using heterogeneous lengths of the loan, at 20 or 30 years). For the agents entering at $t>0$, we assign $n_{i,0}=120$ for simplicity. We calculate the monthly instalment $P_i$ of agent $i$ as the remaining loan principal divided by the remaining number of payments $n_{i,0}$:
\begin{equation}
\label{eqn:installment}
P_i = P_{i,0} = \frac{\LTV_{i,0}\, v_i}{n_{i,0}} = \frac{\ell_{i,0}}{n_{i,0}}.
\end{equation}
The loan principal $\ell_{i,0} = \LTV_{i,0} \cdot v_i$ is initialised by multiplying the agent's property value ($v_i$) and loan-to-value ratio ($\LTV_{i,0}$) at origination. The ratio $\LTV_{i,t}$ is the proportion of the house's value that was financed by the mortgage. The property value is drawn from a distribution fitted to housing market data, and the loan-to-value ratio from a normal distribution whose mean and variance are set as model parameters (\SItab{tab:parameter_distributions}). We assume the instalment has a constant value throughout the life of an agent, that is $P_i = P_{i,0}$. Note that Equation \ref{eqn:installment} adopts a zero-interest (linear amortisation) structure for tractability.
Throughout, $t$ denotes the model month, and agent attributes carry the indices $i,t$. A subscript $0$ on an agent attribute denotes its value at origination, the month $t^0_i$ in which agent $i$ enters the model, which is $t^0_i=0$ for the initial stock; the mortgage age is $a_{i,t}=t-t^0_i$. We initialise the agents' funds as $w_{i,0} = \zeta_{i,0} \ell_{i,0}$, where $\zeta_{i,0}$ is the ratio of liquid funds to the loan at origination; it sets the initial funds and enters the admission scorecard (\SIsec{sec:app:scorecard}).

New agents are initialised in a similar way to agents initialised at the beginning of the simulation. The financial parameters are generated in the same way as described for the agents initialised at epoch zero. The generation of the coordinates for the new agents is designed to match the density map, therefore not modifying the population density; we detail the procedure in \SIsec{sec:app:agent_initialization} (Algorithms~\ref{alg:preprocess} and~\ref{alg:binary_search}). The initialised agents are generated with randomised attributes and are only added to the simulation. The number of applicants per step is given in Equation~\ref{eqn:number_new_join}.

The stock of agents initialised at $t=0$ is placed on the grid directly, without screening, so the credit quality of the initial book is the one implied by \SItab{tab:parameter_distributions} rather than the one selected by the scorecard. The acceptance threshold applies to the agents that enter at $t>0$.
Each candidate's credit score is computed using the scorecard detailed in \SIsec{sec:app:scorecard}, and the agent joins the simulation only if the score clears a minimum acceptance threshold set by the lender. Because the number of applicants does not respond to flood conditions, the effect of a flood shock is measured on the households already holding a mortgage, without a confounding change in loan demand.

Each attribute is drawn independently, so the synthetic population reproduces the fitted marginal distribution of every attribute and not their joint dependence. One grid cell of the Limburg map corresponds to roughly $160$ m, so the baseline neighbourhood radius $\kappa=10$ (Section~\ref{sec:neighbourhood}) spans about $1.6$ km around the household. We repeat the cascade analysis at radii from $\kappa=3$ to $\kappa=18$ (about $0.5$ to $2.9$ km) in \SIfig{fig:tipping_sensitivity}.

\subsection{Income and expenditure}
\label{sec:income_empl_dynamics}
The income in the model is initialised according to a Burr distribution (\SItab{tab:parameter_distributions}) and then evolves with the agent's employment status. An agent who loses their job receives an unemployment benefit equal to $70\%$ of their last income, which is withdrawn after a finite period, and an agent who finds a new job draws a new salary from the initial income distribution. This replacement rate follows the Dutch unemployment insurance scheme (WW) and lies within the range of the main developed countries' schemes. \SIsec{sec:appendix_unemployment} reports the full update rules and the cross-country comparison.

At every step, after the agents' income $I_{i,t}$ is determined, the agents update their consumption (or living expenditures) $s_{i,t}$ as a nonlinear function of income. Empirically, the average propensity to consume (APC), the ratio of consumption to income, declines with income levels. This pattern is well-documented in cross-sectional household surveys \citep{dynan2004rich}. We define the single household consumption share as
\begin{equation}
\label{eqn:income_share}
s_{i,t} = \frac{m}{1 + \varepsilon  I_{i,t}},
\end{equation}
where $\varepsilon$ is the steepness coefficient governing how rapidly the income share declines with $I_{i,t}$, and $m$ is the scale parameter of the consumption function. This functional form ensures that lower-income agents allocate a larger fraction of their income to consumption, a property consistent with empirical consumption microdata and standard heterogeneous-agent macroeconomic models. The total expenditures $E_{i,t}$ for agent $i$ at time $t$ are calculated as $E_{i,t} = \max(s_{i,t} \cdot I_{i,t}, \mathit{exp}_{lb}) + P_i$ (calculated for non-defaulted agents).
The GSA explores the parameters $m$, $\varepsilon$ and $\mathit{exp}_{lb}$ over the ranges in \SItab{tab:sa_ranges}, centred on the baseline values derived in \SIsec{sec:baseline_parametrization}.

\subsection{Neighbourhood utility and foreclosure spillover}
\label{sec:neighbourhood}
The response to neighbourhood flood damage enters through the utility $U^{\kappa}_{i,t}$, the collateral value of nearby properties relative to their undamaged value. Neighbours' foreclosures enter the willingness to pay (Equation~\ref{eqn:wtp}) only through this collateral (Equation~\ref{eqn:collateral}). The radius $\kappa$ sets its spatial extent and the loading $\beta_1$ its weight; both are varied in the GSA (\SItab{tab:sa_ranges}). Neighbourhood delinquency within $5$ miles raises the probability of strategic default \parencite{Goodstein2017}.

The neighbourhood $\mathcal{H}^{\kappa}_{i,t}$ of agent $i$ is the set of households holding an active mortgage in month $t$ within Euclidean distance $\kappa$ (in grid cells) of it, agent $i$ included (\SIsec{sec:app:neighbourhood}). The local utility is
\begin{equation}
\label{eqn:utility}
U^{\kappa}_{i,t} = \frac{\sum_{j \in \mathcal{H}^{\kappa}_{i,t}} c_{j,t}}{\sum_{j \in \mathcal{H}^{\kappa}_{i,t}} v_j} \in [0,1],
\end{equation}
which equals one when the neighbourhood is undamaged and no foreclosure occurred near it in the previous month. The count $n^{\kappa}_{i,t-1}$ in Equation~\ref{eqn:collateral} is the number of households foreclosed at $t-1$ whose last position lies within $\kappa$ of agent $i$. At the start of each step the model computes $n^{\kappa}_{i,t-1}$, $c_{i,t}$ and $U^{\kappa}_{i,t}$ for every household before any household acts, so that all households respond to the same foreclosures.

The spillover follows the evidence on foreclosure externalities (\SIsec{sec:app:distress_probs}). A foreclosed house sold into the local market lowers nearby prices through comparable sales, by about $1\%$ for a house $0.05$ miles away \parencite{Campbell_Giglio_Pathak_2011}, and through competition with nearby listings \parencite{Anenberg_Kung_2014}; it also raises neighbours' default probability through lower prices and a lower perceived cost of default \parencite{Gupta_2019}. The spillover on collateral represents both effects, so $\delta$ measures the combined effect of a foreclosure on its neighbours' willingness to pay rather than the price discount of a single sale. In line with this evidence, the spillover is local (within $\kappa$), cumulative (each foreclosure multiplies collateral by $1-\delta$) and transitory; it lasts one step; nearby prices return to their pre-listing level within $6$ to $12$ weeks of the sale of the foreclosed house \parencite{Anenberg_Kung_2014}.

The empirical moments at $\mathcal{S}_0$ come from a housing market without floods, so the validation at $\mathcal{S}_0$ uses $\delta=0$. After a flood, foreclosed houses are sold into a market with more forced sales and fewer buyers: flood risk lowers both prices and ease of sale \parencite{Turnbull_ZahirovicHerbert_Mothorpe_2013}, the price effect of a foreclosure works mainly through the added supply \parencite{Hartley_2014, Anenberg_Kung_2014}, and it is larger in high-vacancy neighbourhoods \parencite{Turnbull_vanDerVlist_2024}. The GSA therefore explores $\delta\in[0.05,0.5]$ as stress scenarios for flooded markets, without linking $\delta$ to flood severity.

Since the spillover lowers $c_{j,t}$ within $\kappa$ of a foreclosure and $U^{\kappa}_{i,t}$ aggregates $c_{j,t}$ within $\kappa$ of agent $i$, a foreclosure lowers the willingness to pay of households up to $2\kappa$ away and raises foreclosure probabilities in the following step. Section~\ref{sec:cascades} quantifies this feedback.

\subsection{Willingness to pay, distress and foreclosure}
\label{sec:distress}
A household enters distress through the financial or the strategic channel (Section~\ref{sec:overview}). Distress does not by itself end the mortgage; within the same month the household is either retained under forbearance, paying the instalment in full on unchanged terms, or foreclosed. Every default we report is a foreclosure; distress resolved through forbearance is not a default.

Financial distress, the inability to pay \citep{Campbell_Cocco_2015}, occurs when
\begin{equation}
w_{i,t} - \max\left( s_{i,t} \cdot I_{i,t}, \mathit{exp}_{lb}\right) \leq P_i, \quad
\label{eqn:financial_distress}
\end{equation}
where $w_{i,t}$ is liquid funds, $s_{i,t} \cdot I_{i,t}$ is consumption spending with floor $\mathit{exp}_{lb}$, and $P_i$ is the mortgage payment.

Both channels are resolved through the willingness to pay
\begin{equation}
\label{eqn:wtp}
\text{wtp}_{i,t} = \beta_1 \cdot U^{\kappa}_{i,t} - \frac{\beta_2 \cdot X}{\beta_3 + X}, \qquad X = \mathrm{DTI}_{i,t} \cdot (\tau_{i,t}+1).
\end{equation}
The coefficients are sampled over $\beta_1\in[0,1]$, $\beta_2\in[0,1]$ and $\beta_3\in[0.25,3]$ in the GSA, subject to $\beta_1>\beta_2$ (\SItab{tab:sa_ranges}), with baseline values $\beta_1=1$, $\beta_2=0.5$ and $\beta_3=1$. The neighbourhood term $\beta_1 U^{\kappa}_{i,t}$ falls as flood damage lowers nearby collateral $c_{j,t}$ relative to its undamaged value $v_j$. The second term is the debt pressure. $\mathrm{DTI}_{i,t}$ is the outstanding loan $\ell_{i,t}$ relative to annual income (\SIsec{sec:app:dti}), the primary affordability metric in \textcite{Campbell_Cocco_2015}, and $\tau_{i,t}$ counts consecutive months of unemployment (zero for employed agents), so the penalty $\tau_{i,t}+1$ grows with the duration of the income shock. The half-saturation point is set by $\beta_3$, and $\text{wtp}_{i,t}$ lies in $[0,1]$ (\SIsec{sec:app:distress_probs}).

Forbearance involves no concession: no payment is suspended or reduced and no arrears carry over, so the next period is assessed from scratch. A household in financial distress is forborne with probability $\text{wtp}_{i,t}$ and foreclosed with probability $1-\text{wtp}_{i,t}$. For a household that can pay, the outcome probabilities are (\SIsec{sec:app:distress_probs})
\begin{align}
\label{eqn:distress_probs}
\Pr(\text{no distress} \mid \text{wtp}_{i,t}) &= \text{wtp}_{i,t}, \\
\Pr(\text{forbearance} \mid \text{wtp}_{i,t}) &= \min(\text{wtp}_{i,t},\ 1-\text{wtp}_{i,t}), \\
\Pr(\text{foreclosure} \mid \text{wtp}_{i,t}) &= \max(0,\ 1-2\,\text{wtp}_{i,t}),
\end{align}
so strategic foreclosure is possible only when $\text{wtp}_{i,t}<1/2$.

In an undamaged neighbourhood $U^{\kappa}_{i,t}=1$ and the debt pressure is at most $\beta_2$, so $\text{wtp}_{i,t}\geq\beta_1-\beta_2$. Whenever $\beta_1-\beta_2\geq1/2$, strategic foreclosure therefore requires flood damage, whatever the debt pressure; we call this the \textit{damage-required region}. The GSA explores combinations inside and outside it (\SIsec{sec:app:sa}). The baseline $\beta_1=1$, $\beta_2=0.5$ lies on its boundary. With $\beta_1=1$, a positive probability of strategic foreclosure requires $U^{\kappa}_{i,t}<1/2+\beta_2X/(\beta_3+X)$: debt pressure lowers the damage needed but cannot replace it. \SIfig{fig:beta_regions} locates the region in the plane of the gap $\beta_1-\beta_2$ and the spillover intensity $\delta$, and shows how the active mortgage population declines as the gap narrows outside it and as $\delta$ grows (\SIsec{sec:app:distress_probs}).

\subsection{Calibration and sensitivity analysis}
\label{sec:methods_calibration}
The model is calibrated at $\mathcal{S}_0$ with $\delta=0$ by rejection approximate Bayesian computation \parencite{Grazzini_Richiardi_Tsionas_2017}. The reference sample consists of $1{,}001$ parameter combinations of a scrambled Sobol' sequence \citep{Sobol_1967}, uniform on the ranges of \SItab{tab:sa_ranges} for the nine parameters active at $\mathcal{S}_0$ and restricted to $\beta_1-\beta_2\ge1/2$, each averaged over $10$ seeds. The two targets are the annual foreclosure rate, compared with the NHG loss-claim band, and the share of households in financial distress, compared with the BKR arrears band; the $5\%$ of combinations closest to both bands are retained (\SIsec{sec:app:identification}). The global sensitivity analysis computes first-order and total-order Sobol' indices and PAWN indices for $11$ parameters on $3{,}328$ Saltelli samples, each averaged over $10$ seeds, in every flood scenario (Sections~\ref{sec:app:gsa_method} and~\ref{sec:app:gsa_defaults}).

\subsection{Use of generative AI}
\label{sec:genai}
Fig.~\ref{fig:model_overview} was generated with PaperBanana \parencite{Zhu2026PaperBanana}, an AI tool for academic illustration, using the authors' description of the model as input; the authors reviewed the figure for accuracy. No other figure in the manuscript or Supplementary Information was produced with generative AI. The authors also used Anthropic's Claude Opus and Sonnet (versions 4 and 5) and OpenAI ChatGPT (versions 5 and 6) to support manuscript drafting, language editing and data analysis, including assistance with analysis code. The authors reviewed all AI-assisted code and text and take full responsibility for the analyses, the interpretation of results and the final manuscript.

\section*{Data availability}
\label{sec:data_reproducibility}

Public data used in this study are available from Statistics Netherlands (CBS StatLine), the Dutch credit registry (BKR), Nationale Hypotheek Garantie (NHG), the Dutch National Georegister and the Dutch Ministry of Social Affairs and Employment. Detailed source citations, dataset identifiers and descriptions of their use are provided in the main manuscript and SI. The population density map of the Limburg region is based on data from the Dutch Association of Real Estate Agents (NVM), as used by \textcite{Lin2024ESGClimate}.

\section*{Acknowledgements}
We thank Chang Lin, whose MSc thesis \parencite{Lin2024ESGClimate} is the starting point of this contribution.

\noindent The Global Sensitivity Analysis of the model in the results was run on the Dutch Snellius supercomputing system.

\section*{Author contributions}
M.V. contributed to the conceptualisation, methodology, software, data analysis, visualisation and writing. D.K. and C.D. contributed to the conceptualisation, methodology and supervision. D.R. contributed to the conceptualisation, methodology, analysis, writing and supervision.

\printbibliography

\clearpage
\appendix
\renewcommand{\appendixname}{}
\setcounter{section}{0}\setcounter{figure}{0}\setcounter{table}{0}
\setcounter{equation}{0}\setcounter{algocf}{0}
\renewcommand{\thesection}{S\arabic{section}}
\renewcommand{\thefigure}{S\arabic{figure}}
\renewcommand{\thetable}{S\arabic{table}}
\renewcommand{\theequation}{S\arabic{equation}}
\renewcommand{\thealgocf}{S\arabic{algocf}}
\renewcommand{\theHsection}{S\arabic{section}}
\renewcommand{\theHfigure}{S\arabic{figure}}
\renewcommand{\theHtable}{S\arabic{table}}
\renewcommand{\theHequation}{S\arabic{equation}}
\renewcommand{\theHalgocf}{S\arabic{algocf}}
\let\SIorigsubsection\subsection
\renewcommand{\subsection}{\FloatBarrier\SIorigsubsection}
\let\SIorigsubsubsection\subsubsection
\renewcommand{\subsubsection}{\FloatBarrier\SIorigsubsubsection}

\begin{center}
{\LARGE Supplementary Information}
\end{center}
\addcontentsline{toc}{section}{Supplementary Information}

\nocite{NHG_jv2014,NHG_jv2015,NHG_jv2016,NHG_q3_2014,NHG_q1_2016,NHG_q4_2016}

\section{Model Specification}
\label{sec:app:params}
\subsection{Flood Severity and Flood Scenarios}
\label{sec:app:gev_construction}
\begin{figure}[ht]
    \centering
    \includegraphics[width=0.6\linewidth]{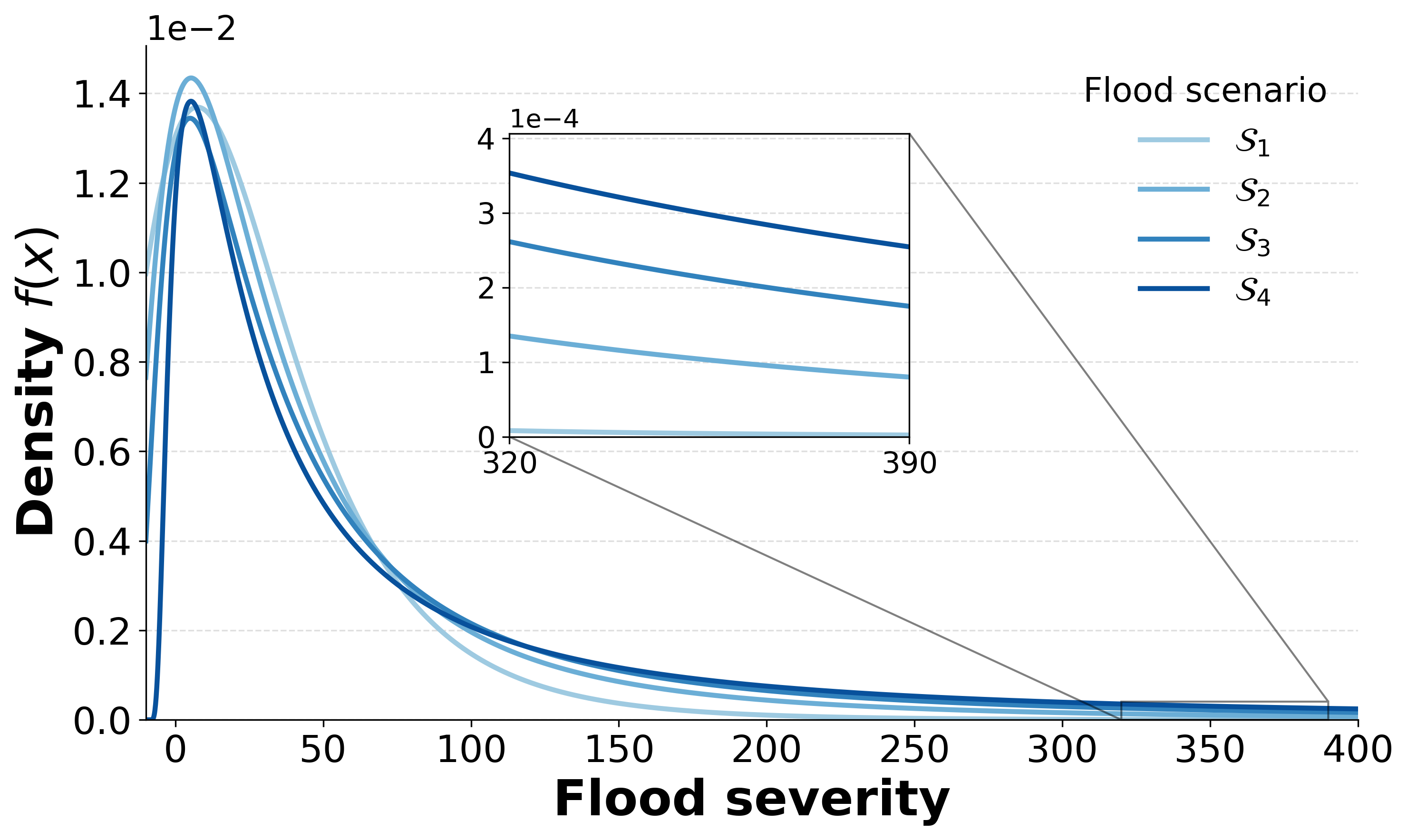}
    \caption{\footnotesize{Flood severity probability density functions $f(x)$ under four distinct parameterisations ($\mathcal{S}$). 
    Note that the shape parameter $\xi$ governs the tail behaviour of the distribution; higher values of $\xi$ (e.g., $\mathcal{S}_4$) result in a ``thicker'' tail, indicating a higher probability of extreme flood events compared to lower values of $\xi$. }}
    \label{fig:gev_distrib}
\end{figure}

We detail the functional form of both the cumulative distribution function $F(x)$ and the probability density function $f(x)$, reported in \SIfig{fig:gev_distrib}, of the GEV distribution used in the study to model the severity of the floods $\mathit{sev} \sim GEV(\mathcal{S})$:
\begin{equation}
\label{eqn:cdf_gev}
    F(x; \mu, \sigma, \xi) = 
    \begin{cases} 
        \exp \left\{ - \left[ 1 + \xi \left( \frac{x - \mu}{\sigma} \right) \right]^{-1/\xi} \right\} & \text{if } \xi \neq 0 \\
        \exp \left\{ - \exp \left( - \frac{x - \mu}{\sigma} \right) \right\} & \text{if } \xi = 0
    \end{cases}
\end{equation}

\begin{equation}
\label{eqn:pdf_gev}
    f(x; \mu, \sigma, \xi) = 
    \begin{cases} 
        \frac{1}{\sigma} \left[ 1 + \xi \left( \frac{x - \mu}{\sigma} \right) \right]^{-(1 + 1/\xi)} \exp \left\{ - \left[ 1 + \xi \left( \frac{x - \mu}{\sigma} \right) \right]^{-1/\xi} \right\} & \text{if } \xi \neq 0 \\
        \frac{1}{\sigma} \exp \left( - \frac{x - \mu}{\sigma} \right) \exp \left[ - \exp \left( - \frac{x - \mu}{\sigma} \right) \right] & \text{if } \xi = 0
    \end{cases}
\end{equation}
We use a more compact notation by setting $\mathcal{S} = [\mu, \sigma, \xi]^\top$.

We construct stylised flood stress scenarios with progressively heavier upper tails while approximately preserving the location and height of the density peak, following \textcite{Lin2024ESGClimate}. The baseline scenario $\mathcal{S}_1$ uses $\xi=0.1$, corresponding to the mean of the geophysical prior proposed by \textcite{MartinsStedinger2000}. These authors parameterise the GEV shape with the opposite sign convention and propose a Beta prior for annual maximum floods which, expressed in terms of $\xi$, has support $\xi\in[-0.5,0.5]$, mean $\xi=0.10$ and standard deviation $0.122$. The prior therefore corresponds to our baseline value $\xi=0.1$ and places its upper limit at $\xi=0.5$.

Empirical estimates can nevertheless indicate substantially heavier tails: \textcite{morrison_smith_2002} report estimates corresponding to $\xi>0.5$ in $28\%$ of 104 central Appalachian basins. This evidence provides one motivation for considering heavier-tailed distributions. A second motivation for adopting the stress-scenario approach described in Section~\ref{sec:flood_method} is the difficulty of obtaining calibrated projections of future flood severity: historical time series are often short and contain few extreme observations, leaving tail-parameter estimates noisy and difficult to pin down. We therefore select $\xi=0.5$, $0.8$, and $1.2$ for $\mathcal{S}_2$, $\mathcal{S}_3$, and $\mathcal{S}_4$, respectively, as progressively more demanding stress scenarios for mortgage credit outcomes, rather than as calibrated projections of future climate conditions.

The density in \SIeq{eqn:pdf_gev} peaks at
\begin{equation}
    x^{*} = \mu + \frac{\sigma}{\xi}\left[(1+\xi)^{-\xi} - 1\right],
    \qquad
    f(x^{*}) = \frac{(1+\xi)^{1+\xi}\, e^{-(1+\xi)}}{\sigma}.
    \label{eqn:gev_peak}
\end{equation}

For each of the shape parameters in $\mathcal{S}_2$--$\mathcal{S}_4$, we select $\mu$ and $\sigma$ on a $20 \times 20$ grid to minimise the scaled distance
\begin{equation}
    d(\mu,\sigma) = \left(\frac{x^{*} - x^{*}_{0}}{x^{*}_{0}}\right)^{2} + \left(\frac{f(x^{*}) - f_0(x^{*}_{0})}{f_0(x^{*}_{0})}\right)^{2}
\end{equation}
between the peak of the candidate density, computed from \SIeq{eqn:gev_peak}, and the peak $(x^{*}_{0}, f_0(x^{*}_{0})) = (5.17, 0.0139)$ of the reference distribution $GEV(10, 27, 0.2)$. The scenario $\mathcal{S}_1$ retains the location and scale of the reference with $\xi = 0.1$. The resulting modes lie between $5.0$ and $7.4$ and the peak densities between $0.0134$ and $0.0143$, so the scenarios differ primarily in their upper tail. The $r$-th moment of the GEV distribution exists only for $\xi < 1/r$, so the variance is infinite under $\mathcal{S}_2$ to $\mathcal{S}_4$ and the mean is infinite under $\mathcal{S}_4$; the scenarios are therefore aligned on the mode of the distribution. The parameter vectors are:
\begin{itemize}
    \item $\mathcal{S}_0$: no-flood baseline. In this scenario $\mathit{sev}_t = 0$ at every time step $t$: no flood event occurs, the damage factor $D(x_i, y_i, \mathcal{S}) = 0$ for all agents, and unemployment remains at the baseline rate $u_t = \eta$. This scenario serves as the counterfactual reference against which flood-affected scenarios $\mathcal{S}_1$--$\mathcal{S}_4$ are compared.
    \item $\mathcal{S}_1 = [10, 27, 0.1]^\top$
    \item $\mathcal{S}_2 = [15.789, 28.579, 0.5]^\top$
    \item $\mathcal{S}_3 = [21.579, 35.421, 0.8]^\top$
    \item $\mathcal{S}_4 = [28.421, 45.421, 1.2]^\top$
\end{itemize}

\subsection{Flood Risk Maps}
\label{sec:appendix_map_preprocessing}

\begin{figure}[htbp]
    \centering
    \includegraphics[width=\textwidth]{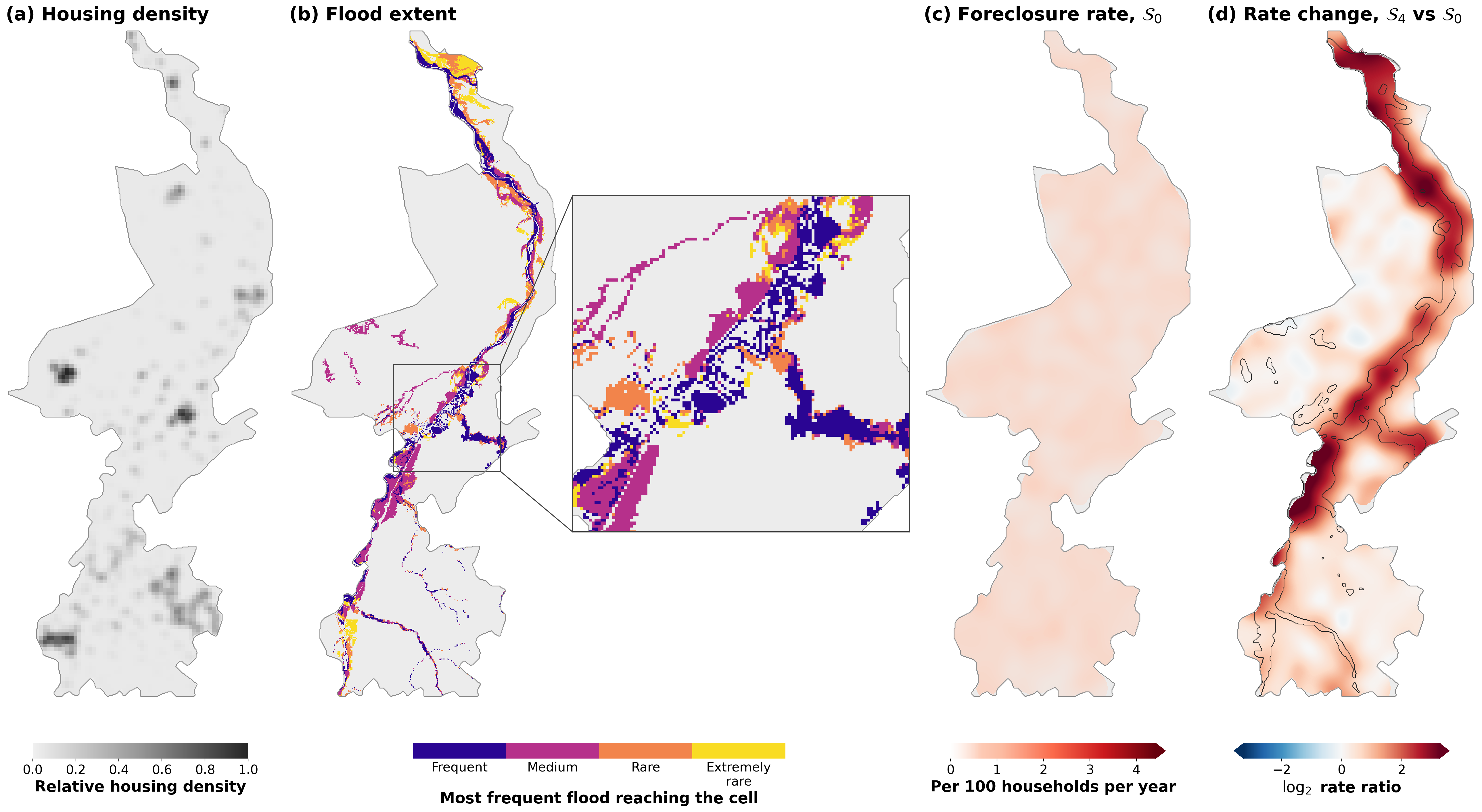}
    \caption{\footnotesize{Model geography and the spatial distribution of foreclosures in the Limburg region. (a) Relative housing density, based on spatial data from the Dutch Association of Real Estate Agents (NVM), rescaled to $[0,1]$. At initialisation each cell receives a number of agents proportional to this density, and entrants are placed in a cell with probability proportional to it (\SIsec{sec:app:agent_initialization}). (b) Extent of the four flood damage maps of Table~\ref{tab:flood_severity}, with a magnified section of the river Meuse. The extents are nested: every cell covered by a more frequent map is also covered by all rarer ones. Each cell is coloured by the most frequent map that covers it, so a cell floods in that event and in every rarer one; grey cells lie outside all four extents ($D=0$). The flooded share of the region grows from $4.4\%$ for the most frequent map to $9.1\%$, $12.1\%$ and $14.3\%$ for the extremely rare map. (c) Local annual foreclosure rate without floods ($\mathcal{S}_0$), in foreclosures per $100$ households per year. The rate is the calibration target of \SIsec{sec:app:identification} computed cell by cell: in each run, a Gaussian kernel density estimate of the foreclosure locations (standard deviation $10$ cells) is divided by the run's mean active population, distributed over cells in proportion to the smoothed housing density, and by the number of years; the map is the mean over runs. Averaged over the population, the map returns the regional rate of $0.42\%$. The colour scale ends at the $99.5$th percentile of the $\mathcal{S}_4$ rate. (d) Ratio of the local foreclosure rate under the most severe flood scenario ($\mathcal{S}_4$) to the rate of panel (c), on a $\log_2$ scale; the black line marks the outer flood extent. The median ratio over grid cells is $5.2$ inside the flood extent and $1.4$ outside it. In (c) and (d), cells where the smoothed housing density is below $5\%$ of its maximum ($2.1\%$ of the region) are masked. Calibrated parametrisation with $\delta=0$, $\kappa=10$, $\beta_1-\beta_2=0.5$ and $\vartheta=0.05$ (\SItab{tab:model_params}), over the $300$ months after a burn-in of $100$ months; $20$ seeds for $\mathcal{S}_0$ and $20$ seeds for each $\lambda\in\{0.25,0.5,0.75\}$ for $\mathcal{S}_4$.}}
    \label{fig:combined_maps}
\end{figure}

The national GIS flood risk maps \citep{IPO_2019_overstromingsrisicokaarten} are aggregated to PNG via a mean convolution kernel in QGIS, reducing resolution. The PNG is read as a greyscale array, pixel values are inverted ($M[i,j] \gets 255 - M[i,j]$, since white encodes no flood in standard greyscale), and normalised by the array maximum so that $M \in [0,1]$. Following \textcite{Lin2024ESGClimate}, these normalised values are used directly as the spatial damage factors $D(x_i,y_i,\mathcal{S})$: they provide a depth-derived proxy for relative damage, without a separate depth--damage curve transformation. The extents of the four maps are shown in \SIfig{fig:combined_maps}, panel (b), and the damage factor enters the property loss of Equation~\ref{eqn:delta_p}. Agents outside the flood zone receive $D=0$.

\subsection{Agent Variable Initialisation}
\label{sec:app:agent_initialization}
\begin{table}[!htbp]
\centering
\small
\caption{Global parameters. The calibrated value is that of the parameter combination whose outcomes lie inside both empirical bands (\SIsec{sec:app:identification}), reported for the three parameters the calibration constrains; the remaining parameters keep their baseline value. The column Eq.\ gives the equation in which the parameter enters.}
\label{tab:model_params}
\renewcommand{\arraystretch}{1.17}
\setlength{\tabcolsep}{5pt}
\begin{tabularx}{\textwidth}{@{} l >{\raggedright\arraybackslash}X c >{\raggedright\arraybackslash}p{2.7cm} c @{}}
\toprule
Parameter & Description & Eq. & Baseline value & \begin{tabular}[b]{@{}c@{}}Calibrated\\ value\end{tabular} \\
\midrule
\multicolumn{5}{@{}l}{\textbf{Parameters varied in the GSA (\SItab{tab:sa_ranges})}} \\
\addlinespace[2pt]
$\kappa$ & Radius for defining an agent's neighbourhood utility & (\ref{eqn:utility}) & $10$ & \\
$\beta_1$ & Coefficient of the willingness to pay & (\ref{eqn:wtp}) & $1$ & \\
$\beta_2$ & Coefficient of the willingness to pay & (\ref{eqn:wtp}) & $0.5$ & \\
$\beta_3$ & Coefficient of the willingness to pay & (\ref{eqn:wtp}) & $1.0$ & \\
$\lambda$ & Share of house value lost at $D=1$ & (\ref{eqn:delta_p}) & $\{0.25,\,0.5,\,0.75\}$ & \\
$\delta$ & Spillover intensity: share of collateral lost per foreclosure within $\kappa$ in the previous month & (\ref{eqn:collateral}) & $0$\,$^{a}$ & \\
$\eta$ & Baseline unemployment rate & (\ref{eqn:unemployment}) & $0.037$ & $0.0383$ \\
$\vartheta$ & Maximum flood-induced increase in unemployment & (\ref{eqn:unemployment}) & $0.05$ & \\
$m$ & Scale parameter of the consumption function & (\ref{eqn:income_share}) & $6.2$ & $5.25$ \\
$\varepsilon$ & Velocity of decay of the income share dedicated to living expenses $s_{i,t}$ (velocity w.r.t.\ increases in $I_{i,t}$) & (\ref{eqn:income_share}) & $0.00198$ & $0.00250$ \\
$\mathit{exp}_{lb}$ & Minimum floor for an agent's living expenses & (\ref{eqn:income_share}) & $1{,}087$ & \\
\midrule
\multicolumn{5}{@{}l}{\textbf{Other parameters}} \\
\addlinespace[2pt]
$\mathcal{S}$ & Selects the flood scenario ($\mathit{sev}$ distribution parameters) & & $\mathcal{S}_0,\dots,\mathcal{S}_4$ & \\
$T$ & Number of simulated time steps & & $400$ & \\
\begin{tabular}[t]{@{}l@{}}$R_{grote}$,\\ $R_{middel}$,\\ $R_{klein}$,\\ $R_{eklein}$\end{tabular} & Flood risk maps & & & \\
$\mathcal{A}$ & Minimum credit score for new agents to be accepted in the simulation & & $200$ & \\
$\alpha$ & Seniority cap & SI~(\ref{eqn:sen}) & $150$ & \\
$u_t$ & Rate of unemployment & (\ref{eqn:unemployment}) & $u_0 = \eta$\,$^{b}$ & \\
\midrule
\multicolumn{5}{@{}l}{\textbf{New joiners parameters}} \\
\addlinespace[2pt]
$N_{nj}$ & Base number of potential new joiners per step & (\ref{eqn:number_new_join}) & $35$ & \\
$\rho$ & Density in agents generation & (\ref{eqn:number_new_join}) & $4$ & \\
\bottomrule
\end{tabularx}

\vspace{4pt}
\begin{minipage}{\textwidth}
\footnotesize
$^{a}$ Section~\ref{sec:cascades} uses $12$ values in $[0,\,0.25]$ (\SIsec{sec:app:cascade_method}).\\
$^{b}$ At $t=0$; afterwards $u_t$ follows Equation~\ref{eqn:unemployment}.
\end{minipage}
\end{table}

The distributions in \SItab{tab:parameter_distributions} builds on \textcite{Lin2024ESGClimate}. Income and property value use the fitted distributions directly, fitted to mortgage data and to housing market data. The loan-to-value distribution is not fitted to data; its mean and variance follow the model parametrisation of \textcite{Lin2024ESGClimate}. The model draws every attribute independently from its marginal distribution, so that the initialisation can be reproduced without access to the underlying data. The ratio of liquid funds to the loan enters with scale $0.015$, which places the median applicant at $0.65$ of the loan. To test this choice, we simulate the baseline scenario at the calibrated parametrisation with $\delta = 0$ for 101 values of the scale between $0.005$ and $0.03$, with five seeds each. For scales between $0.008$ and $0.02$, a range that includes the scale $0.01$ at which the fitted ratio is expressed as a fraction, the annual foreclosure rate stays between $0.39\%$ and $0.46\%$ and the annual share of households in financial distress between $0.46\%$ and $0.54\%$. A larger scale delays foreclosure rather than preventing it: the median loan age at financial foreclosure is 51, 61 and 69 months at scales $0.01$, $0.015$ and $0.02$.

Agents entering at $t>0$ draw their attributes from the same distributions as the agents placed at $t=0$, with $n_{i,0}=120$: the property value is drawn from its fitted distribution and the loan follows as $\ell_{i,0}=\LTV_{i,0}\,v_i$. Affordability is not imposed through the size of the loan; it is assessed by the admission scorecard, which scores the income and the monthly installment of each applicant (\SItab{tab:score_card}). The installment is constant, $P_i=\ell_{i,0}/n_{i,0}$, and repays the principal over the remaining term without interest. The model therefore abstracts from interest-rate variation, and the debt pressure of a household changes only through its income. Because an annuity with interest has a higher installment for the same principal, $P_i$ is a lower bound on the payment burden. New agents are located with the same population density map as the initial stock (Algorithms~\ref{alg:preprocess} and~\ref{alg:binary_search}), and flood losses scale with the property value (Equation~\ref{eqn:delta_p}), so the property enters the model only through $v_i$.

\begin{table}[ht]
    \centering
    \caption{Agent initialisation: distributions of the household, property and mortgage attributes, and assumed mortgage maturities. Income and property value follow the fits of \textcite{Lin2024ESGClimate}. $\mathrm{Burr}(c,\,d,\,\mathrm{loc},\,\mathrm{scale})$ denotes the Burr Type III distribution with cumulative distribution function $F(x)=\left(1+z^{-c}\right)^{-d}$, $z=(x-\mathrm{loc})/\mathrm{scale}>0$ (the parametrisation of \texttt{scipy.stats.burr}); draws are generated by inverse transform sampling, $x=\mathrm{loc}+\mathrm{scale}\,\left(u^{-1/d}-1\right)^{-1/c}$ with $u\sim U(0,1)$.
    The draws are independent.}
    \label{tab:parameter_distributions}
    \footnotesize
    \begin{tabular}{@{}lll@{}}
        \toprule
        Attribute & Role in the model & Distribution or value \\
        \midrule
        \multicolumn{3}{@{}l}{\textit{Property}} \\
        $v_{i}$ & Initial property value (\EUR{}) & $10^{3}\cdot\mathrm{Burr}(3.38,\,0.92,\,-1.34,\,195.25)$ \\
        \addlinespace
        \multicolumn{3}{@{}l}{\textit{Household}} \\
        $I_{i,0}$ & Monthly income (\EUR{} per month) & $\mathrm{Burr}(3.30,\,0.45,\,-12.76,\,3101.46)$ \\
        $\mathit{sen}_{i,0}$ & Work seniority (months) & $1+106.34\cdot\mathrm{Beta}(0.88,\,2.79)$ \\
        $\zeta_{i,0}$ & Liquid funds over loan size, $w_{i,0}/\ell_{i,0}$ & $0.015\cdot\mathrm{Burr}(9.42,\,0.14,\,-0.11,\,73.40)$ \\
        \addlinespace
        \multicolumn{3}{@{}l}{\textit{Mortgage}} \\
        $\LTV_{i,0}$ & Loan-to-value ratio, $\ell_{i,0}/v_{i}$ & $\mathcal{N}(0.57,\,0.1)$, truncated to positive values \\
        $n_{i,0}$ & Remaining instalments (months) & \begin{tabular}[t]{@{}l@{}}Discrete uniform on $\{1,\dots,120\}$ at $t=0$;\\ $120$ for entrants at $t>0$\end{tabular} \\
        \bottomrule
    \end{tabular}
\end{table}

\begin{landscape}
\begin{table}[p]
\centering
\footnotesize
\caption{Agent-Specific Parameters}
\label{tab:agent_attributes}
\setlength{\tabcolsep}{5pt}
\begin{tabularx}{\linewidth}{@{} l >{\hsize=1.2\hsize}L >{\hsize=0.9\hsize}L >{\hsize=0.9\hsize}L @{}}
\toprule
Attribute name & Description & Value at origination & Evolution over $t$ \\
\midrule
$(x_i, y_i)$ & Agent's spatial coordinates on the grid & Based on population density map, see \SIsec{sec:app:agent_initialization} & Constant \\
$D(x_i,y_i,\mathcal{S})$ & Damage factor based on the flood risk maps & & See Equation \ref{eqn:delta_p} \\
$a_{i,t}$ & Mortgage age, months since origination, $a_{i,t}=t-t^0_i$ & 0 & Increased by one at every time step \\
$I_{i,t}$ & Agent's monthly income & Burr distribution see \SItab{tab:parameter_distributions}  \\
$v_i$ & Initial House value & Burr distribution see \SItab{tab:parameter_distributions}  & Constant over life time of the agent \\
$c_{i,t}$ & Collateral value of the house & $c_{i,0}=v_i$ & Evolution according to Equation \ref{eqn:collateral} \\
$\LTV_{i,t}$ & Loan-to-Value ratio of the mortgage & Normal distribution see \SItab{tab:parameter_distributions} & \\
$\ell_{i,t}$ & Total amount of the mortgage loan & $\ell_{i,0} = \LTV_{i,0} \cdot v_i$ & Reduced by $P_i$ at every step $t$\\
$P_i$ & mortgage payment & Calculated as $\frac{\ell_{i,0}}{n_{i,0}}$ & Constant over life time of the agent \\
$n_{i,t}$ & Number of remaining mortgage payments & for agents joining at $t=0$ $n_{i,0} \sim U[1,120]$, else  $n_{i,0}=120$ & Reduced by 1 for every step if agent not defaulted \\
$w_{i,t}$ & Agent's liquid savings or financial buffer & Initialisation as $\zeta_{i,0} \cdot \ell_{i,0}$ & Section \ref{sec:income_empl_dynamics} \\
$\zeta_{i,0}$ & Ratio of liquid funds to loan size at origination, $w_{i,0}/\ell_{i,0}$ & Burr distribution with parameters specified in \SItab{tab:parameter_distributions} & Used at origination only \\
$\mathit{sen}_{i,t}$ & Agent's work seniority, entering the admission scorecard & Beta distribution with parameters specified in \SItab{tab:parameter_distributions} & Increased by 1 at every step $t$, See \SIeq{eqn:sen} \\
$\psi^{e}_{i}$ & Agent-specific layoff threshold, Equation \ref{eqn:employment} & Uniform $U[0, 1]$ & Constant \\
$\psi^{d}_{i,t}$ & Agent-specific reservation threshold, \SIeq{eqn:strategic_trigger} & Uniform $U[0, 1]$ & Redrawn independently at every step $t$ \\
$\text{score}_{i,0}$ & Agent's credit score for risk assessment & Calculated from financial attributes see \SItab{tab:score_card} & Constant \\
$\tau_{i,t}$ & Unemployment time & Set to $0$ &  Increased by $1$ for every unemployment timestep $t$\\
$s_{i,t}$ & Share of income dedicated to living expenses & Calculated according to Equation \ref{eqn:income_share} & Calculated according to Equation \ref{eqn:income_share} \\
$\text{wtp}_{i,t}$ & Willingness to pay & See Equation \ref{eqn:wtp} & Calculated according to Equation \ref{eqn:wtp} \\
$E_{i,t}$ & living expenditures & See Section \ref{sec:income_empl_dynamics} & \\
\bottomrule
\end{tabularx}
\tablenote{Attributes are indexed by agent $i$ and model month $t$; a subscript $0$ denotes the value at origination, the month $t^0_i$ in which the agent enters the model. Attributes without the index $t$ are constant over the life of the mortgage.}
\end{table}
\end{landscape}

\SIalg{alg:coordinate_agents} details the initialisation of the agents' coordinates $(x_i,y_i)$ so that they closely match the empirical density map. The empirical density map is based on spatial data from the Dutch Association of Real Estate Agents (NVM). The map is processed in the same way as the flood risk maps (\SIsec{sec:appendix_map_preprocessing}), so that $M \in [0,1]$: each cell value acts as the activation level of that area, which the density scaling factor $\rho$ converts into a number of agents. $\rho$ is the same parameter that governs the number of agents joining at every time step $t\neq 0$ (see Eqn. \ref{eqn:number_new_join}).\\
\smallskip
\begin{algorithm}[H]
    \LinesNotNumbered %
    \SetKwInOut{Input}{Input}

    \caption{Populate agents according to the density map}

    \Input{
        $M$: density map  \\
        $\rho$: density scaling factor
    }

    $n, m = \text{shape}(M)$\;

    \For{$i \text{ in range}(n)$}{
        \For{$j \text{ in range}(m)$}{

            $c = \text{int}(\text{round}(\rho \cdot M[i][j]))$\;

            \For{$a \text{ in range}(c)$}{
                Initialise agent and add to system\;
            }
        }
    }
    \label{alg:coordinate_agents}
\end{algorithm}
\smallskip
For agents initialised at epochs other than zero we follow Algorithms~\ref{alg:preprocess} and \ref{alg:binary_search}. The coordinates of new agents are determined by converting the discretised density map into a normalised cumulative distribution function, a procedure commonly referred to as inverse transform sampling. Valid coordinates are assigned intervals on the $[0,1]$ scale proportional to their contribution to the total density, while zero-density cells are excluded so that no agent is initialised at a coordinate with zero density. 
This method ensures that the spatial distribution of generated agents is directly weighted by the density in the empirical density map.
\smallskip
\begin{algorithm}[H]
    \LinesNotNumbered
    \SetKwInOut{Input}{Input}
    \SetKwInOut{Output}{Output}
    
    \caption{Populate agents following the empirical density map}
    \label{alg:preprocess}
    
    \Input{Density map $M$ (numpy array)}
    \Output{
        $D$: list of cumulative densities \\
        $C$: list of corresponding coordinates $[x, y]$
    }
    
    $n, m = \text{shape}(M)$\;
    $S = 0$\;
    $D = [\ ]$\; 
    $C = [\ ]$\;
    
    \For{$i$ in $range(n)$}{
        \For{$j$ in $range(m)$}{
            \If{$M[i][j] > 0$}{
                $S \ += M[i][j]$\;
                $D.\text{append}(S)$\;
                $C.\text{append}([i,j])$\;
            }
        }
    }
    
    $D \ /= S$\;
    
    \Return{$D, C$}
\end{algorithm}
\smallskip
\begin{algorithm}[H]
    \LinesNotNumbered
    \SetKwInOut{Input}{Input}
    \SetKwInOut{Output}{Output}

    \caption{Binary search for coordinate retrieval}
    \label{alg:binary_search}

    \Input{
        $r$: random value sampled from $U(0,1)$ \\
        $D$: list of cumulative densities (sorted) \\
        $C$: list of corresponding coordinates $[x, y]$
    }
    \Output{Coordinate pair corresponding to $r$}

    $low = 0$\;
    $high = \text{length}(D)$\;

    \While{$low \le high$}{
        $mid = \lfloor (low + high) / 2 \rfloor$\; 
        
        \If{$r < D[mid]$}{
            $high = mid - 1$\;
        }
        \uElseIf{$r > D[mid]$}{
            $low = mid + 1$\;
        }
        \lElse{
            \Return $C[mid]$
        }
    }

    \If{$mid == low$}{
        \Return $C[mid]$\;
    }
    \lElse{
        \Return $C[low]$
    }

\end{algorithm}

\subsection{Income Dynamics and Unemployment Benefit Scheme}
\label{sec:appendix_unemployment}
Income is determined by employment status. In this section we describe the seniority dynamics and the unemployment benefit scheme; the benefit dynamics closely match the unemployment insurance schemes of the euro area.

The layoff threshold $\psi^{e}_{i}$ is drawn once at entry to represent persistent differences in employment vulnerability associated with education, skills and job characteristics. \textcite{Cairo_Cajner_2018} document lower and less volatile job-separation rates among more educated workers and relate this pattern to investment in job-specific human capital. \textcite{Chapuis_Coglianese_2024} also find substantial heterogeneity in unemployment risk among employed workers, with a broadly stable ratio of high to low risk over the business cycle. We therefore interpret $\psi^{e}_{i}$ as a latent resilience rank, without explicitly modelling these characteristics.

The seniority of each agent evolves based on the employment status. For each time step an agent remains employed, their seniority increases. If they become unemployed, their seniority freezes until they find new employment. This dynamic is capped at a maximum value to represent a realistic career progression plateau. The update rule for an employed agent is as follows:
\begin{equation}
\label{eqn:sen}
\mathit{sen}_{i,t} = \min\left(\mathit{sen}_{i,t-1} + 1,\; \alpha \right)
\end{equation}
Seniority increases by one at each employed time step and is capped at $\alpha$ (maximum seniority). The seniority of unemployed agents does not change.
Income dynamics are linked to employment status. We model unemployment benefits following the Dutch unemployment benefit scheme administered by the Employee Insurance Agency (UWV) \citep{UWV_ww,GovNL_ww}, which reflects a tiered structure common across European unemployment benefit systems. The agents' income is updated at each time step according to one of four cases:
\begin{enumerate}
\item \textbf{Continuous Employment:} If an agent remains employed, their income is constant: $I_{i,t} = I_{i,t-1}$.

\item \textbf{Job Loss:} If an agent becomes unemployed, their income is replaced by an unemployment benefit, calculated as 70\% of their last income, but capped at a maximum of \EUR 5,000: $I_{i,t} = \min\left(I_{i,t-1} \cdot 0.7, 5000\right)$. 
\item \textbf{Long-term Unemployment:} If an agent remains unemployed for more than 12 months, their benefits cease, and their income drops to zero ($I_{i,t}=0$).
\item \textbf{New Employment:} When an unemployed agent finds a new job, their new income is drawn from the initial income distribution (see \SItab{tab:parameter_distributions}), representing a new salary ($I_{i,t} \sim Burr(c=3.30, d=0.45, \text{loc}=-12.76, \text{scale}=3101.46)$).
\end{enumerate}

Unemployment insurance in the Netherlands, France, Germany, Spain, Italy and Belgium follows the same design: benefits replace between roughly half and three quarters of the last wage, decline over the unemployment spell and end after a finite duration \citep{UWV_ww,ServicePublic_ARE,BA_arbeitslosengeld,SEPE_benefits,INPS_unemployment,ONEM_site,OECD_benefits_unemployment}. The model uses a single replacement rate of 70\% and a benefit duration of 12 months, and can be adapted to another country by changing these two values.

\subsection{Credit Scorecard}
\label{sec:app:scorecard}

The lender screens each applicant with a points-based scorecard over five risk drivers: the loan-to-value ratio, income, work seniority, the ratio of liquid funds to the loan, and the monthly instalment, following the choice of drivers in \textcite{Lin2024ESGClimate}. Each driver is divided into bands, and the points awarded rise in equal steps from the riskiest band to the safest (\SItab{tab:score_card}). The band limits are stylised and are placed across the bulk of the applicant distribution, so that each driver ranks applicants over the range in which they differ. An applicant is admitted if the total score reaches the threshold $\mathcal{A} = 200$, out of a maximum of 300.

\begin{table}[ht]
    \centering
    \caption{Admission scorecard.}
    \label{tab:score_card}
    \begin{tabular}{l S[table-format=2.1] l}
        \toprule
        Risk driver & {Score} & Band \\
        \midrule
        $\LTV_{i}$ & 50   & $\LTV \leq 0.4$ \\
                           & 37.5 & $0.4 < \LTV \leq 0.6$ \\
                           & 25   & $0.6 < \LTV \leq 0.8$ \\
                           & 12.5 & $\LTV > 0.8$ \\
        \addlinespace
        \midrule
        \addlinespace
        $I_{i}$ (euro per month) & 13 & $I \leq 1500$ \\
                           & 26 & $1500 < I \leq 2000$ \\
                           & 39 & $2000 < I \leq 2500$ \\
                           & 52 & $2500 < I \leq 3000$ \\
                           & 65 & $I > 3000$ \\
        \addlinespace
        \midrule
        \addlinespace
        $\mathit{sen}_{i}$ (months) & 13 & $\mathit{sen} \leq 100$ \\
                           & 26 & $100 < \mathit{sen} \leq 200$ \\
                           & 39 & $200 < \mathit{sen} \leq 300$ \\
                           & 52 & $300 < \mathit{sen} \leq 400$ \\
                           & 65 & $\mathit{sen} > 400$ \\
        \addlinespace
        \midrule
        \addlinespace
        $\zeta_{i}$        & 20 & $\zeta \leq 0.4$ \\
                           & 40 & $0.4 < \zeta \leq 0.6$ \\
                           & 60 & $0.6 < \zeta \leq 0.8$ \\
                           & 80 & $\zeta > 0.8$ \\
        \addlinespace
        \midrule
        \addlinespace
        $P_{i}$ (euro per month) & 40 & $P \leq 800$ \\
                           & 30 & $800 < P \leq 1000$ \\
                           & 20 & $1000 < P \leq 1200$ \\
                           & 10 & $P > 1200$ \\
        \bottomrule
    \end{tabular}
    \tablenote{The total score is the sum across the five risk drivers, evaluated on the applicant's attributes at entry. Applicants with a total score of at least $\mathcal{A} = 200$ are admitted to the simulation; all others are rejected.}
\end{table}

Admission is a deterministic function of the applicant's attributes at entry, so its effect on the population follows directly from the distributions in \SItab{tab:parameter_distributions}. It does not depend on the flood scenario or on any of the parameters varied in the sensitivity analysis. The lender admits 16.5\% of applicants. \SItab{tab:accepted_population} compares the applicants with the admitted households.

\begin{table}[ht]
    \centering
    \caption{Applicants and admitted households at entry: median and interquartile range of each attribute, computed from $2 \cdot 10^{6}$ applicant draws.}
    \label{tab:accepted_population}
    \begin{tabular}{l l l}
        \toprule
        Attribute & Applicants & Admitted \\
        \midrule
        Income $I_i$ (euro per month)         & 2080 [1224, 3196] & 3521 [2866, 4620] \\
        Funds over loan $\zeta_i$             & 0.65 [0.38, 0.90] & 0.92 [0.80, 1.06] \\
        Instalment over income $P_i/I_i$     & 0.43 [0.24, 0.84] & 0.19 [0.13, 0.27] \\
        Instalment $P_i$ (euro per month)    & 877 [608, 1254]   & 686 [502, 899] \\
        Loan-to-value $\LTV_i$        & 0.57 [0.50, 0.64] & 0.54 [0.47, 0.59] \\
        Property value $v_i$ (thousand euro)  & 187 [133, 261]    & 155 [115, 202] \\
        Seniority $\mathit{sen}_i$ (months)            & 21 [9, 39]        & 21 [9, 39] \\
        \bottomrule
    \end{tabular}
\end{table}

Selection operates mainly through income and liquid funds. Among admitted households, 70\% have an income above 3000 euro per month and 75\% hold liquid funds above 80\% of their loan, against 28\% and 35\% of applicants. The instalment and the loan-to-value ratio select more weakly, towards smaller loans on less expensive properties. As a result the instalment absorbs 19\% of income for the median admitted household, against 43\% for the median applicant, and 95\% of admitted households pay less than 40\% of their income, against 46\% of applicants. Seniority does not discriminate at admission: 99.95\% of applicants fall in its first band, so the driver adds the same 13 points to almost every score. Seniority enters the model only through this score, so it has no effect on income, distress or foreclosure after admission.

Since liquid funds decide admission, the scale of $\zeta$ sets the admission rate and, through it, the size of the population. In the baseline scenario the share of applicants admitted rises from $5.9\%$ at a scale of $0.01$ to $16.5\%$ at $0.015$ and $22.6\%$ at $0.02$, and the mean number of active households from about 960 to 2,700 and 3,700. Over the same range the foreclosure and distress rates per household stay within the bounds given above ($0.39\%$ to $0.46\%$ and $0.46\%$ to $0.54\%$).

\subsection{Willingness to Pay, Distress and Foreclosure}
\label{sec:app:wtp}
\subsubsection{Debt-to-Income Ratio in the Debt Pressure}
\label{sec:app:dti}
The debt pressure of Equation~\ref{eqn:wtp} uses the debt-to-income ratio
\begin{equation}
\label{eqn:dti}
\mathrm{DTI}_{i,t} = \frac{\ell_{i,t}}{12\, I_{i,t}},
\end{equation}
which relates the outstanding loan $\ell_{i,t}$, a stock, to annual income $12\,I_{i,t}$, where $I_{i,t}$ is the monthly income of Section~\ref{sec:income_empl_dynamics}. Expressed on this annual scale, the ratio takes the values of a few multiples of income at which debt-to-income ratios are commonly reported, and the half-saturation parameter $\beta_3$ has a direct reading: at the baseline $\beta_3=1$, an employed household ($\tau_{i,t}=0$) experiences half the maximum debt pressure $\beta_2$ when its outstanding loan equals one year of income.

A household whose unemployment benefits have ended has $I_{i,t}=0$ (\SIsec{sec:appendix_unemployment}), and \SIeq{eqn:dti} is unbounded. In this case the debt pressure takes its limiting value, $\beta_2 X/(\beta_3+X) = \beta_2$, the maximum debt pressure a household can experience.

\subsubsection{Neighbourhood and Foreclosure Count}
\label{sec:app:neighbourhood}
Each household occupies a cell $p_i$ of the grid. Two households are neighbours when the straight-line distance between their cells is at most $\kappa$ cells, $\lVert p_i - p_j \rVert \leq \kappa$. In each month $t$ the model runs two radius-$\kappa$ queries around agent $i$:
\begin{enumerate}
\item over the households with an active mortgage, which returns the neighbourhood $\mathcal{H}^{\kappa}_{i,t}$ of Equation~\ref{eqn:utility} (agent $i$ included);
\item over the households foreclosed in month $t-1$, each at its last cell, which returns the foreclosure count $n^{\kappa}_{i,t-1}$ of Equation~\ref{eqn:collateral}.
\end{enumerate}
A foreclosed household leaves the model, so the two queries never return the same household. The grid does not wrap: a neighbourhood near the map edge is cut off at the boundary.

\subsubsection{Distress and Foreclosure Probabilities}
\label{sec:app:distress_probs}

There are two main mechanisms that have been proposed to understand the spillovers that ultimately trigger a stop in the payments. They are separated by whether the effect requires a price effect or not. The first mechanism operates through market prices. In the \textit{fire-sale} route, a lender that must liquidate quickly sells into a thin local market at a deep discount, and that discounted transaction enters the neighbourhood price level through the comparable sales on which buyers, sellers and appraisers rely \parencite{Campbell_Giglio_Pathak_2011, Mian_Sufi_Trebbi_2015}. In the \textit{competition} route, the additional listing bids for the same buyers as the houses already for sale nearby, so their sellers must cut asking prices until the inventory clears \parencite{Anenberg_Kung_2014}. Both require a transaction, both operate through the price at which a house sells, and both are transitory: \textcite{Anenberg_Kung_2014} date the decline to the listing of the foreclosed house rather than to the foreclosure itself, and find that nearby prices return to their pre-listing level within $6$ weeks of the sale of a single foreclosed house and within $12$ weeks of the sale of several.

The second mechanism does not operate through prices: the household observes what is happening around it, and this alone changes behaviour. A damaged property lowers the value and liveability of its neighbours directly \parencite{Gerardi_Rosenblatt_Willen_Yao_2015, Fisher_LambieHanson_Willen_2015}, while observing distress next door also lowers the reputational and informational cost of ceasing to pay oneself, an effect documented in survey attitudes and in default hazards that respond to neighbours' behaviour independently of local prices \parencite{Guiso_sapienza_zingales2017, Towe_Lawley_2013, Goodstein2017, Gupta_2019}. The model includes both mechanisms. Flood damage to neighbouring properties enters $U^{\kappa}_{i,t}$ directly through the collateral $c_{j,t}$ (Equation~\ref{eqn:utility}), and nearby foreclosures enter through the foreclosure spillover of intensity $\delta$ (Equation~\ref{eqn:collateral}), which represents both their price effect and their effect on the perceived cost of default \parencite{Gupta_2019}. The spillover applies within the same radius $\kappa$ as the neighbourhood utility. The compounding form $(1-\delta)^{n}$ keeps the collateral positive for any number of foreclosures. The spillover lasts one step, the shortest horizon of the model, so that every foreclosure induced by the spillover follows a foreclosure of the previous step and can be traced to it.

The debt pressure in Equation~\ref{eqn:wtp} is $\beta_2 X/(\beta_3+X)$, where $X=\mathrm{DTI}_{i,t}(\tau_{i,t}+1)$.
We adopt this rational form for three reasons. First, and decisively, a debt-free household carries no financial-stress penalty. Second, the financial penalty is monotone increasing in $X$ and approaches $\beta_2$ as $X\to\infty$, with $\beta_3$ marking the value of $X$ at which the penalty reaches half its maximum. Third, the term stays bounded in $[0,\beta_2]\subset[0,1]$, so that $\text{wtp}_{i,t}$, clipped to $[0,1]$, can be read as a probability: the probability that the agent does not enter strategic distress. 

An agent who can pay enters strategic distress when:
\begin{equation}
\label{eqn:strategic_trigger}
\text{wtp}_{i,t} < \psi^{d}_{i,t}
\end{equation}

where $\psi^{d}_{i,t} \sim U[0,1]$ is an agent specific reservation threshold, redrawn independently each period and independent of the layoff threshold $\psi^{e}_{i}$. As with the financial condition, this only marks a distress event: it does not by itself end the mortgage.
A household in distress this period, whether through Equation~\ref{eqn:financial_distress} or \SIeq{eqn:strategic_trigger}, is either granted forbearance or foreclosed within the period, according to its willingness to pay and the same draw $\psi^{d}_{i,t}$:
\begin{equation}
\label{eqn:strategic_distress}
\begin{cases}
\text{Forbearance:} &
  \text{wtp}_{i,t} > 1-\psi^{d}_{i,t}
  \;\Rightarrow\; \text{forbearance is granted, the household is retained} \\[4pt]
\text{Foreclosure:} &
  \text{wtp}_{i,t} \leq 1-\psi^{d}_{i,t}
  \;\Rightarrow\; \text{the mortgage is foreclosed.}
\end{cases}
\end{equation}

\SIeq{eqn:strategic_distress} resolves distress from either financial (Equation~\ref{eqn:financial_distress}) or strategic (\SIeq{eqn:strategic_trigger}) origin, against the draw of $\psi^{d}_{i,t}$. 
The two origins are not symmetric with respect to that draw: the financial condition does not involve $\psi^{d}_{i,t}$ at all, so a household triggered financially carries no information about where its draw falls, while a household triggered strategically is there precisely because $\psi^{d}_{i,t}$ exceeded $\text{wtp}_{i,t}$. A household in financial distress is therefore forborne with probability $\text{wtp}_{i,t}$ and ends in foreclosure with probability $1-\text{wtp}_{i,t}$, so $\text{wtp}_{i,t}$ governs the outcome of the financial channel as well as that of the strategic one. The strategic channel admits a closed form in terms of $\text{wtp}_{i,t}$. 

Taking the strategic channel on its own, that is, treating $\psi^{d}_{i,t}$ as the household's only potential source of distress this period, $\psi^{d}_{i,t}$ can be integrated out of Equations~SI~(\ref{eqn:strategic_trigger})--SI~(\ref{eqn:strategic_distress}), turning the trigger-and-resolution pair into a closed-form function of $\text{wtp}_{i,t}$ alone, as reported in Equation~\ref{eqn:distress_probs}.
The three probabilities sum to one and split the channel into two regimes at $\text{wtp}_{i,t}=1/2$. In the \textit{foreclosure-possible} regime ($\text{wtp}_{i,t}<1/2$) an agent faces $\Pr(\text{no distress})=\Pr(\text{forbearance})=\text{wtp}_{i,t}$ and $\Pr(\text{foreclosure})=1-2\,\text{wtp}_{i,t}$: it is exactly as likely to avoid distress entirely as it is to fall into distress and be retained under forbearance, and its foreclosure risk declines linearly from certainty at $\text{wtp}_{i,t}=0$ to zero at $\text{wtp}_{i,t}=1/2$. In the \textit{retention-only} regime ($\text{wtp}_{i,t}\ge1/2$) it faces $\Pr(\text{no distress})=\text{wtp}_{i,t}$, $\Pr(\text{forbearance})=1-\text{wtp}_{i,t}$, and $\Pr(\text{foreclosure})=0$: the strategic condition can still trigger distress, but it cannot foreclose the mortgage. These three probabilities are the content of the strategic channel. The draw $\psi^{d}_{i,t}$ is only the device that generates them: it contributes the randomisation that converts $\text{wtp}_{i,t}$ into a probability, consistent with its role as a redrawn, non-persistent draw, and it carries no interpretation of its own as a trait of the household. The economic content is carried entirely by $\text{wtp}_{i,t}$, through the neighbourhood and debt pressure components of Equation~\ref{eqn:wtp}.

The $1{,}162$ sampled parameter vectors with $\beta_1-\beta_2\ge1/2$ sustain between $514$ and $2{,}709$ active mortgages at $\mathcal{S}_0$. Among them the population falls as the spillover intensity grows, from a median of $2{,}550$ for $\delta\le 0.15$ to $1{,}566$ for $\delta\ge 0.35$. Below $\beta_1-\beta_2=1/2$ the population shrinks as the gap narrows, from a median of $1{,}229$ for $0.4<\beta_1-\beta_2<0.5$ to $16$ for $\beta_1-\beta_2\le 0.1$. The spillover has its largest effect near $\beta_1-\beta_2=1/2$: for $0.4<\beta_1-\beta_2<0.5$ the median falls from $2{,}341$ for $\delta\le 0.15$ to $622$ for $\delta\ge 0.35$. The spread is wide at every gap because the population also depends on the remaining parameters.

\begin{figure}[htbp]
    \centering
    \includegraphics[width=0.6\textwidth, keepaspectratio]{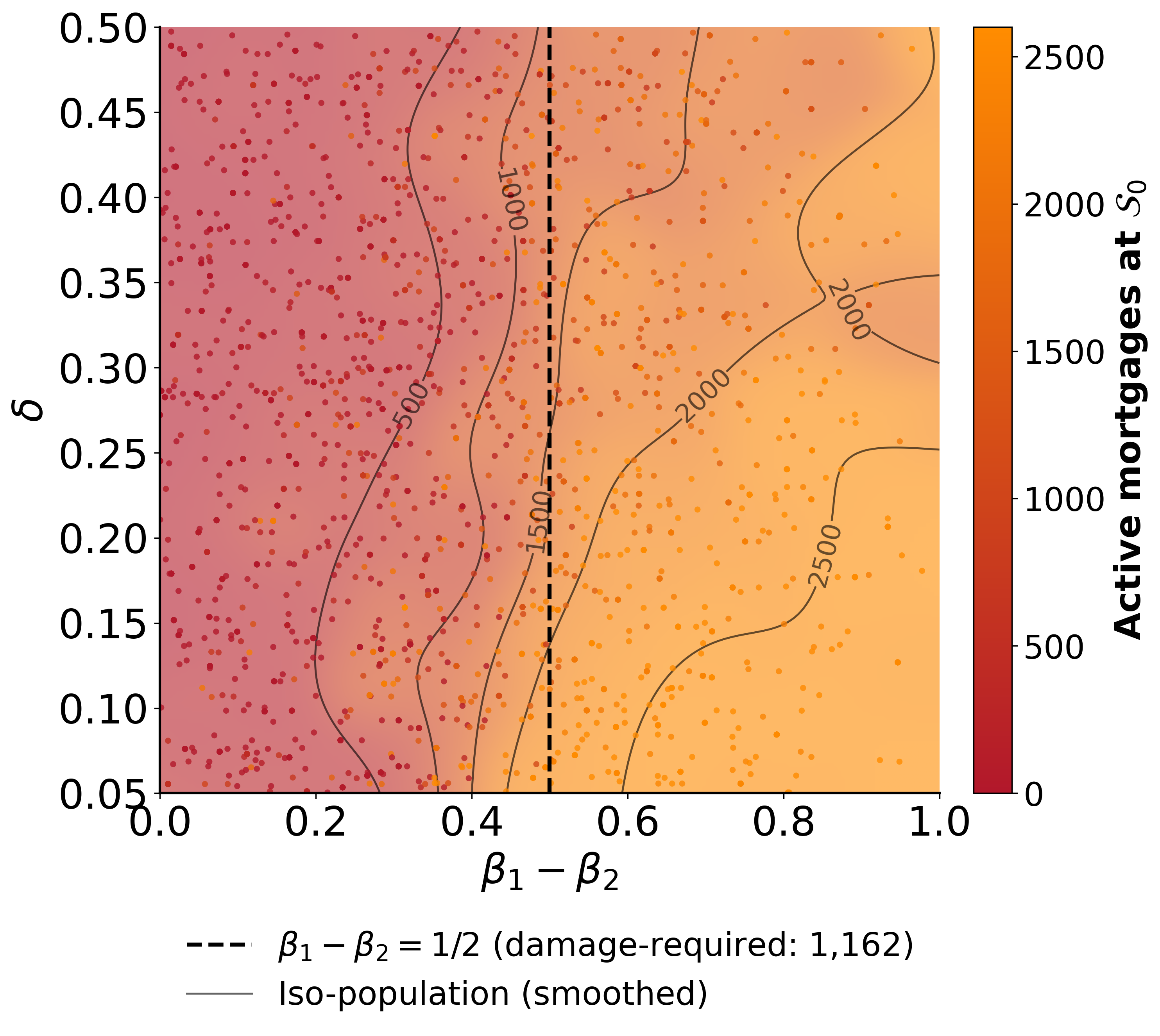}
    \caption{\small Mean number of active mortgages at $\mathcal{S}_0$ across the plane of the gap $\beta_1-\beta_2$ and the spillover intensity $\delta$, over the $3{,}328$ parameter vectors of the GSA design (\SIsec{sec:app:sa}). Points are coloured by the mean active population over seeds. The background is a Gaussian kernel smoothing of the same quantity, with bandwidth $0.045$ on the gap and $0.045$ times the sampled range of $\delta$ on $\delta$; the solid lines are iso-population contours of a smoothing with bandwidth $0.07$. The dashed line marks $\beta_1-\beta_2=1/2$; $1{,}162$ parameter vectors lie on or above it. The design samples $\beta_2<\beta_1$ only. The baseline $(\beta_1-\beta_2,\delta)=(0.5,0)$ lies below the sampled range of $\delta$.}
    \label{fig:beta_regions}
\end{figure}

\clearpage
\section{Calibration}
\label{sec:app:calibration}
In this section we set the baseline parametrisation of the model and calibrate the model to empirical credit outcomes. The baseline values of the parameters with a direct empirical counterpart come from data on the model inputs only (household expenditure and unemployment), and they centre the parameter ranges of the Global Sensitivity Analysis (GSA) in \SItab{tab:sa_ranges}. The calibration then uses two credit outcomes without floods, the annual foreclosure rate and the share of households in financial distress, to determine which parameter combinations within these ranges reproduce the empirical evidence (\SIsec{sec:app:identification}). The GSA of \SIsec{sec:app:sa} explores the same ranges.
\subsection{Baseline parametrisation}
\label{sec:baseline_parametrization}

We set the baseline values of three parameters governing the non-housing expenditure rule of household agents: the intercept $m$, the income-share decay coefficient $\varepsilon$, and the subsistence floor $\mathit{exp}_{lb}$ (see Equation~\ref{eqn:income_share}). We use CBS StatLine data (Table 85887NED), which reports average household income and consumption expenditure by income decile for the Netherlands over the period 2021--2023. For each year and each of the ten income deciles we extract: average gross disposable income per household (\textit{BrutoBeschikbaarInkomen}), total consumptive expenditure (\textit{TotaalConsumptieveBestedingen}), and housing expenditure under COICOP classification 04 (\textit{HuisvestingWaterEnEnergie}), all reported in thousands of EUR per household per year. Pooling three years of decile observations yields $n = 30$ data points spanning a monthly income range of approximately \EUR{550} to \EUR{12{,}900}. We define the non-housing expenditure share for each decile observation $k$ as $ s_k = \frac{C_k - H_k}{Y_k}$,
where $C_k$ is total consumption, $H_k$ housing expenditure, and $Y_k$ gross disposable income; the thousand-euro units cancel in the ratio. Monthly income is recovered as $I_k = Y_k \cdot 1000 / 12$ (EUR per month). $s_k$ is the sample-level analogue of the agent share $s_{i,t}$ in Equation~\ref{eqn:income_share}.

We estimate $m$ and $\varepsilon$ by nonlinear least squares (NLS), fitting the hyperbolic Engel curve directly to the $n=30$ pairs $(I_k,\, s_k)$.
To ensure numerical stability we reparametrise income in units of thousands of EUR ($\tilde{I}_k = I_k / 1000$), fitting a scaled coefficient $\tilde{\varepsilon}$ and post-transforming as $\varepsilon = \tilde{\varepsilon}/1000$. We use initial values and impose non-negativity constraints. The NLS estimator minimises residuals in the share space $s$, which is the space the model directly uses, and assigns equal weight to all decile observations. As a robustness check, a linearised OLS estimator (regressing $1/s_{i,t}$ on $I_{i,t}$) yields $m_{\mathrm{OLS}} = 1.55$ and $R^2 = 0.48$ in share space, confirming that NLS with $R^2 = 0.94$ provides substantially better fit in the economically relevant space. The estimated parameters are:
\begin{align*}
m &= 6.22 \quad (s.e. = 0.79), \\
\varepsilon &= 1.98 \cdot 10^{-3} \text{ per EUR} \quad (s.e. = 3.61 \cdot 10^{-4} ).
\end{align*}
These estimates are the baseline values of $m$ and $\varepsilon$ and the centre of their GSA ranges, which extend $30\%$ on either side (\SItab{tab:sa_ranges}). The CBS deciles describe all Dutch households, while $m$ and $\varepsilon$ govern the non-housing expenditure of mortgage holders in Limburg. The estimates therefore centre the ranges explored by the GSA rather than fix the two parameters, and the calibration of \SIsec{sec:app:identification} determines which values within these ranges reproduce the empirical credit outcomes.

We set $\mathit{exp}_{lb}$ to the Dutch statutory social assistance norm (\textit{bijstandsnorm}), the legally defined minimum household income floor in the Netherlands. The norm stood at \EUR{1{,}021} per month in 2021, \EUR{1{,}046} in 2022, and \EUR{1{,}195} in 2023 \citep{SZW_2021jul,SZW_2022jan,SZW_2023jul}. These amounts are set uniformly at the national level and therefore apply to the Limburg region. We adopt the mean of the three years, \EUR{1{,}087} per month, as our baseline.

We set the parameters of Equation \ref{eqn:unemployment}, that is $\eta$ and $\vartheta$ in line with the data of the Dutch historical unemployment with a reported peak of 8.3\% in 2014 and a lower bound of 3.4\% in 2019 \citep{Salverda2025,CBS2023arbeidsmarkt}. The values of $\eta$ and $\vartheta$ in the GSA are set to conservative upper bounds providing additional margin to account for broad labour underutilisation above any observed Dutch value (See \SItab{tab:sa_ranges}).
We estimate the baseline of $\eta$, the steady-state unemployment rate in Equation~\ref{eqn:unemployment}, as the average monthly seasonally-adjusted unemployment rate over the period 2021--2025 from CBS data (table 80590NED, \textcite{CBS80590NED}), obtaining 3.77\%. We adopt $\eta = 0.037$ (3.7\%) as the empirically grounded baseline. In the GSA, $\eta$ is varied over the range in \SItab{tab:sa_ranges}.

\subsection{Calibration to empirical credit outcomes}
\label{sec:app:identification}
The credibility of an ABM rests on its ability to match empirical outcomes quantitatively, not only qualitatively \parencite{pangallo2024datadriven}. We calibrate the model on the two credit outcomes of Section~\ref{sec:empirical_validation} without floods ($\mathcal{S}_0$): the annual foreclosure rate, against the NHG band ($0.27$--$0.42\%$), and the share of households in financial distress, against the BKR band ($0.87$--$3.54\%$). Following \textcite{Lamperti_Roventini_Sani_2018}, we define calibration as the search for the parameter combinations whose output satisfies a condition on its distance to the data, rather than as the estimation of a single parameter vector. We implement the search as rejection approximate Bayesian computation (ABC) \citep{Beaumont_Zhang_Balding_2002,Grazzini_Richiardi_Tsionas_2017}: parameter combinations are drawn from a prior distribution and simulated, a combination is retained when the distance between its simulated outcomes and the empirical ones falls within a tolerance, and the retained combinations approximate the posterior distribution of the parameters.

The prior is uniform on the GSA ranges (\SItab{tab:sa_ranges}) of the nine parameters that act at $\mathcal{S}_0$ when $\delta=0$ ($\kappa$, $\beta_1$, $\beta_2$, $\beta_3$, $\eta$, $\vartheta$, $m$, $\varepsilon$ and $\mathit{exp}_{lb}$), restricted to $\beta_1-\beta_2\ge 1/2$, where every foreclosure at $\mathcal{S}_0$ with $\delta=0$ is financial (Section~\ref{sec:empirical_validation}). The damage share $\lambda$ has no effect without floods, and $\delta=0$ as in Section~\ref{sec:empirical_validation}. We draw the combinations from a scrambled Sobol' sequence \citep{Sobol_1967} of $2^{13}$ points and keep the $1{,}001$ that satisfy this restriction. Each combination is simulated with $10$ seeds and summarised by its mean over seeds; among the retained combinations, the standard error of the mean foreclosure rate is $0.008$ percentage points, small relative to the width of the NHG band. Each combination is marginally distributed as the prior.

For combination $j$ and outcome $k$, let $y_{jk}$ be the simulated outcome and $B_k=[a_k,b_k]$ its empirical band. The distance of combination $j$ to the data is
\begin{equation}
d_j=\sqrt{\sum_{k}\left(\frac{g(\ln y_{jk};\ln a_k,\ln b_k)}{\sigma_k}\right)^2}, \qquad g(x;a,b)=\max(a-x,0)+\max(x-b,0),
\label{eqn:abc_distance}
\end{equation}
where $\sigma_k$ is the standard deviation of $\ln y_{jk}$ across combinations. The distance is zero when both outcomes fall inside their bands. Weighting each outcome by the inverse of its standard deviation is the usual choice in ABC \citep{Grazzini_Richiardi_Tsionas_2017}; logarithms account for both outcomes being positive and right-skewed. We retain the combinations with $d_j\le h$, with the tolerance $h$ set to the $5\%$ quantile of $d_j$ ($51$ combinations) and, as a robustness check, to the $1\%$ quantile ($11$ combinations).

A parameter is constrained by the calibration when the distribution of its retained values differs from its prior distribution. Regional sensitivity analysis compares the distribution of each input among the runs that meet a condition on the output with its distribution among the remaining runs through the Kolmogorov distance \citep{Spear_Hornberger_1980}. We adapt it to the calibration: for each parameter we compute the Kolmogorov--Smirnov (KS) statistic between its retained values and its values across all $1{,}001$ combinations. Since the retained combinations are a subset of the reference, the standard KS $p$-values do not apply. We obtain the null distribution from $2{,}000$ random subsets of the reference of the same size as the retained set, record for each subset the maximum KS statistic over the nine parameters, and declare a parameter constrained when its KS statistic exceeds the $95\%$ quantile of this maximum, a single-step max-$T$ procedure \citep{Westfall_Young_1993}. The threshold is $0.23$ at the $5\%$ tolerance and $0.49$ at the $1\%$ tolerance. The comparison is with the prior sample and not with the uniform distribution on the sampled ranges, because the condition $\beta_1-\beta_2\ge 1/2$ already concentrates $\beta_1$ at high and $\beta_2$ at low values. Sampling uncertainty is quantified by resampling the combinations with replacement $1{,}000$ times with $h$ fixed.

Three parameters are constrained by the calibration (Figure~\ref{fig:abc_identification} and \SItab{tab:abc_ks}): the steady-state unemployment rate $\eta$ (KS statistic $0.71$, bootstrap $90\%$ interval $0.68$--$0.75$), retained between $0.032$ and $0.056$ (5th to 95th percentile) of its range $[0.03,0.13]$; the expenditure intercept $m$ ($0.55$, interval $0.51$--$0.61$), retained between $4.38$ and $5.90$ of $[4.34,8.06]$; and the income-share decay coefficient $\varepsilon$ ($0.33$, interval $0.28$--$0.43$), retained between $1.78\cdot10^{-3}$ and $2.55\cdot10^{-3}$ of $[1.39\cdot10^{-3},2.57\cdot10^{-3}]$. All three exceed the threshold in every bootstrap resample, and the KS statistics computed separately on the two halves of the Sobol' sequence differ by less than $0.05$. At the $1\%$ tolerance $\eta$ and $m$ remain above the threshold ($0.80$ and $0.56$), while $\varepsilon$ falls below it ($0.42$). The remaining six parameters stay below the threshold (KS statistics $0.10$--$0.14$), in at least $83\%$ of the bootstrap resamples. Values of $\beta_1$ close to $1$ among the retained combinations therefore reflect the restriction of the prior to $\beta_1-\beta_2\ge1/2$ rather than the data. Among the retained combinations, $m$ and $\varepsilon$ are positively correlated (Spearman correlation $0.54$, zero in the prior): a higher intercept of the expenditure share is compensated by a faster decline of the share with income.

\begin{table}[htbp]
\centering
\small
\caption{Kolmogorov--Smirnov (KS) statistic between the retained and the reference values of each parameter at the $5\%$ tolerance, with its bootstrap $90\%$ interval and the share of the $1{,}000$ resamples in which it exceeds the threshold. The max-$T$ threshold is the $95\%$ quantile of the maximum KS statistic over the nine parameters across $2{,}000$ random subsets of the reference of the same size as the retained set. Statistics in bold exceed it. The adjusted $P$ value is the share of these subsets whose maximum KS statistic is at least the observed one, computed as $(1+\#)/(2{,}001)$ \citep{Phipson_Smyth_2010} (one-sided, family-wise over the nine parameters); $<0.001$ marks the smallest attainable value, $1/2{,}001$.}
\label{tab:abc_ks}
\begin{tabular}{l c c c c}
\toprule
Parameter & KS & Bootstrap $90\%$ interval & Share above threshold & Adjusted $P$ \\
\midrule
$\kappa$ & $0.14$ & $[0.09,\,0.27]$ & $0.17$ & $0.87$ \\
$\beta_1$ & $0.11$ & $[0.10,\,0.24]$ & $0.07$ & $1.00$ \\
$\beta_2$ & $0.10$ & $[0.09,\,0.23]$ & $0.05$ & $1.00$ \\
$\beta_3$ & $0.11$ & $[0.09,\,0.24]$ & $0.07$ & $1.00$ \\
$\eta$ & $\mathbf{0.71}$ & $[0.68,\,0.75]$ & $1.00$ & $<0.001$ \\
$\vartheta$ & $0.10$ & $[0.09,\,0.22]$ & $0.03$ & $1.00$ \\
$m$ & $\mathbf{0.55}$ & $[0.51,\,0.61]$ & $1.00$ & $<0.001$ \\
$\varepsilon$ & $\mathbf{0.33}$ & $[0.28,\,0.43]$ & $1.00$ & $<0.001$ \\
$\mathit{exp}_{lb}$ & $0.12$ & $[0.11,\,0.25]$ & $0.08$ & $0.99$ \\
\midrule
Max-$T$ threshold & $0.23$ & & & \\
\bottomrule
\end{tabular}
\end{table}

The two outcomes carry largely the same information. For $\beta_1-\beta_2\ge1/2$ with $\delta=0$ no strategic foreclosure occurs at $\mathcal{S}_0$, so every foreclosure follows financial distress, and the logarithms of the two outcomes have a correlation of $0.93$ across combinations. Financial distress falls inside the BKR band for $512$ combinations, the foreclosure rate inside the NHG band for $6$, and both outcomes inside their bands for $1$. The retained combinations are therefore those with the lowest foreclosure rates, $0.43$--$0.65\%$ per year (5th to 95th percentile), with financial distress between $0.56\%$ and $1.49\%$.

The calibrated expenditure parameters lie outside the joint $95\%$ confidence region of the estimates of \SIsec{sec:baseline_parametrization}. None of the retained combinations falls inside it; inside it, the foreclosure rate is at least $0.66\%$ per year, above the NHG band, while financial distress is at least $1.56\%$, inside the BKR band. The baseline $m=6.2$ lies above the retained range, whose upper end has a bootstrap $90\%$ interval of $5.69$ to $6.11$. The baseline values $\eta=0.037$ and $\varepsilon=1.98\cdot10^{-3}$ lie inside the retained ranges.

\clearpage
\section{Global Sensitivity Analysis}
\label{sec:app:sa}
\subsection{Method}
\label{sec:app:gsa_method}
In this section we detail the GSA on the key parameters of the model. We perform the GSA using Sobol' indices \citep{Sobol_2001} to characterise contributions to output variance and PAWN indices $T_i$ to characterise changes in the full output distribution. The starting point of both analyses is a common sample of the parameter space, for which we use the Saltelli scheme \citep{Saltelli_2002,Saltelli_Annoni_Azzini_Campolongo_Ratto_Tarantola_2010}.
Both analyses are performed over the region of the parameter space defined by $\beta_1>\beta_2$. Parameter combinations violating this condition are excluded because near-total default depletes the mortgage population, making them unsuitable for the analysis. This sampling restriction is distinct from the restriction $\beta_1-\beta_2\geq1/2$ used in the validation and calibration (Section~\ref{sec:empirical_validation}): the GSA includes combinations on both sides of $\beta_1-\beta_2=1/2$.
To enforce $\beta_1 > \beta_2$ in every row of the Saltelli matrix, we adopt a rejection-and-reconstruction procedure. Because the constraint discards a large share of the base points, more points must be drawn than the $N = 256$ the design requires. We therefore generate $M = 2^{13} = 8{,}192$ points of a scrambled Sobol' sequence in $2k = 22$ dimensions, twenty times $N$ rounded up to the next power of two, at which the Sobol' sequence is balanced. Following \textcite{Saltelli_Annoni_Azzini_Campolongo_Ratto_Tarantola_2010}, the first $k$ coordinates of each point form a row of the base matrix $\mathbf{A}$ and the last $k$ the corresponding row of $\mathbf{B}$. We then retain only the rows $i$ for which $\beta_1 > \beta_2$ holds in both $\mathbf{A}$ and $\mathbf{B}$ and in the two cross-substitutions ($B_{\beta_1,i} > A_{\beta_2,i}$ and $A_{\beta_1,i} > B_{\beta_2,i}$), so that the constraint holds in every row of the extended matrix. About $17\%$ of the points satisfy all four conditions ($1{,}383$ of $8{,}192$), and we keep the first $N$ of them in sequence order. From these rows we reconstruct the Saltelli matrix as $[\mathbf{A};\, \mathbf{B};\, \mathbf{AB}_0;\, \ldots;\, \mathbf{AB}_{k-1}]$, where each $\mathbf{AB}_j$ is $\mathbf{A}$ with column $j$ replaced by $\mathbf{B}_{:,j}$, for $N(k+2) = 3{,}328$ parameter combinations. Since the filter removes most of the sequence, the retained rows are not guaranteed to be as evenly spread as an unfiltered Sobol' sample; we therefore generate five candidate designs and retain the one with the lowest centred $L_2$-discrepancy \citep{Hickernell_1998}.
For the GSA we denote the vector of parameters as $\mathbf{x} = (x_1, \ldots, x_{11})$, we sample using the Saltelli Method within the bounds specified in \SItab{tab:sa_ranges}. The Saltelli sampling scheme generates $N(k+2)$ model evaluations per scenario-seed combination, where $k=11$ is the number of sampled parameters (\SItab{tab:sa_ranges}) and $N=256$ is the base sample size, yielding $N(k+2) = 256 \cdot 13 = 3{,}328$ evaluations per combination. Running these across the five flood scenarios and $R=10$ random seeds gives a total of $N(k+2) \cdot 5 \cdot R = 3{,}328 \cdot 5 \cdot 10 = 166{,}400$ model runs.

\begin{table}[h!]
\centering
\caption{Parameters included in the GSA and their sampling ranges. The ranges of $m$ and $\varepsilon$ extend $30\%$ on either side of their estimates from CBS household income and expenditure data by income decile, and the range of $\mathit{exp}_{lb}$ extends $30\%$ on either side of the Dutch social assistance norm (\SIsec{sec:baseline_parametrization}).}
\label{tab:sa_ranges}
\begin{tabular}{lll}
\toprule
Parameter & Lower bound & Upper bound \\
\midrule
$\kappa$ & 1 & 25 \\
$\beta_1$ & 0.01 & 1 \\
$\beta_2$ & 0.01 & 1 \\
$\beta_3$ & 0.25 & 3.0 \\
$\lambda$ & 0 & 0.5 \\
$\delta$ & 0.05 & 0.5 \\
$\eta$ & 0.03 & 0.13 \\
$\vartheta$ & 0.03 & 0.13 \\
$m$ & 4.34 & 8.06 \\
$\varepsilon$ & 0.001386 & 0.002574 \\
$\mathit{exp}_{lb}$ & 760 & 1413 \\
\bottomrule
\end{tabular}
\end{table}

We compute the first-order and total-order Sobol' indices as:
\[
S_i = \frac{V_{x_i}\left[\mathbb{E}_{\mathbf{x}_{\sim i}}[Y|x_i]\right]}{V[Y]}, \qquad
S_{Ti} = 1 - \frac{V_{\mathbf{x}_{\sim i}}\left[\mathbb{E}_{x_i}[Y|\mathbf{x}_{\sim i}]\right]}{V[Y]},
\]
We denote by $S_i$ the first-order Sobol' index for parameter $x_i$ quantifying the direct effect of $x_i$ on $Y$. $Y$ is the model output and $V[\cdot]$ denotes the variance operator. We use different statistics of the model as output measures, in particular the percentage of agents defaulted due to strategic default. We consider the first 100 steps of the model as the warm-up period, and we compute the statistics on the steps $t>100$. All the simulations included in the GSA are run for $T=400$ time steps. The parameter ranges for the simulation are reported in \SItab{tab:sa_ranges}. We start from the baseline value of the model and as a general rule we add +/-30\% from the baseline values. We use a different rule for the behavioural parameters, as those parameters are the most difficult ones to pin down with data. The damage share $\lambda$ (Equation~\ref{eqn:delta_p}) is sampled over $[0, 0.5]$, from no loss of collateral to the loss of half the house value at the maximum damage factor. The spillover intensity $\delta$ (Equation~\ref{eqn:collateral}) is sampled over $[0.05, 0.5]$. Its lower bound lies above the empirical magnitude of the foreclosure externality ($\delta \approx 0.01$, Section~\ref{sec:neighbourhood}), and its upper bound is fifty times that magnitude, so the GSA describes the model with an active foreclosure feedback. The baseline $\delta = 0$ lies outside the sampled range.

We complement the variance-based indices with the PAWN index $T_i$ of \textcite{Pianosi_Wagener_2015}, in the generic-sample formulation of \textcite{Pianosi_Wagener_2018}, reported alongside the Sobol' indices in Figs.~\ref{fig:sa_perc_fin}, \ref{fig:sa_n_financial} and~\ref{fig:sa_n_strategic}. PAWN characterises the influence of $x_i$ by the distance between the unconditional cumulative distribution function of the output, $F_Y(y)$, and the conditional one obtained when $x_i$ is restricted to a sub-interval $\mathcal{I}_j$ of its range, $F_{Y|x_i \in \mathcal{I}_j}(y)$, measured by the Kolmogorov--Smirnov (KS) statistic
\[
\text{KS}(x_i, \mathcal{I}_j) = \max_{y} \left| F_Y(y) - F_{Y|x_i \in \mathcal{I}_j}(y) \right|,
\qquad
T_i = \operatorname{median}_{j=1,\ldots,J} \text{KS}(x_i, \mathcal{I}_j),
\]
where the $J$ intervals $\mathcal{I}_1,\ldots,\mathcal{I}_J$ partition the range of $x_i$ and each
holds approximately the same number of samples, being delimited by the empirical $J$-quantiles of
$x_i$. The choice of $J$, that is the number of intervals, is a tunable parameter that influences the construction of the $T_i$ indices. Conceptually PAWN asks how the distribution of $Y$ changes when $x_i$ is held at a fixed value $x_i^{*}$, which would require estimating $F_{Y|x_i = x_i^{*}}$. Estimating it directly would require many runs sharing one value of $x_i$ while the remaining inputs vary, and a generic sample like ours does not provide them. A low-discrepancy design like the one introduced above places at most a handful of runs at any single value of a parameter, far too few to estimate a distribution from. For this reason we opt for interval-conditioning: all draws whose $x_i$ falls in $\mathcal{I}_j$ are treated as an approximate sample from the conditional distribution, and $J$ is the number of such intervals. This substitution is what makes PAWN computable on the same sample used for the Sobol' analysis. Therefore the choice of $J$ acts as a bias--variance trade-off, and $J=10$ is a compromise between its two failure modes. A large $J$ makes each interval narrow, so interval-conditioning approximates point-conditioning closely, but it leaves few draws per interval and the KS statistic computed on a small subsample is unstable. A small $J$ puts many draws in each interval and stabilises the estimate, but each interval then spans a wide range of $x_i$, so the conditional distribution is diluted towards the unconditional one, reducing the KS distance. With $2N = 512$ draws and $J = 10$ each interval holds ${\approx}51$ draws.

We show in \SItab{tab:pawn_rank_vs_S} how the conclusions do not depend on the choice of $J$ by recomputing $T_i$ at $J \in \{5, 10, 20\}$. $\beta_1$ ranks first for all three outcomes at every $J$, and $m$ second for the share of financial defaults and for $n_{\text{financial}}$. For $n_{\text{strategic}}$, $\delta$ and $\kappa$ exchange the second and third positions across $J$, while $m$ ranks fourth to sixth.

The two measures answer different questions. Sobol' indices are variance-based: they attribute the variance of the output to each input, so a parameter registers only to the extent that it moves the second moment of $Y$. PAWN is moment-independent: because $T_i$ compares whole cumulative distribution functions rather than second moments, it is sensitive to a parameter that shifts or reshapes the distribution of $Y$, including its tails, without changing its variance, which a Sobol' index does not register \citep{KhorashadiZadeh_Nossent_Sarrazin_Pianosi_vanGriensven_Wagener_Bauwens_2017}. Since the outcomes of interest are the composition and the levels of a heavy-tailed default distribution rather than its variance alone, the CDF-based reading is the more appropriate primary diagnostic. The price is that $T_i$ is not a share of anything: indices do not sum to one and are not comparable in level to $S_i$ or $S_{Ti}$. Their magnitudes are KS distances and are interpreted through their relative ordering and comparison with the permutation reference.

The Sobol' indices are computed on the full Saltelli matrix, including the $\mathbf{AB}_j$ cross-blocks that the variance decomposition requires. The PAWN indices are computed on the $\mathbf{A}$ and $\mathbf{B}$ blocks alone, each of which is on its own a constraint-respecting draw of the full parameter space. The vertical bars in the PAWN panels of Figs.~\ref{fig:sa_perc_fin}, \ref{fig:sa_n_financial} and~\ref{fig:sa_n_strategic} report the min--max spread of $KS(x_i, \mathcal{I}_j)$ across the $J$ intervals. The $J$ values they summarise are not repeated estimates of one quantity (as confidence intervals would be) but the association of $x_i$ measured separately in each part of its range, so their spread quantifies how uniform that association is over the support of $x_i$. A parameter whose association is confined to one subsection of its sampling range therefore produces a long bar for substantive rather than statistical reasons, as $\beta_1$ does here.
The $\beta_1 > \beta_2$ sampling constraint of \SIsec{sec:app:sa} makes the two parameters dependent, with a sample correlation of $0.36$. Under this dependence, the classical Sobol' indices no longer retain their standard variance-decomposition interpretation, while PAWN remains computable but describes marginal association under the constrained joint input distribution rather than an isolated parameter effect. To examine the sensitivity of the PAWN ranking to this dependence, we recompute $T_i$ over the rectangular sub-region $\beta_1 > 0.5$, $\beta_2 < 0.5$. This retains 322 of the 512 design points and reduces their correlation to $-0.02$. Within this sub-region $\beta_1$ remains ahead of $\beta_2$ for all three outcomes at every flood scenario, by $0.09$--$0.13$: for the share of financial defaults, $T_{\beta_1}=0.32$--$0.33$ against $T_{\beta_2}=0.22$--$0.23$. $T_{\beta_2}$ rises to $0.20$--$0.23$ for all three outcomes, the level of $\kappa$ ($T_{\kappa}=0.21$--$0.22$), $\delta$ ($T_{\delta}=0.18$--$0.26$) and, for the share of financial defaults and $n_{\text{financial}}$, $m$ ($T_m=0.20$--$0.27$). The position of $\beta_2$ relative to these parameters is therefore sensitive to the sampling constraint, and we make no claim about their ordering.

Finite samples can produce positive estimated PAWN statistics even when inputs and outputs are unrelated \citep{Pianosi_Wagener_2018}. We therefore use $10{,}000$ output permutations to construct a finite-sample null distribution representing no association between the inputs and the outcome, while keeping the input matrix, the conditioning intervals and $J=10$ fixed. Let $T_{ih}$ denote the PAWN median calculated from the original data for parameter $i$ in flood scenario $\mathcal{S}_h$. In each permutation $b$, we keep the 512 parameter vectors fixed, randomly shuffle their associated output values among the rows, and recompute the PAWN median, denoted by $T_{ih}^{(b)}$. Shuffling removes the observed input--output association while preserving the empirical output distribution. Adapting the single-step max-$T$ procedure of \textcite{Westfall_Young_1993}, for each outcome and permutation we record
\[
M_Y^{(b)}=
\max_{\substack{i=1,\ldots,11\\ h=0,\ldots,4}}
T_{ih}^{(b)},
\]
where the maximum covers the eleven parameters and all five flood scenarios, $h=0,\ldots,4$. The red line in the PAWN panels is the 95th percentile of this maximum-statistic null distribution. Under the joint permutation null of no association between the inputs and the outcome, this single-step procedure controls the probability of at least one false rejection across the 55 comparisons displayed for each outcome.

\begin{table}[htbp]
\centering
\small
\caption{Rank of each parameter under the PAWN index $T_i$ for three choices of $J$, the number of conditioning intervals, across the three outcomes of Figs.~\ref{fig:sa_perc_fin}, \ref{fig:sa_n_financial} and~\ref{fig:sa_n_strategic}. Indices are averaged over the five flood scenarios before ranking. Parameters that exceed the outcome-wide permutation threshold at every flood scenario and whose rank is the same in every column are shown in bold: only these are identified independently of $J$, while the remaining ranks permute freely, which is why no claim is made about the ordering within that group.}
\label{tab:pawn_rank_vs_S}
\sisetup{text-series-to-math=true, reset-text-series=false}
\begin{tabular}{l S[table-format=2.0] S[table-format=2.0] S[table-format=2.0] S[table-format=2.0] S[table-format=2.0] S[table-format=2.0] S[table-format=2.0] S[table-format=2.0] S[table-format=2.0]}
\toprule
 & \multicolumn{3}{c}{\% Financial defaults} & \multicolumn{3}{c}{$n_{\text{financial}}$} & \multicolumn{3}{c}{$n_{\text{strategic}}$} \\
\cmidrule(lr){2-4}\cmidrule(lr){5-7}\cmidrule(lr){8-10}
Parameter & {$J{=}5$} & {$J{=}10$} & {$J{=}20$} & {$J{=}5$} & {$J{=}10$} & {$J{=}20$} & {$J{=}5$} & {$J{=}10$} & {$J{=}20$} \\
\midrule
$\kappa$ & 6 & 4 & 5 & 5 & 5 & 4 & 2 & 3 & 2 \\
$\beta_1$ & \bfseries 1 & \bfseries 1 & \bfseries 1 & \bfseries 1 & \bfseries 1 & \bfseries 1 & \bfseries 1 & \bfseries 1 & \bfseries 1 \\
$\beta_2$ & 5 & 5 & 3 & 3 & 3 & 3 & 6 & 5 & 5 \\
$\beta_3$ & 4 & 6 & 6 & 4 & 4 & 6 & 5 & 6 & 4 \\
$\lambda$ & 10 & 10 & 11 & 8 & 9 & 9 & 10 & 11 & 10 \\
$\delta$ & 3 & 3 & 4 & 7 & 6 & 7 & 3 & 2 & 3 \\
$\eta$ & 11 & 11 & 9 & 11 & 11 & 10 & 7 & 8 & 8 \\
$\vartheta$ & 9 & 9 & 8 & 10 & 10 & 8 & 8 & 7 & 7 \\
$m$ & \bfseries 2 & \bfseries 2 & \bfseries 2 & \bfseries 2 & \bfseries 2 & \bfseries 2 & 4 & 4 & 6 \\
$\varepsilon$ & 7 & 7 & 7 & 6 & 7 & 5 & 11 & 10 & 9 \\
$\mathit{exp}_{lb}$ & 8 & 8 & 10 & 9 & 8 & 11 & 9 & 9 & 11 \\
\bottomrule
\end{tabular}
\end{table}

\subsection{Composition and Counts of Defaults}
\label{sec:app:gsa_defaults}

Figure~\ref{fig:defaults_gap} shows default composition across $\beta_1-\beta_2$ for $786$ GSA parameter vectors with $\delta\le0.15$. Below $0.46$, more than $92\%$ of defaults are strategic in every flood scenario, with fewer than $1{,}600$ active mortgages on average. At wider gaps financial defaults dominate: in the widest-gap bin, the strategic share falls to $1.5$--$3\%$ and the active population reaches about $2{,}500$. Floods raise the strategic share near $\beta_1-\beta_2=1/2$ by up to $13$ percentage points within a bin; for $\beta_1-\beta_2\ge1/2$, it rises from $20\%$ at $\mathcal{S}_0$ to $28\%$ at $\mathcal{S}_4$. In this range the strategic defaults at $\mathcal{S}_0$ come from the spillover alone, since with $\delta=0$ there are none, and a stronger spillover sustains more of them: for $\beta_1-\beta_2\ge1/2$ at $\mathcal{S}_0$, the strategic share is $63\%$ for $0.15<\delta\le0.35$ and $83\%$ for $\delta>0.35$.

Averaged over seeds, the total annual foreclosure rate is higher at $\mathcal{S}_4$ than at $\mathcal{S}_0$ for $3{,}286$ of the $3{,}328$ GSA parameter vectors. At the calibrated parametrisation ($\beta_1-\beta_2=1/2$) without foreclosure spillovers ($\delta=0$), strategic foreclosures rise from zero at $\mathcal{S}_0$ to $0.55\%$ per year at $\mathcal{S}_4$. Financial foreclosures fall from $0.42\%$ at $\mathcal{S}_0$ to $0.36\%$ in every flood scenario. The total annual rate rises from $0.42\%$ to $0.91\%$, with about half of the increase occurring between $\mathcal{S}_0$ and $\mathcal{S}_1$ (Figure~\ref{fig:annual_rate_channel}).

\begin{figure}[htbp]
    \centering
    \captionsetup[subfigure]{font=footnotesize}
    \begin{subfigure}[b]{0.49\textwidth}
        \centering
        \includegraphics[width=\textwidth, keepaspectratio]{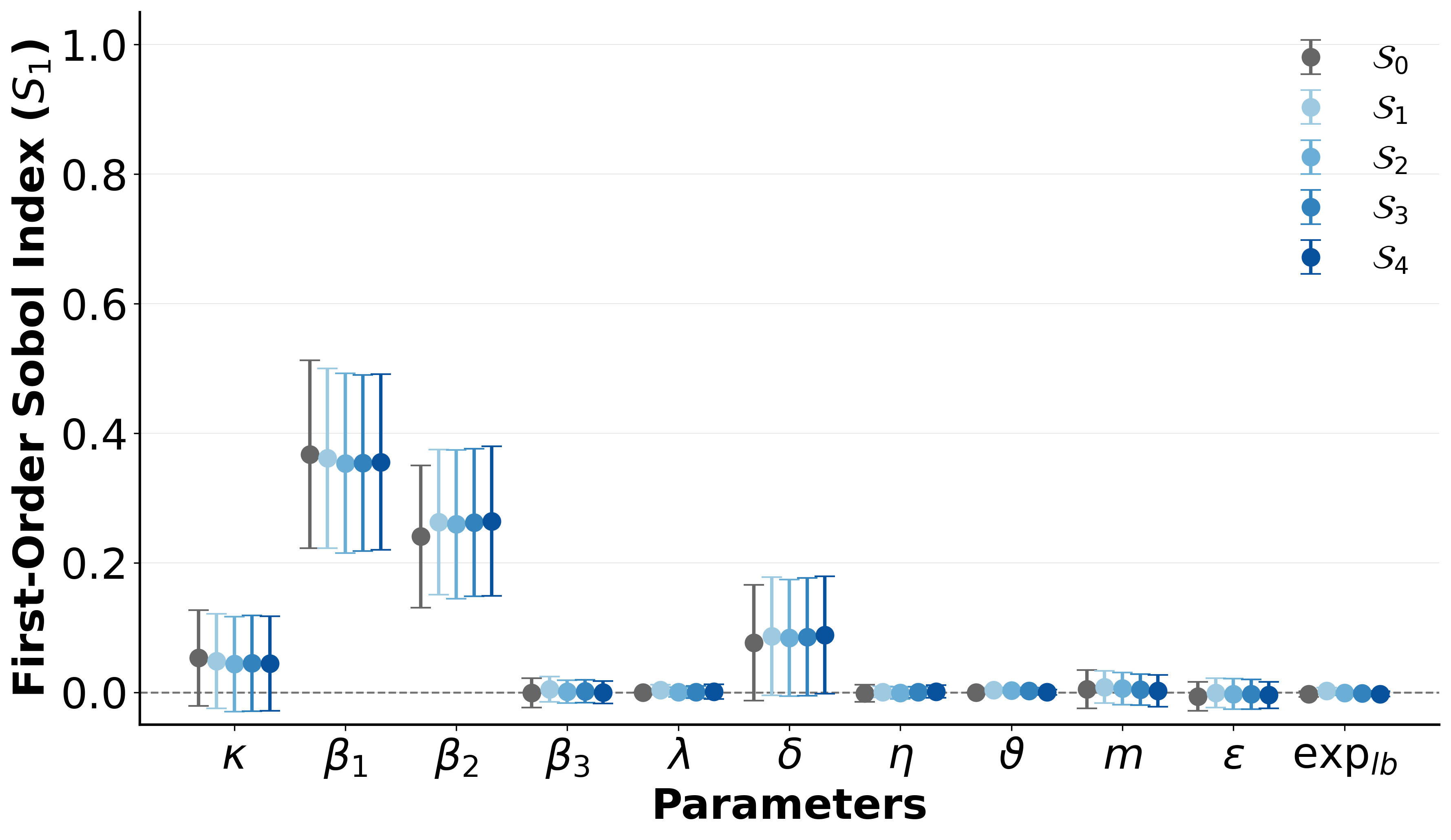}
        \caption{$S_i$ --- $Y=$ \% Financial defaults}
        \label{fig:si_perc}
    \end{subfigure}
    \hfill
    \begin{subfigure}[b]{0.49\textwidth}
        \centering
        \includegraphics[width=\textwidth, keepaspectratio]{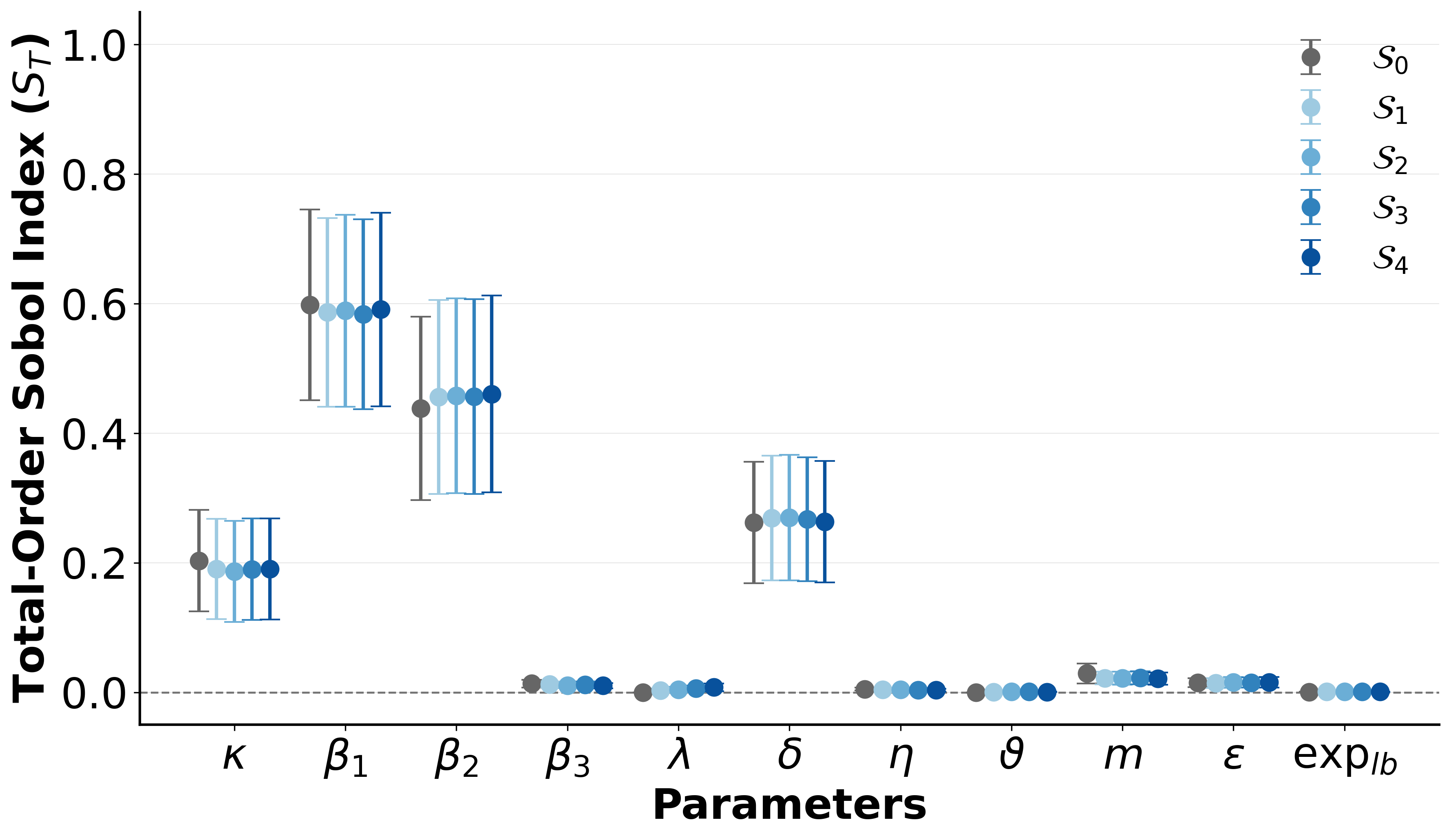}
        \caption{$S_{Ti}$ --- $Y=$ \% Financial defaults}
        \label{fig:sti_perc}
    \end{subfigure}

    \vspace{0.3cm}

    \begin{subfigure}[b]{0.49\textwidth}
        \centering
        \includegraphics[width=\textwidth, keepaspectratio]{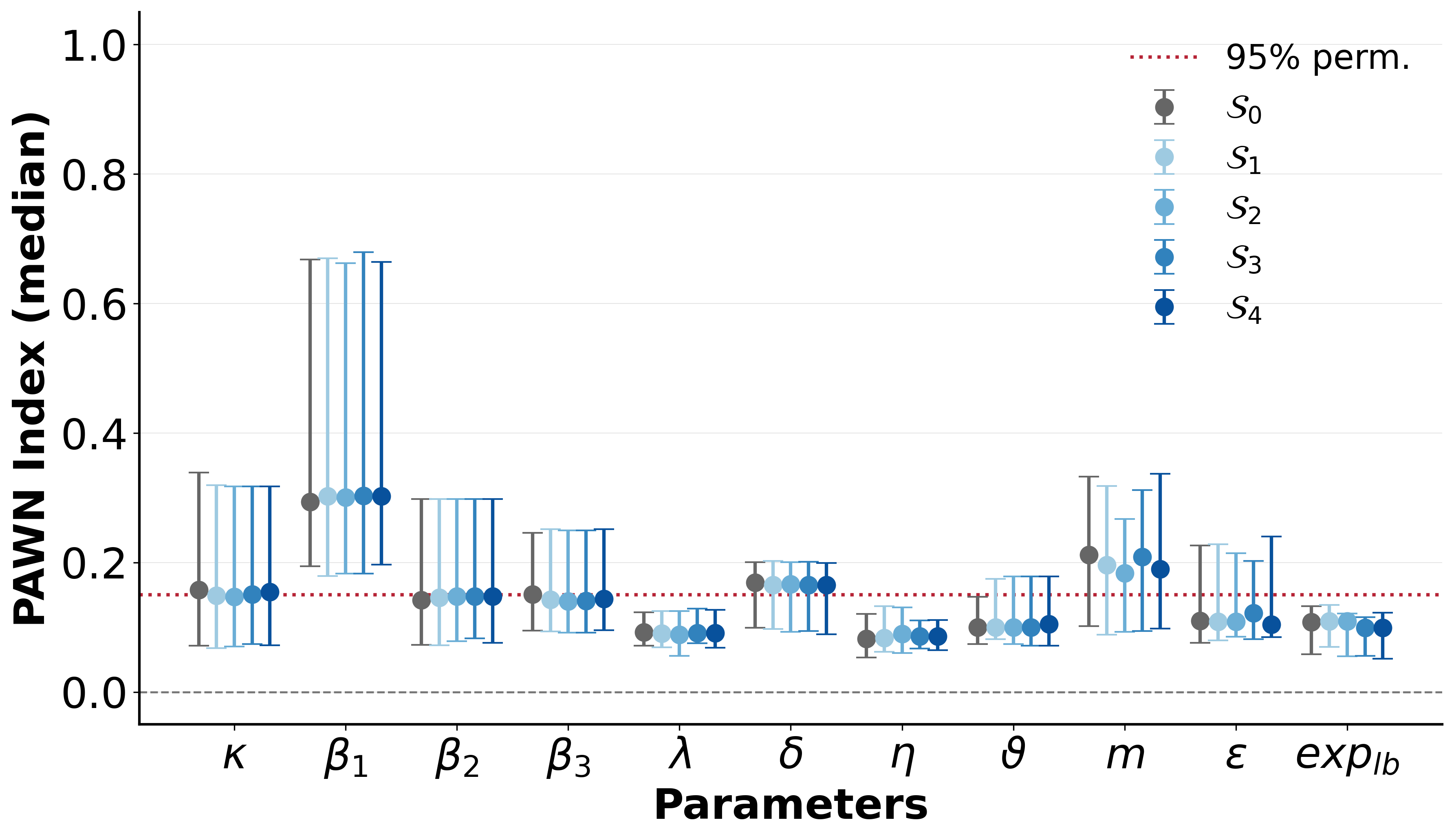}
        \caption{$T_i$ --- $Y=$ \% Financial defaults}
        \label{fig:pawn_perc}
    \end{subfigure}
    \caption{\small Sensitivity of the share of financial defaults across scenarios $\mathcal{S}_0$--$\mathcal{S}_4$ (3{,}328 Saltelli samples, 10 seeds each, per flood scenario): (a) first-order Sobol' indices $S_i$, (b) total-order Sobol' indices $S_{Ti}$, (c) PAWN indices $T_i$. In panel (c), markers report $T_i$, the median KS distance between the conditional and the unconditional distribution of $Y$ over the $J$ conditioning intervals of $x_i$, and bars span the minimum and maximum KS distance across the intervals. The red dotted line is the outcome-wide 95\% permutation max-\(T\) threshold for the 55 parameter--scenario comparisons (\SIsec{sec:app:gsa_method}).}
    \label{fig:sa_perc_fin}
\end{figure}

\begin{figure}[htbp]
    \centering
    \captionsetup[subfigure]{font=footnotesize}
    \begin{subfigure}[b]{0.49\textwidth}
        \centering
        \includegraphics[width=\textwidth, keepaspectratio]{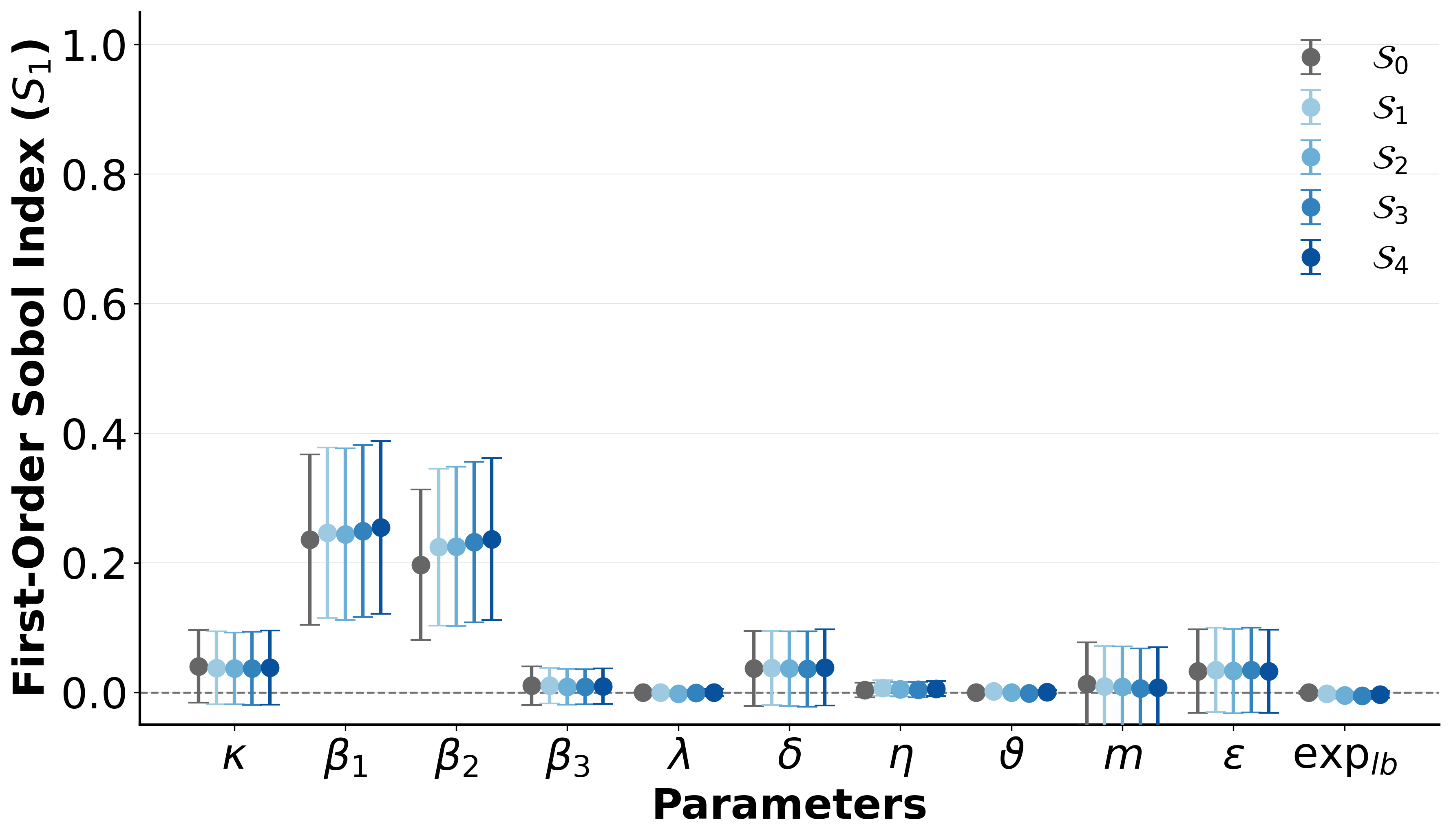}
        \caption{$S_i$ --- $Y=n_{\text{financial}}$}
        \label{fig:si_n_financial}
    \end{subfigure}
    \hfill
    \begin{subfigure}[b]{0.49\textwidth}
        \centering
        \includegraphics[width=\textwidth, keepaspectratio]{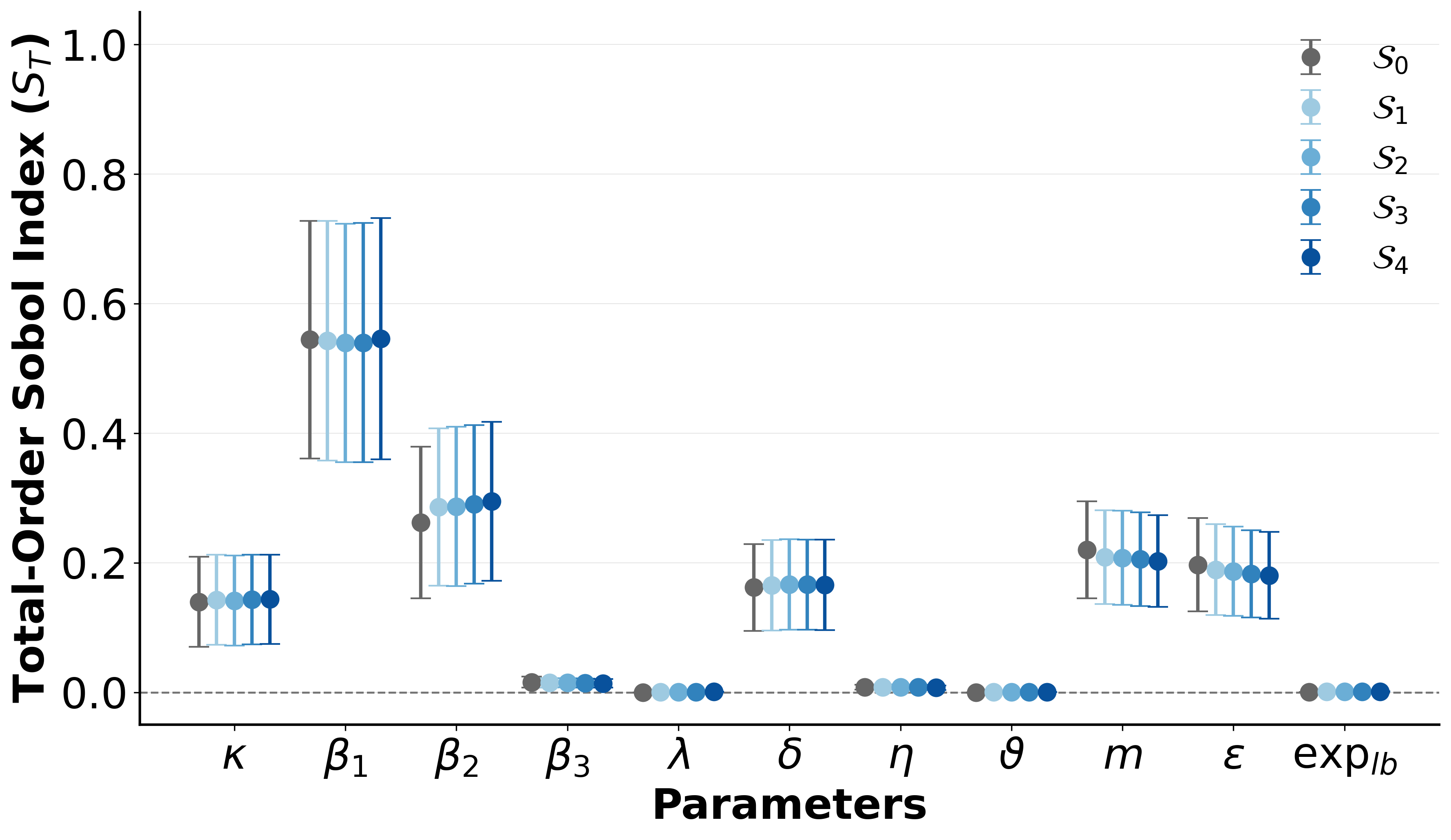}
        \caption{$S_{Ti}$ --- $Y=n_{\text{financial}}$}
        \label{fig:sti_n_financial}
    \end{subfigure}

    \vspace{0.3cm}

    \begin{subfigure}[b]{0.49\textwidth}
        \centering
        \includegraphics[width=\textwidth, keepaspectratio]{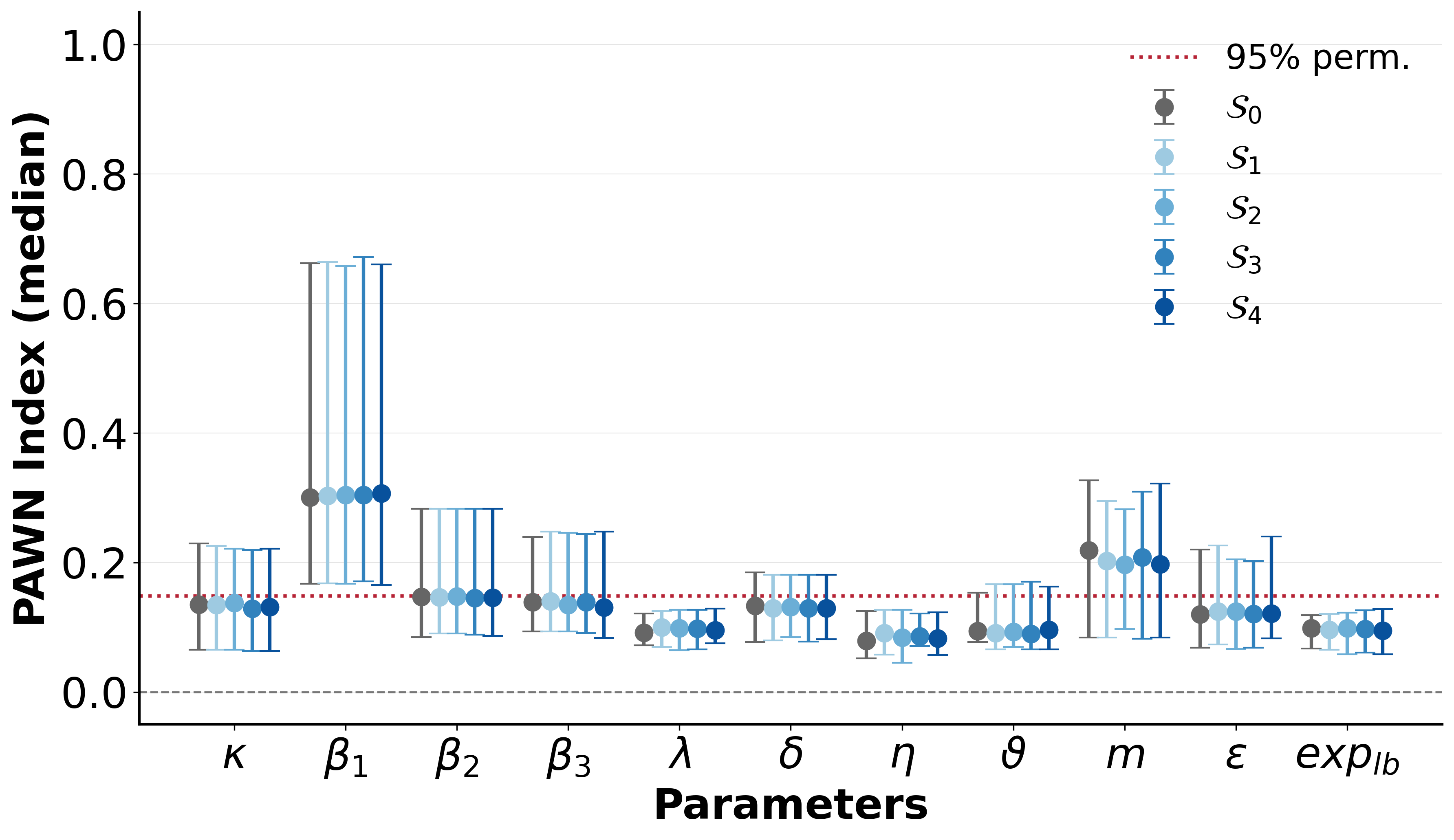}
        \caption{$T_i$ --- $Y=n_{\text{financial}}$}
        \label{fig:pawn_n_financial}
    \end{subfigure}
    \caption{\small Sensitivity of the absolute number of financial defaults, $n_{\text{financial}}$, across scenarios $\mathcal{S}_0$--$\mathcal{S}_4$: (a) first-order Sobol' indices $S_i$, (b) total-order Sobol' indices $S_{Ti}$, (c) PAWN indices $T_i$, constructed as in \SIfig{fig:sa_perc_fin}.}
    \label{fig:sa_n_financial}
\end{figure}

\begin{figure}[htbp]
    \centering
    \captionsetup[subfigure]{font=footnotesize}
    \begin{subfigure}[b]{0.49\textwidth}
        \centering
        \includegraphics[width=\textwidth, keepaspectratio]{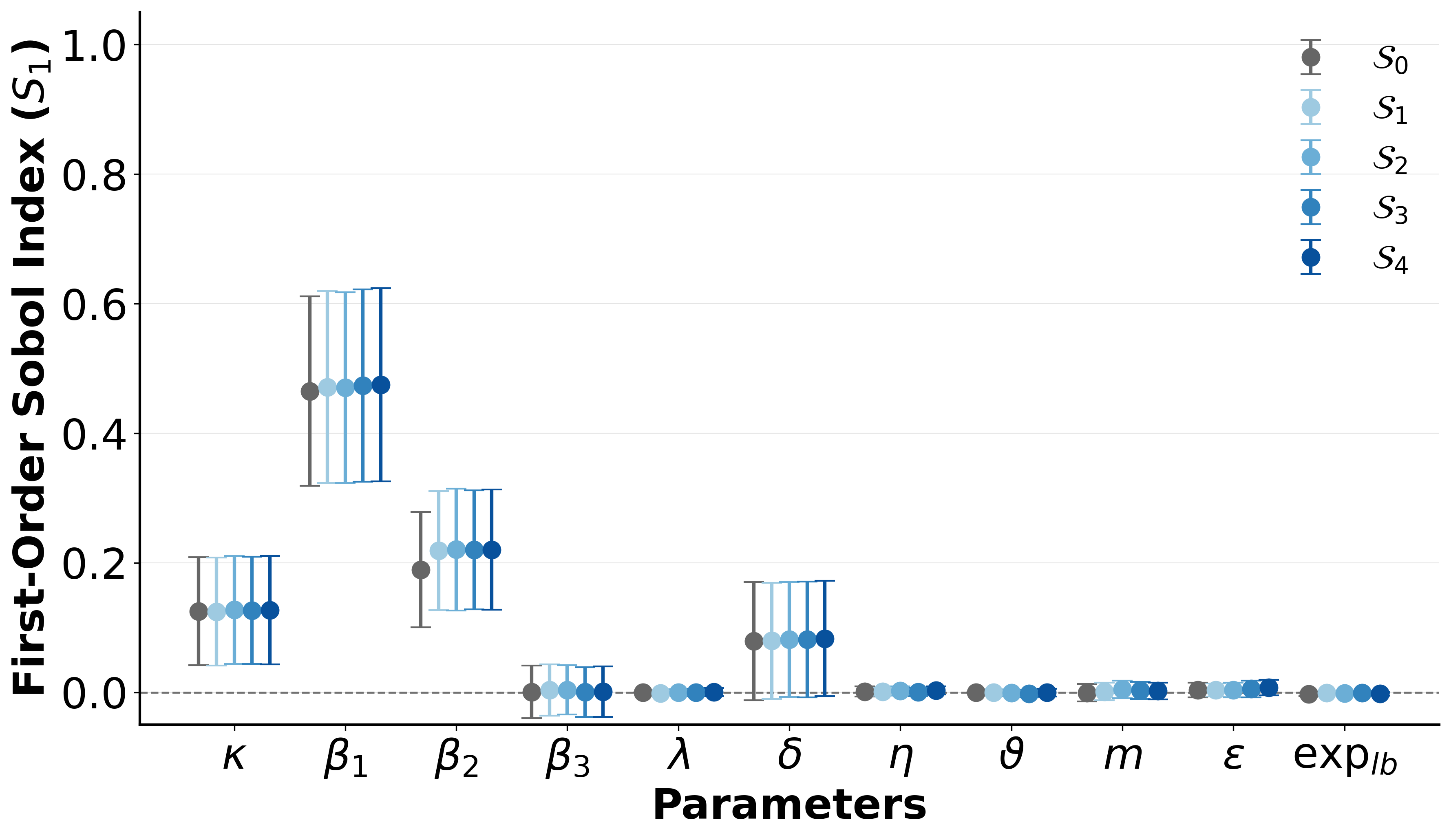}
        \caption{$S_i$ --- $Y=n_{\text{strategic}}$}
        \label{fig:si_n_strategic}
    \end{subfigure}
    \hfill
    \begin{subfigure}[b]{0.49\textwidth}
        \centering
        \includegraphics[width=\textwidth, keepaspectratio]{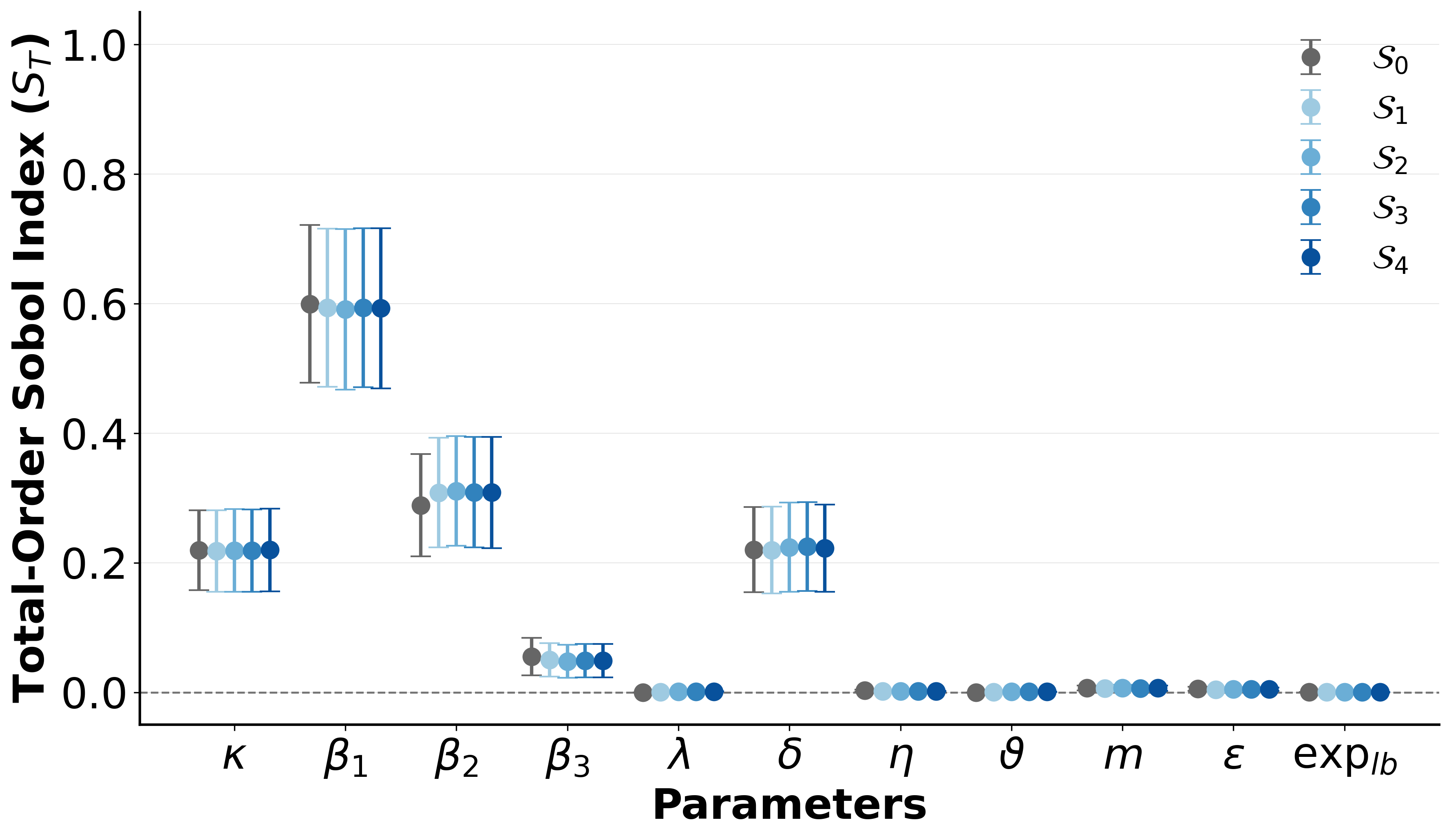}
        \caption{$S_{Ti}$ --- $Y=n_{\text{strategic}}$}
        \label{fig:sti_n_strategic}
    \end{subfigure}

    \vspace{0.3cm}

    \begin{subfigure}[b]{0.49\textwidth}
        \centering
        \includegraphics[width=\textwidth, keepaspectratio]{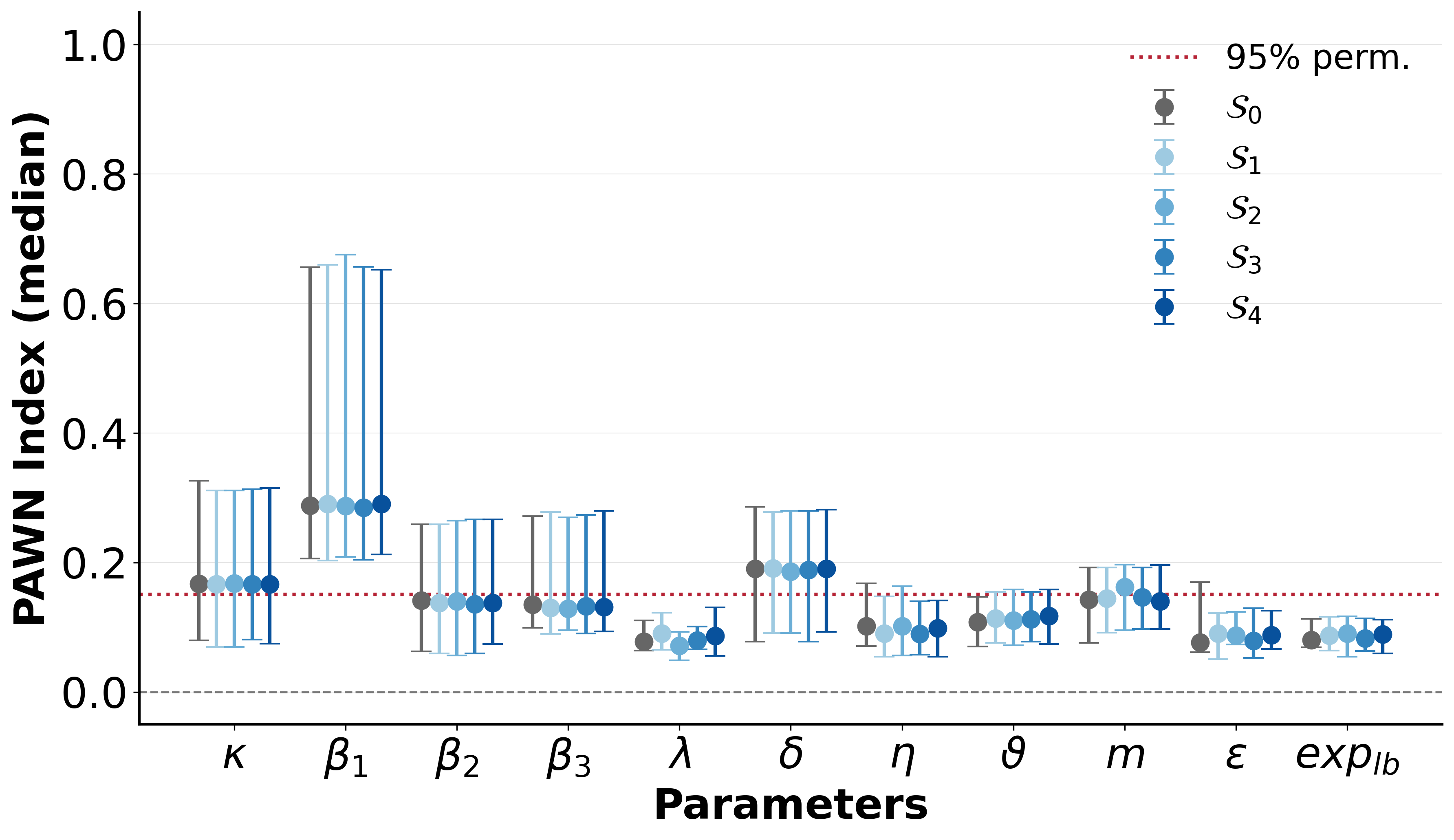}
        \caption{$T_i$ --- $Y=n_{\text{strategic}}$}
        \label{fig:pawn_n_strategic}
    \end{subfigure}
    \caption{\small Sensitivity of the absolute number of strategic defaults, $n_{\text{strategic}}$, across scenarios $\mathcal{S}_0$--$\mathcal{S}_4$: (a) first-order Sobol' indices $S_i$, (b) total-order Sobol' indices $S_{Ti}$, (c) PAWN indices $T_i$, constructed as in \SIfig{fig:sa_perc_fin}.}
    \label{fig:sa_n_strategic}
\end{figure}

\SIfig{fig:sa_perc_fin} reports the Sobol' indices and the PAWN indices $T_i$ for the share of financial defaults, and Figs.~\ref{fig:sa_n_financial} and~\ref{fig:sa_n_strategic} those of the absolute default counts $n_{\text{financial}}$ and $n_{\text{strategic}}$. For the share of financial defaults, $\beta_1$ and $\beta_2$ have first-order indices $S_i=0.35$--$0.37$ and $0.24$--$0.26$ across $\mathcal{S}_0$--$\mathcal{S}_4$ (panel~\subref{fig:si_perc}), the only two first-order indices whose intervals clear zero at every scenario. The first-order indices sum to $0.73$--$0.78$, so about a quarter of the variance is carried by interactions (panel~\subref{fig:sti_perc}): $\beta_2$ has $S_{Ti}=0.44$--$0.46$, and the neighbourhood radius $\kappa$ acts mainly through them ($S_i=0.04$--$0.05$, with an interval that includes zero, and $S_{Ti}=0.19$--$0.20$). The half-saturation constant $\beta_3$ and $\varepsilon$ stay below ${\approx}0.03$ at either order. Under PAWN (panel~\subref{fig:pawn_perc}), $\kappa$ exceeds the permutation threshold at three of the five scenarios and $\beta_3$ at one.

The variance of both counts is explained mainly by $\beta_1$ and $\beta_2$, but the split differs: $\beta_1$ explains more of the variance of $n_{\text{strategic}}$ ($S_i\approx0.47$; Figs.~\ref{fig:si_n_strategic} and~\ref{fig:sti_n_strategic}) than of $n_{\text{financial}}$ ($S_i\approx0.24$--$0.25$, $S_{Ti}\approx0.54$--$0.55$; Figs.~\ref{fig:si_n_financial} and~\ref{fig:sti_n_financial}), while $\beta_2$ has similar total-order indices for the two counts ($S_{Ti}\approx0.26$--$0.29$ for $n_{\text{financial}}$ and $0.29$--$0.31$ for $n_{\text{strategic}}$). $\beta_3$ remains small for both ($S_{Ti}\le0.06$). The PAWN indices of the counts are shown in Figs.~\ref{fig:pawn_n_financial} and~\ref{fig:pawn_n_strategic}.

The two methods agree on the leading parameter and differ below it. For the share of financial defaults and for both counts (Figs.~\ref{fig:pawn_perc}, \ref{fig:pawn_n_financial} and~\ref{fig:pawn_n_strategic}), the neighbourhood-utility loading $\beta_1$ has the largest total-order Sobol' index and the largest PAWN index at every flood scenario: for the share of financial defaults $S_{T\beta_1}=0.58$--$0.60$ and $T_{\beta_1}=0.29$--$0.30$, for $n_{\text{strategic}}$ $S_{T\beta_1}=0.59$--$0.60$ and $T_{\beta_1}=0.28$--$0.29$. No parameter overtakes another as flood severity rises. The measures differ on the financial-stress weight $\beta_2$. The Sobol' indices place it second for all three outcomes ($S_i=0.19$--$0.26$, $S_{Ti}=0.26$--$0.46$), while its PAWN index, $T_{\beta_2}=0.14$--$0.15$, stays below the outcome-wide permutation max-$T$ threshold \citep{Westfall_Young_1993} at every flood scenario. This difference is tied to the sampling constraint (\SIsec{sec:app:gsa_method}).

The two measures agree on the spillover intensity. $\delta$ has total-order indices $S_{Ti}=0.26$--$0.27$ for the share of financial defaults and $0.22$--$0.23$ for $n_{\text{strategic}}$, mostly through interactions ($S_i=0.08$--$0.09$), and exceeds the permutation threshold at every flood scenario for both outcomes ($T_{\delta}=0.17$ and $0.19$). The damage share $\lambda$ has no detectable influence on any of the three outcomes ($S_{Ti}\le0.01$, $T_{\lambda}=0.07$--$0.10$, ranking eighth to eleventh at every $J$): over the sampled range of $\delta$, the composition and the counts of defaults respond to the foreclosure feedback and not to the scale of the flood loss on collateral. The measures differ again on the consumption intercept $m$, in the opposite direction to $\beta_2$: for the share of financial defaults it has near-zero Sobol' indices ($S_{Ti}=0.02$--$0.03$) but exceeds the permutation threshold at every flood scenario ($T_m=0.18$--$0.21$), so it changes the shape of the distribution of the share without contributing to its variance. Of the 55 comparisons per outcome, 19 are significant for the share of financial defaults, 10 for $n_{\text{financial}}$ and 16 for $n_{\text{strategic}}$; the parameters below the threshold have $T_i=0.07$--$0.15$.

For the two counts, both measures detect an asymmetry between the channels, with a different group of parameters on each side. For $n_{\text{financial}}$, the consumption intercept $m$ exceeds the permutation threshold at every flood scenario ($T_m=0.20$--$0.22$) and ranks second at every $J$ (\SItab{tab:pawn_rank_vs_S}), while for $n_{\text{strategic}}$ it falls to $T_m=0.14$--$0.16$, above the threshold in one scenario only, and ranks fourth to sixth. The Sobol' indices show the same asymmetry: for $n_{\text{financial}}$, $m$ and $\varepsilon$ have total-order indices $S_{Ti}=0.20$--$0.22$ and $0.18$--$0.20$, almost entirely through interactions ($S_i\le0.04$), and both are near zero for $n_{\text{strategic}}$. Under PAWN, $\varepsilon$ stays below the threshold for both counts ($T_{\varepsilon}=0.12$ for $n_{\text{financial}}$). Conversely, $\delta$ and $\kappa$ exceed the threshold at every flood scenario for $n_{\text{strategic}}$ ($T_{\delta}=0.19$, $T_{\kappa}=0.17$) and occupy the second and third ranks at every $J$, with total-order indices $S_{Ti}=0.22$--$0.23$ and a first-order index of $\kappa$, $S_i=0.12$--$0.13$, whose interval clears zero. For $n_{\text{financial}}$ both stay below the threshold ($T_i=0.13$--$0.14$, $S_{Ti}=0.14$--$0.17$). Both measures therefore support the same conclusion: $\beta_1$ leads for both counts, as expected from its role in $\text{wtp}_{i,t}$, which governs the outcome of either channel. Below it, the consumption parameters move the financial count, while the spillover intensity and the radius $\kappa$ over which it acts move the strategic count.

\subsection{Distress months}
\label{sec:appendix_sobol_distress}

We count the months after warm-up in which a household meets either the financial or strategic distress condition defined in Section~\ref{sec:distress}. These are accumulated months, including forborne months and the terminal foreclosure, rather than the duration of an uninterrupted delinquency spell. Histories crossing the observation-window boundary are truncated. The GSA uses the mean count among households with at least one distress month; the foreclosure-history comparisons of \SIsec{sec:app:fcltime} use households that foreclose.

\begin{figure}[htbp]
    \centering
    \captionsetup[subfigure]{font=footnotesize}
    \begin{subfigure}[b]{0.49\textwidth}
        \centering
        \includegraphics[width=\linewidth]{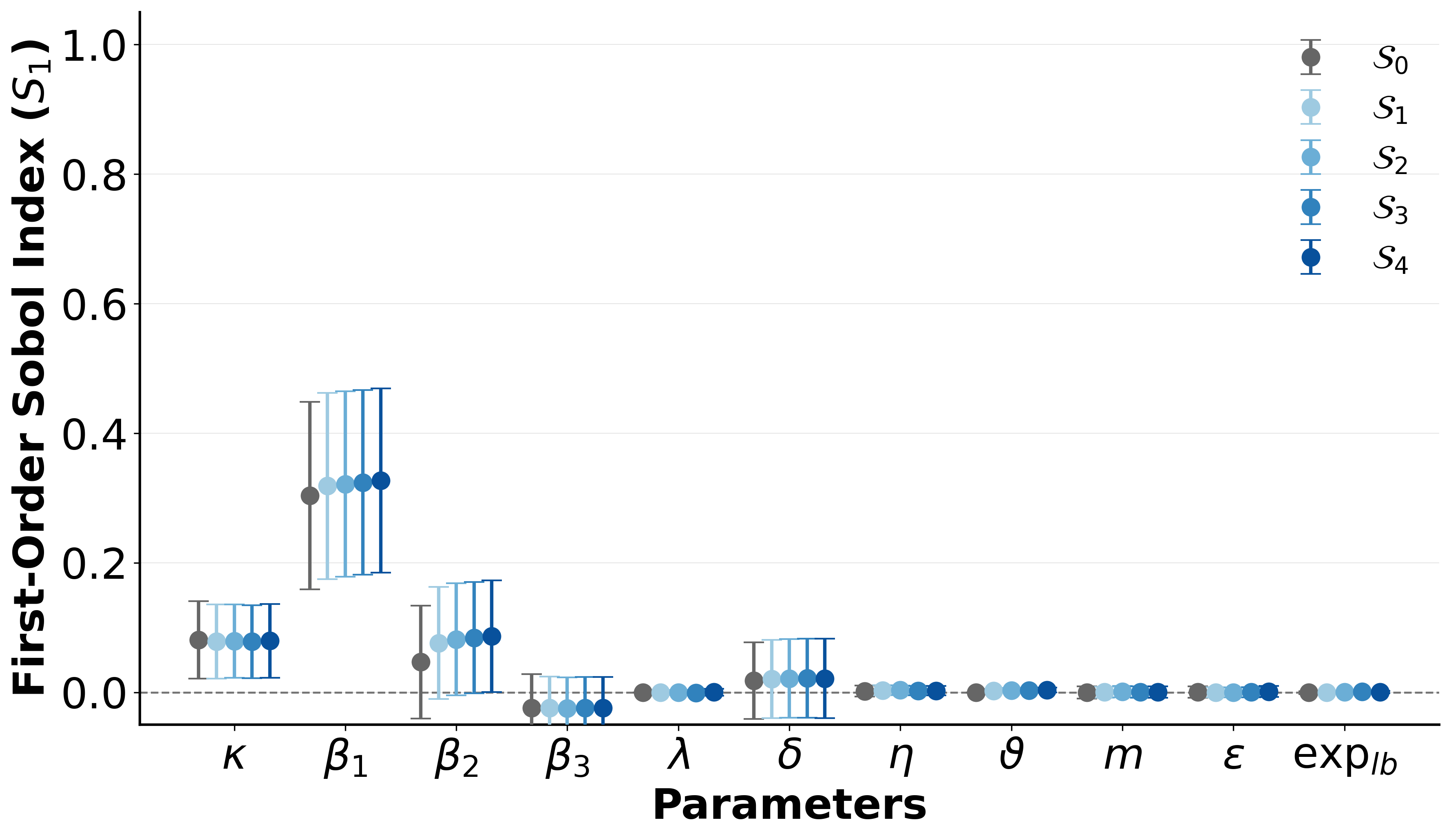}
        \caption{First-order Sobol' indices $S_i$}
        \label{fig:forbearance_sobol}
    \end{subfigure}
    \hfill
    \begin{subfigure}[b]{0.49\textwidth}
        \centering
        \includegraphics[width=\linewidth]{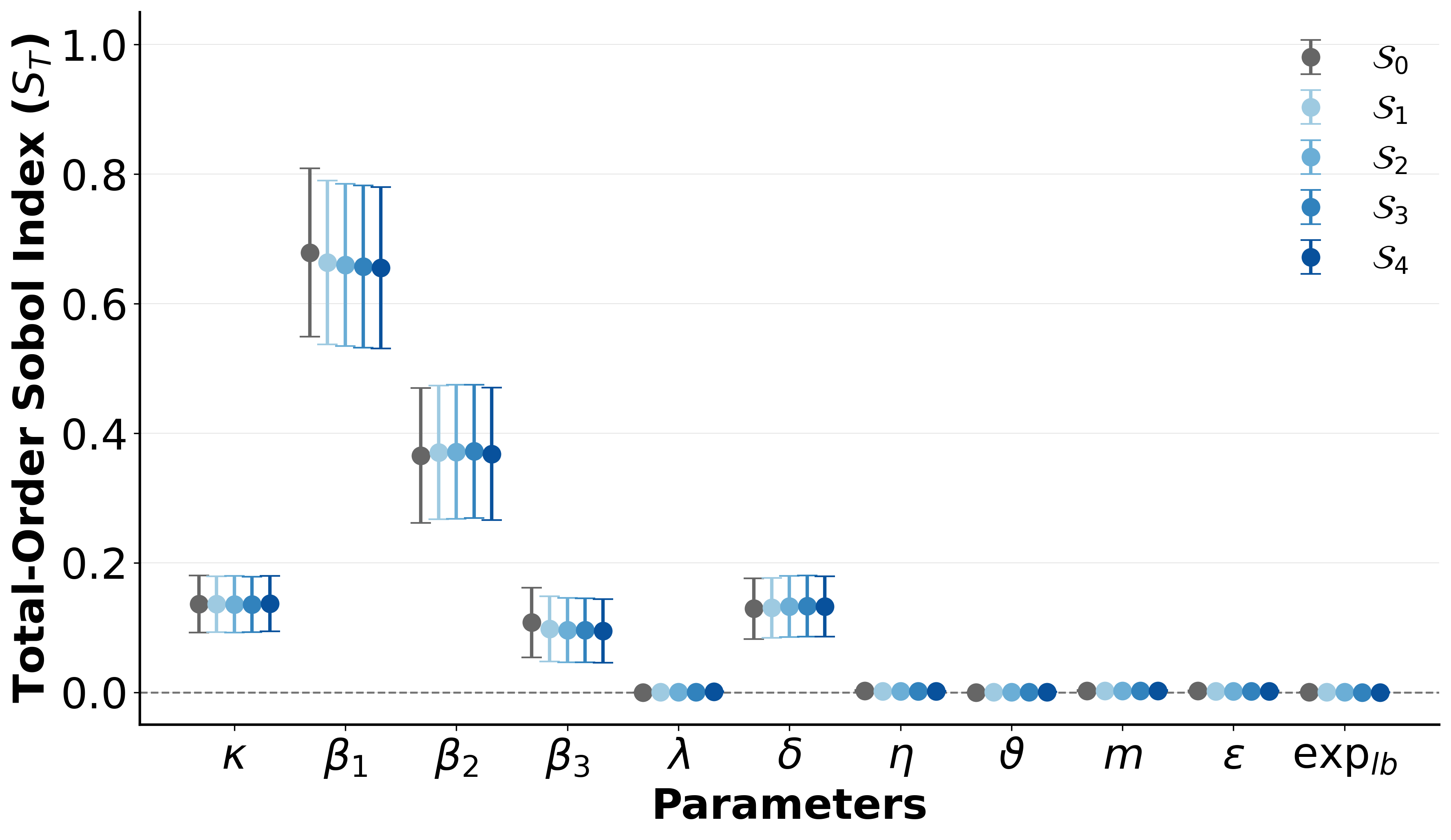}
        \caption{Total-order Sobol' indices $S_{Ti}$}
        \label{fig:forbearance_sobol_total}
    \end{subfigure}

    \vspace{0.3cm}

    \begin{subfigure}[b]{0.49\textwidth}
        \centering
        \includegraphics[width=\linewidth]{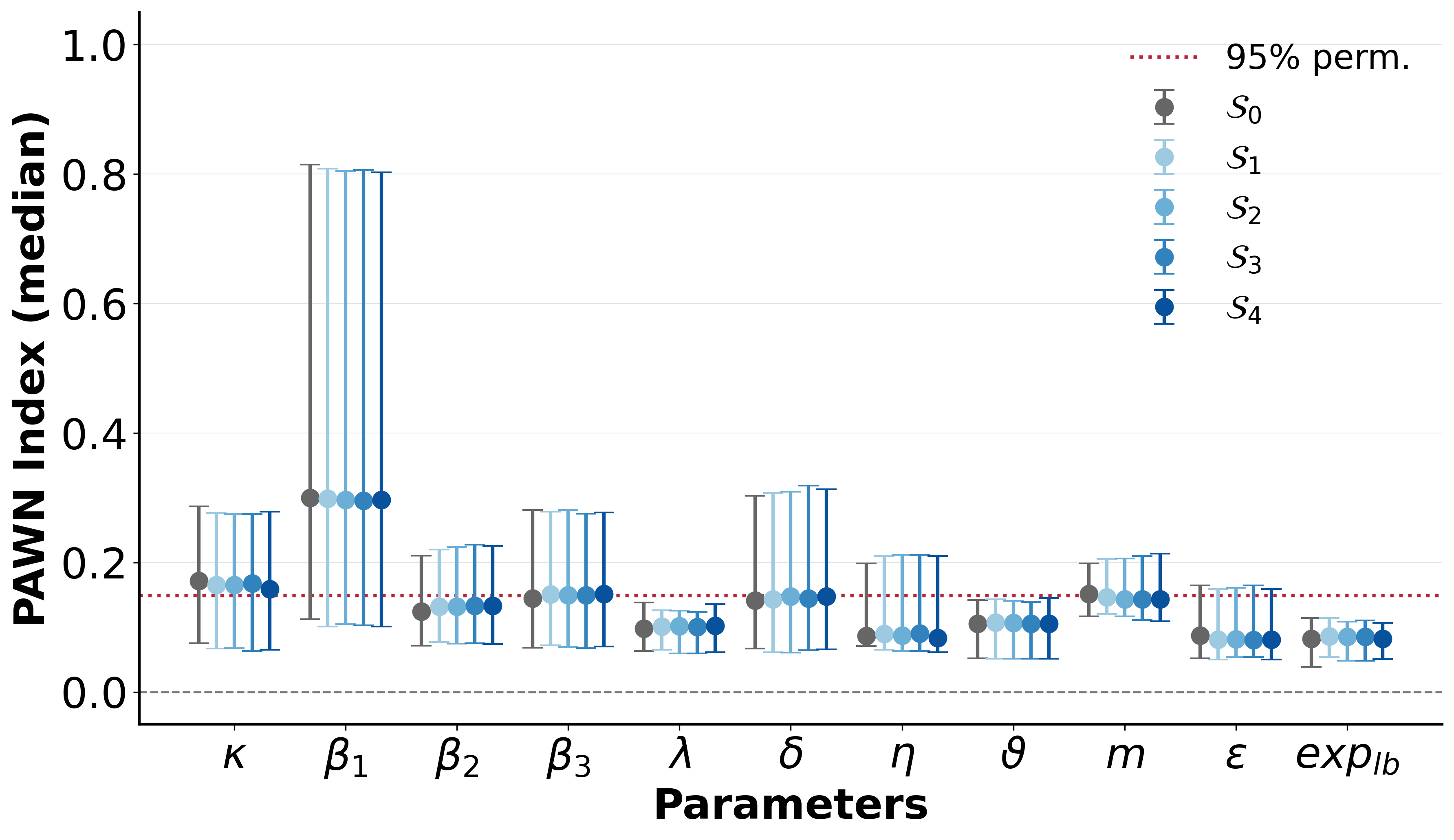}
        \caption{PAWN indices $T_i$}
        \label{fig:forbearance_pawn}
    \end{subfigure}
    \caption{Sensitivity of the \textit{mean number of distress months per household}, among households with at least one, across flood scenarios $\mathcal{S}_0$--$\mathcal{S}_4$. Panels (a) and (b) report first-order and total-order Sobol' indices with bootstrap confidence intervals from the full $3{,}328$-sample Saltelli design. Panel (c) reports the median PAWN indices $T_i$ with $J=10$ conditioning intervals. Bars span the minimum and maximum Kolmogorov--Smirnov statistic across conditioning intervals, rather than confidence intervals. The dotted line shows the $95$th percentile of the outcome-specific max-$T$ null from $10{,}000$ permutations, calibrated jointly over the five scenarios and eleven parameters; significance is assessed by comparing $T_i$ with this threshold.}
    \label{fig:forbearance_indices}
\end{figure}

\SIfig{fig:forbearance_indices} reports the first-order Sobol' indices (panel~\subref{fig:forbearance_sobol}), the total-order Sobol' indices (panel~\subref{fig:forbearance_sobol_total}) and the PAWN indices $T_i$ (panel~\subref{fig:forbearance_pawn}) for the mean number of distress months among households with at least one. Both the Sobol' indices and the PAWN indices $T_i$ rank $\beta_1$ as the main influence on the mean number of distress months ($S_i=0.30$--$0.33$, $T_{\beta_1}=0.30$). The debt pressure parameter $\beta_2$ also matters, with a larger total-order than first-order Sobol' index ($S_{Ti}=0.37$ against $S_i=0.05$--$0.09$) indicating an important role for interactions (panel~\subref{fig:forbearance_sobol_total}). The neighbourhood radius $\kappa$ is the only other parameter with a first-order index whose interval clears zero at every scenario ($S_i=0.08$), and under PAWN $\beta_1$ and $\kappa$ ($T_{\kappa}=0.16$--$0.17$) exceed the permutation threshold at every scenario, $\beta_3$ at four of the five. The radius sets how many neighbours enter $U^{\kappa}_{i,t}$ and, with $\delta > 0$, over how wide an area a foreclosure lowers the collateral of the households around it. Willingness to pay governs both whether a household enters distress and whether it is forborne, linking these sensitivities to the accumulation of distress months over time.

At low values of $\beta_1$, almost all households in distress foreclose during the observation window, leaving little time to accumulate further distress months. At higher values, more households are retained under forbearance and can experience repeated distress. Equation~\ref{eqn:distress_probs} explains this pattern: willingness to pay at or above $1/2$ rules out strategic foreclosure in the current month, while a binding financial constraint can still lead to foreclosure. When willingness to pay remains in this range, every distress event triggered by the strategic condition ends in forbearance. This holds month by month and is not a permanent status: subsequent changes in neighbourhood conditions or debt pressure can lower willingness to pay. A household may also accumulate distress months mainly through the strategic condition before a financial constraint leads to its foreclosure. The counts therefore depend on both the frequency of distress and the time households remain in the model. We assess them alongside the share of households in distress that foreclose.

\clearpage
\section{Foreclosure Cascades}
\label{sec:app:cascades}
\subsection{Reproduction numbers of the foreclosure feedback}
\label{sec:app:cascade_method}

A foreclosure at time $t$ lowers the collateral of every property within $\kappa$ at $t+1$ (Equation~\ref{eqn:collateral}). Through $U^{\kappa}_{i,t+1}$ it then lowers the willingness to pay of households up to $2\kappa$ away (Section~\ref{sec:neighbourhood}). We summarise this feedback with two reproduction numbers, one net and one gross.

The net measure borrows from branching processes \parencite{Hawkes_Oakes_1974}. If background events arrive independently and every event triggers on average $R<1$ further events, the expected number of events is the number of background events divided by $1-R$, and $R$ is the share of events that are triggered \parencite{Helmstetter_Sornette_2003}. Taking the foreclosure rate at $\delta=0$ as the background, we define
\begin{equation}
\label{eqn:r_net}
R_{\text{net}} = 1 - \frac{\Phi(0)}{\Phi(\delta)}, \qquad M = \frac{1}{1-R_{\text{net}}} = \frac{\Phi(\delta)}{\Phi(0)},
\end{equation}
where $\Phi$ is the annual number of foreclosures per $100$ active mortgage holders over the $300$ months after burn-in, and $M$ is the cascade multiplier, the foreclosure rate with spillovers relative to the paired run without them. Each run with $\delta>0$ is paired with a run at $\delta=0$ with the same parameters and seed. The two runs share the flood series, and the households' draws diverge after the first decision that the spillover changes.

$R_{\text{net}}$ is the excess foreclosure rate caused by the spillover, as a share of the rate with the spillover. Reading it as a reproduction number relies on the branching approximation, which the ABM does not impose. The paired runs can also differ in the size and composition of their mortgage population, so $R_{\text{net}}$ does not identify which individual foreclosures would not have happened without the feedback.

The gross measure works at the level of single foreclosures. At every household and step the model also computes a counterfactual $\text{wtp}^{0}_{i,t}$, from Equation~\ref{eqn:wtp} with the same debt pressure but with $U^{\kappa}_{i,t}$ computed from the collateral without the spillover, $v_j(1-\lambda D(x_j,y_j,\mathcal{S})_t)$. The counterfactual enters no decision. A foreclosure is attributed to the spillover when the same draw $\psi^{d}_{i,t}$ would not have led to foreclosure at $\text{wtp}^{0}_{i,t}$: a household in financial distress would have been forborne, $\text{wtp}^{0}_{i,t} > 1-\psi^{d}_{i,t}$ (\SIeq{eqn:strategic_distress}), and a household that can pay would either not have entered strategic distress, $\text{wtp}^{0}_{i,t} \geq \psi^{d}_{i,t}$ (\SIeq{eqn:strategic_trigger}), or would have been forborne. Then
\begin{equation}
\label{eqn:r_gross}
R_{\text{gross}} = \frac{\text{number of attributed foreclosures}}{\text{number of foreclosures}},
\end{equation}
over the same $300$ months.

The attribution compares outcomes within a single month. A household that the spillover pushes into foreclosure a few months before it would have foreclosed anyway counts as triggered, so $R_{\text{gross}}$ includes foreclosures that are only brought forward. Epidemiologists call this harvesting: part of the deaths associated with air pollution are deaths advanced by days or weeks \parencite{Zeger_Dominici_Samet_1999}. Without floods, $R_{\text{net}}$ cannot be distinguished from zero up to $\delta=0.125$ while $R_{\text{gross}}$ reaches $0.15$, as expected if at low intensity the spillover mostly advances foreclosures. The two measures rest on different counterfactuals, so their difference is not an exact split into additional and advanced foreclosures.

The runs used in Section~\ref{sec:cascades} are at the calibrated parametrisation (\SItab{tab:model_params}). They cover the five flood scenarios, $\lambda\in\{0.25,0.5,0.75\}$, $\vartheta\in\{0.03,0.05,0.13\}$ and twelve values of $\delta$ between $0$ and $0.25$, at $\kappa\in\{3,7,10,15,18\}$ with $\beta_1-\beta_2=0.5$ and at $\beta_1-\beta_2\in\{0.6,0.7\}$ with $\kappa=10$, each with $20$ seeds. Unless stated otherwise, results are at $\vartheta=0.05$, averaged over $\lambda$ within each seed, with $95\%$ Student-$t$ intervals over the $20$ seeds.

Up to $\delta=0.05$, floods do not change $R_{\text{net}}$ detectably. Between $\delta=0.1$ and $0.175$ they raise it: the interval of $\Delta R_{\text{net}}$ lies above zero in $17$ of the $20$ combinations of scenario and $\delta$, and in $5$ after a Holm correction over all $44$ combinations with $\delta>0$. The largest increase is at $\mathcal{S}_4$ and $\delta=0.15$, where $R_{\text{net}}$ is $0.48$ against $0.23$ without floods. At $\delta=0.25$ the effect reverses, because floods raise the foreclosure rate without spillover proportionally more than the rate with it.

\begin{figure}[htbp]
    \centering
    \includegraphics[width=\textwidth]{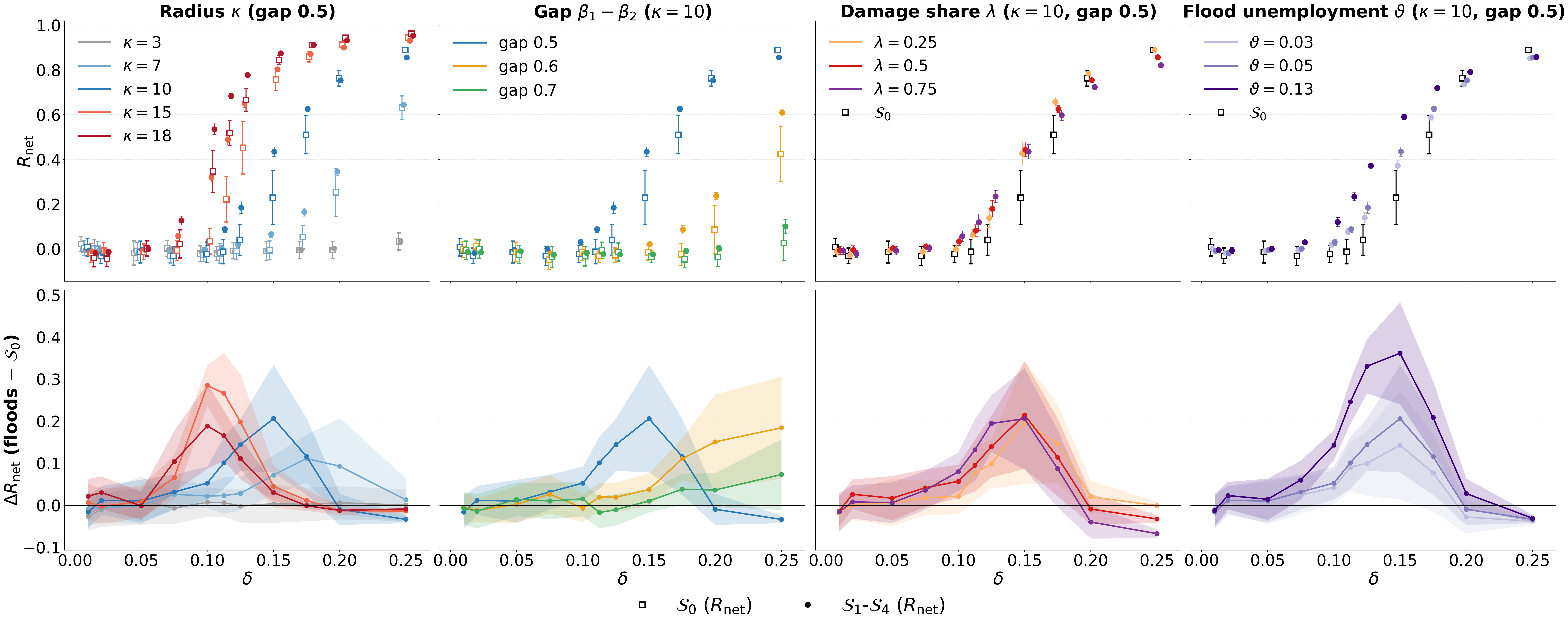}
    \caption{\small Net reproduction number of the foreclosure feedback with and without floods, by radius $\kappa$ (first column, $\beta_1-\beta_2=0.5$), gap $\beta_1-\beta_2$ (second column, $\kappa=10$), damage share $\lambda$ (third column, $\kappa=10$, $\beta_1-\beta_2=0.5$) and flood-induced unemployment $\vartheta$ (fourth column, $\kappa=10$, $\beta_1-\beta_2=0.5$). Top: $R_{\text{net}}$ against the spillover intensity $\delta$ at $\mathcal{S}_0$ (open squares) and under $\mathcal{S}_1$--$\mathcal{S}_4$ (filled circles). In the third and fourth columns $\mathcal{S}_0$ is shown once, since $\lambda$ and $\vartheta$ act only in flood months. Points are means over $20$ seeds with $95\%$ Student-$t$ intervals; under floods the runs of each seed are first averaged over $\mathcal{S}_1$--$\mathcal{S}_4$ and $\lambda$ ($12$ runs per seed; over $\mathcal{S}_1$--$\mathcal{S}_4$ only in the third column, $4$ runs per seed). $\vartheta=0.05$ outside the fourth column. Bottom: $\Delta R_{\text{net}}=R_{\text{net}}(\mathcal{S}_1\text{--}\mathcal{S}_4)-R_{\text{net}}(\mathcal{S}_0)$, computed per seed as the mean over the flooded runs of the seed minus the $\mathcal{S}_0$ run of the same seed, with $95\%$ Student-$t$ bands over $20$ seeds. Other parameters at their calibrated values (\SItab{tab:model_params}).}
    \label{fig:tipping_sensitivity}
\end{figure}

\SIfig{fig:tipping_sensitivity} varies one factor at a time. The flood effect always sits in the band of $\delta$ where $R_{\text{net}}$ without floods starts to rise, and fades on either side of it. At $\kappa=3$ it is not detectable at any $\delta$, and at a gap of $0.7$ only at $\delta=0.175$. The three values of $\lambda$ give similar $R_{\text{net}}$ up to $\delta=0.175$.

\begin{figure}[htbp]
    \centering
    \includegraphics[width=0.6\textwidth]{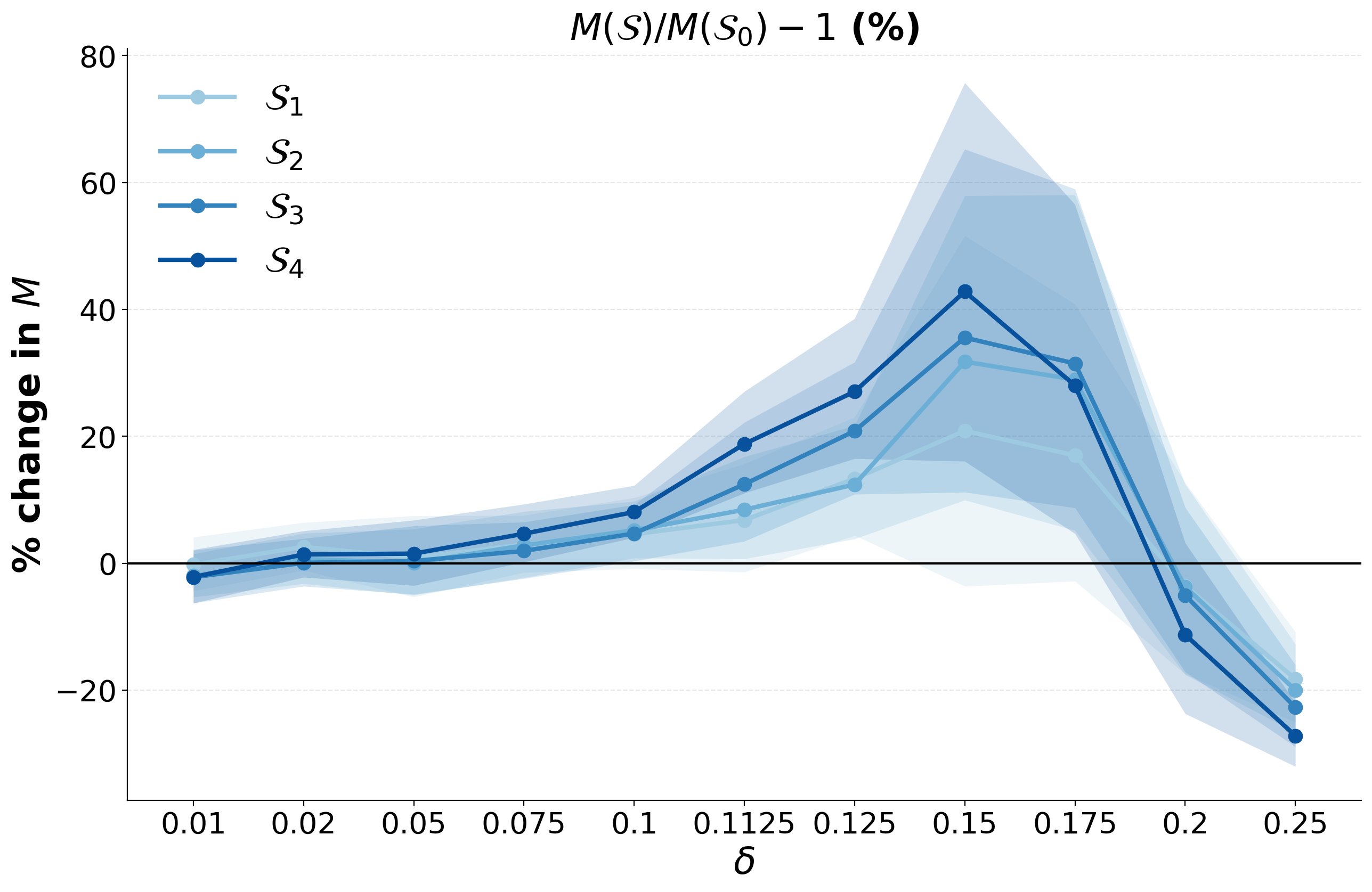}
    \caption{\small Percentage change in the cascade multiplier $M=1/(1-R_{\text{net}})=\Phi(\delta)/\Phi(0)$ (\SIeq{eqn:r_net}) relative to the no-flood scenario, $M(\mathcal{S})/M(\mathcal{S}_0)-1$, against the spillover intensity $\delta$, by flood scenario. Each seed of a flooded scenario is paired with the same seed at $\mathcal{S}_0$; the paired log ratios $\ln[M(\mathcal{S})/M(\mathcal{S}_0)]$ are averaged over the $20$ seeds, and the mean and its $95\%$ Student-$t$ band are transformed to percentage changes as $100[\exp(\cdot)-1]$. $\lambda\in\{0.25,0.5,0.75\}$ is averaged within seed. The levels of $M$ are in Figure~\ref{fig:r_net}, panel (a). Calibrated parametrisation (\SItab{tab:model_params}) with $\kappa=10$, $\beta_1-\beta_2=1/2$ and $\vartheta=0.05$; rates over the $300$ months after burn-in.}
    \label{fig:multiplier}
\end{figure}

The cascade multiplier $M$ in \SIeq{eqn:r_net} expresses the same comparison. Under the branching approximation a background foreclosure triggers on average $R$ foreclosures, each of which triggers $R$ more, so the expected size of the cascade it starts is $1+R+R^2+\dots=1/(1-R)$. Because $M$ is a ratio, we compare flooded and unflooded runs through $M(\mathcal{S})/M(\mathcal{S}_0)$ for each seed, average the logarithm over seeds and report the result as a percentage change. Floods raise $M$ by $12$ to $27\%$ at $\delta=0.125$ and by $21$ to $43\%$ at $\delta=0.15$, and lower it by $18$ to $27\%$ at $\delta=0.25$ (\SIfig{fig:multiplier}), the same pattern as $\Delta R_{\text{net}}$.

\subsection{Distribution of default cascade sizes}
\label{sec:app:cascade_size}

$R_{\text{net}}$ and $M$ describe the mean of the cascade process. They cannot tell whether the extra foreclosures are spread over many short chains or concentrated in a few large cascades, which matters for a lender with exposure in one neighbourhood.

To reconstruct the cascades, a foreclosure is a child when it is attributed to the spillover by the rule behind $R_{\text{gross}}$ (\SIeq{eqn:r_gross}), and a root otherwise. A child at step $t$ can only descend from a foreclosure at step $t-1$ within $2\kappa$. In weighted trees, each child gets one of these candidates as parent, drawn with weight equal to the number of grid cells within $\kappa$ of both \parencite{Zhuang_Ogata_VereJones_2002}, and a cascade is a root with all its descendants. In connected components, each child is linked to all its candidates \parencite{Reasenberg_1985}, and a cascade is a connected group. The first can split a cascade fed from several sides and the second can merge cascades that touch, so the two bracket the cascade structure. The size distribution has a mode at size one and, from intermediate $\delta$, a second mode of large cascades; we separate the two at the minimum of the distribution smoothed with a Gaussian kernel (bandwidth $0.3$ in log size). Cascades are kept when they start in the $300$ months after burn-in, excluding the last $12$.

Large cascades peak between $113$ and $175$ foreclosures in every scenario and under both reconstructions, so $\delta$ changes how many foreclosures belong to large cascades rather than how large they are. Floods raise the number of cascades per run and, at intermediate $\delta$, the chance that a cascade grows large (\SItab{tab:cascade_interplay}): at $\delta=0.15$, $\mathcal{S}_4$ produces about seven times as many large cascades per run as $\mathcal{S}_0$. Near the onset this comparison rests on few large cascades without floods, $4$ in $20$ runs at $\delta=0.125$.

\begin{table}[htbp]
\centering
\small
\caption{\small Number and growth of default cascades (weighted trees) at the calibrated parametrisation with $\kappa=10$, $\beta_1-\beta_2=1/2$ and $\vartheta=0.05$. A large cascade is a cascade above the minimum between the two modes of the size distribution of its cell. Pooled over $20$ runs without floods and $60$ runs per flood scenario.}
\label{tab:cascade_interplay}
\begin{tabular}{l rrr rrr rrr}
\toprule
 & \multicolumn{3}{c}{\begin{tabular}[b]{@{}c@{}}Cascades\\per run\end{tabular}} & \multicolumn{3}{c}{\begin{tabular}[b]{@{}c@{}}Large cascades\\per $1{,}000$ cascades\end{tabular}} & \multicolumn{3}{c}{\begin{tabular}[b]{@{}c@{}}Large cascades\\per run\end{tabular}} \\
\cmidrule(lr){2-4}\cmidrule(lr){5-7}\cmidrule(lr){8-10}
$\delta$ & $\mathcal{S}_0$ & $\mathcal{S}_1$ & $\mathcal{S}_4$ & $\mathcal{S}_0$ & $\mathcal{S}_1$ & $\mathcal{S}_4$ & $\mathcal{S}_0$ & $\mathcal{S}_1$ & $\mathcal{S}_4$ \\
\midrule
$0.125$ & $241$ & $385$ & $496$ & $0.8$ & $2.0$ & $5.3$ & $0.20$ & $0.77$ & $2.63$ \\
$0.15$  & $238$ & $373$ & $483$ & $3.6$ & $6.8$ & $12.7$ & $0.85$ & $2.55$ & $6.12$ \\
$0.2$   & $222$ & $349$ & $453$ & $33.5$ & $30.4$ & $37.9$ & $7.45$ & $10.63$ & $17.18$ \\
$0.25$  & $192$ & $327$ & $431$ & $86.7$ & $67.7$ & $74.2$ & $16.65$ & $22.13$ & $32.00$ \\
\bottomrule
\end{tabular}
\end{table}

The attribution also counts foreclosures that the spillover only brings forward. As a check that uses no attribution, we link every foreclosure at step $t$ to every foreclosure at step $t-1$ within $2\kappa$ and take the connected groups as clusters. At $\delta=0$, where no foreclosure can cause another, these clusters stay small. At $\delta>0$ they recover the large cascades, with a share of foreclosures in large cascades within two percentage points of both reconstructions, so the large cascades are not a product of the attribution.

The mean size of a weighted tree, $1/(1-R_{\text{gross}})$, is the gross counterpart of $M$. The gross and net measures diverge only where large cascades are rare (\SItab{tab:cascade_gross_net}), so the foreclosures that the spillover only brings forward are mostly isolated ones near the onset of the feedback.

\begin{table}[htbp]
\centering
\small
\caption{\small Large default cascades and the gross and net feedback measures without floods ($\mathcal{S}_0$), at the calibrated parametrisation with $\kappa=10$, $\beta_1-\beta_2=1/2$ and $\vartheta=0.05$. Share in large cascades and mean tree size from the weighted trees, pooled over $20$ runs; $R_{\text{gross}}$, $R_{\text{net}}$ and $M=\Phi(\delta)/\Phi(0)$ are means over $20$ seeds.}
\label{tab:cascade_gross_net}
\begin{tabular}{l rrrrrr}
\toprule
$\delta$ & Share in large cascades & $R_{\text{gross}}$ & $R_{\text{net}}$ & $R_{\text{net}}/R_{\text{gross}}$ & $M$ & Mean tree size \\
\midrule
$0.125$ & $8\%$  & $0.15$ & $0.04$ & $0.27$ & $1.07$ & $1.20$ \\
$0.15$  & $31\%$ & $0.33$ & $0.23$ & $0.70$ & $1.50$ & $1.67$ \\
$0.175$ & $55\%$ & $0.57$ & $0.51$ & $0.89$ & $2.33$ & $2.57$ \\
$0.2$   & $77\%$ & $0.79$ & $0.76$ & $0.96$ & $4.59$ & $5.22$ \\
$0.25$  & $89\%$ & $0.91$ & $0.89$ & $0.98$ & $9.27$ & $10.99$ \\
\bottomrule
\end{tabular}
\end{table}

The same pattern holds at the other radii. Large cascades grow with $\kappa$, but far more slowly than the area a foreclosure can reach, and none form at $\kappa=3$. The runs at $\delta=0$ have fewer foreclosures than those at high $\delta$, which makes the check without attribution less strict there; near the onset, however, the foreclosure rate is within $8\%$ of its value at $\delta=0$ while large cascades already appear.
\clearpage
\section{Foreclosure Timing}
\label{sec:app:fcltime}
This section uses the distress months defined in \SIsec{sec:appendix_sobol_distress}, counted among households that foreclose.

\begin{figure}[htbp]
    \centering
    \includegraphics[width=\textwidth, keepaspectratio]{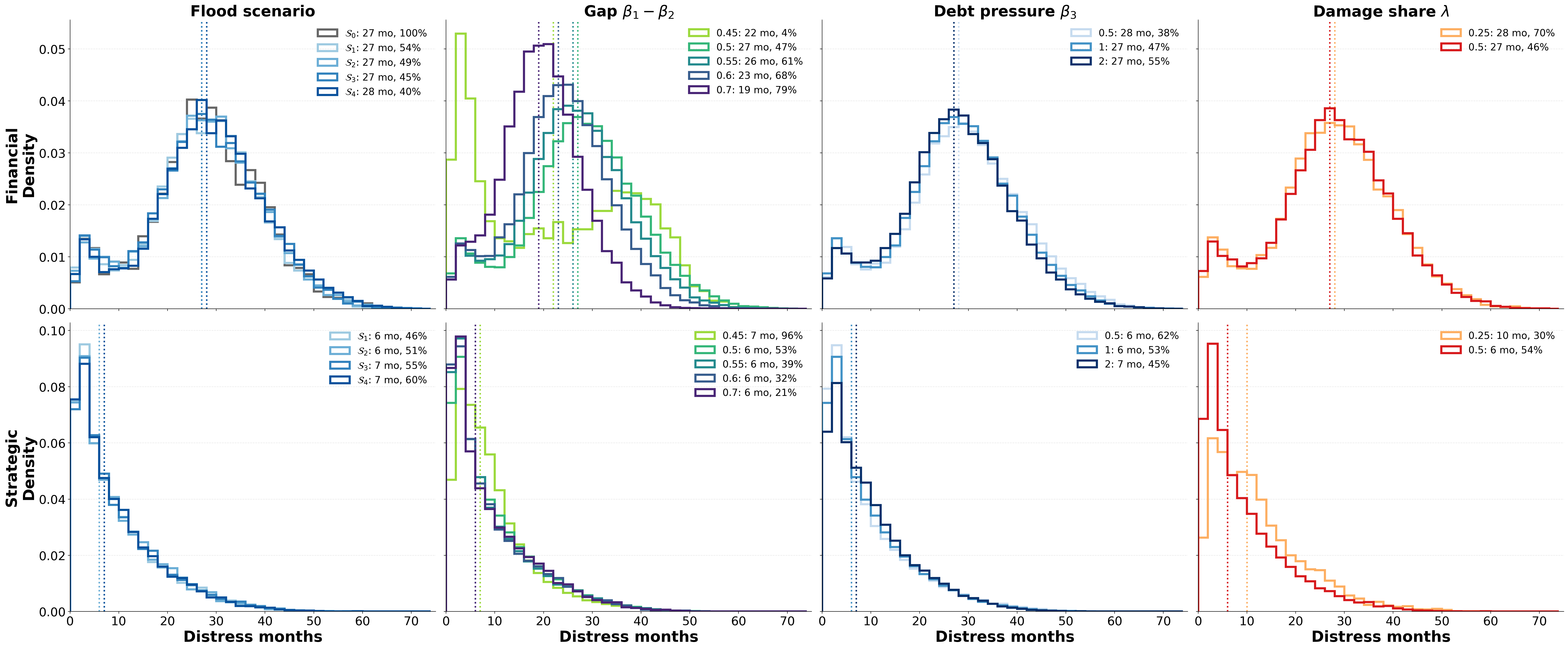}
    \caption{Distress months among foreclosed households, including forborne months and the terminal foreclosure. Both distress months and foreclosures are restricted to steps $100$--$399$; histories crossing the cutoff are truncated. The top row shows financial foreclosures and the bottom row strategic foreclosures. Each column varies one factor, with the others at the calibrated parametrisation ($\beta_1=1.0$, $\beta_2=0.5$, $\beta_3=1.0$). The first column separates flood scenarios and contains the same $68{,}927$ foreclosures as Figure~\ref{fig:foreclosure_time}. The second column varies the gap $\beta_1-\beta_2 \in \{0.45, 0.5, 0.55, 0.6, 0.7\}$ through $\beta_2$, with $\beta_1=1.0$ ($293{,}442$ foreclosures). The third column varies $\beta_3 \in \{0.5, 1.0, 2.0\}$ ($187{,}068$ foreclosures). The fourth column varies the damage share over $\lambda \in \{0.25, 0.5\}$, the values inside the GSA range ($34{,}091$ foreclosures). The second to fourth columns pool $\mathcal{S}_1$--$\mathcal{S}_4$. Legend values are the median number of distress months and the share of foreclosures in the row's channel at that level; dotted lines mark medians and histograms are normalised separately. All samples use ten seeds for each scenario and each value of $\lambda \in \{0.25, 0.5, 0.75\}$, with the three values pooled except in the fourth column, and $\delta=0$, $\kappa=10$, $\eta=0.0383$, $\vartheta=0.05$, $m=5.25$, $\varepsilon=0.00250$ and $\mathit{exp}_{lb}=1087$ (\SItab{tab:model_params}). Counts describe the observation window, not complete lifetime histories or continuous spell duration.}
    \label{fig:distress_time}
\end{figure}

\begin{figure}[htbp]
    \centering
    \includegraphics[width=\textwidth, keepaspectratio]{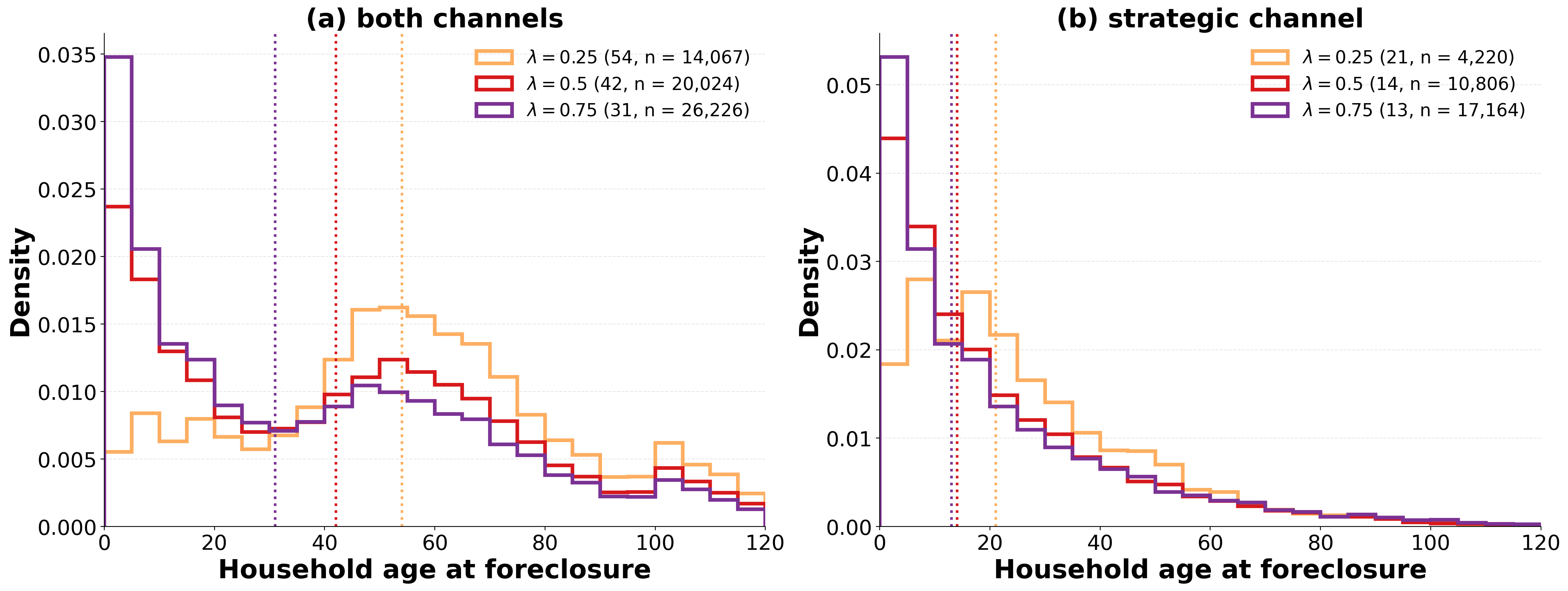}
    \caption{\small Household age at foreclosure by damage share $\lambda \in \{0.25, 0.5, 0.75\}$, pooling $\mathcal{S}_1$--$\mathcal{S}_4$. Panel (a) pools both default channels; panel (b) shows strategic foreclosures only. Legend values are medians and sample sizes; dotted lines mark medians and histograms are normalised separately. Foreclosures are recorded at steps $100$--$399$, from ten seeds for each scenario and value of $\lambda$. Parameters are at the calibrated parametrisation: $\beta_1=1.0$, $\beta_2=0.5$, $\beta_3=1.0$, $\delta=0$, $\kappa=10$, $\eta=0.0383$, $\vartheta=0.05$, $m=5.25$, $\varepsilon=0.00250$ and $\mathit{exp}_{lb}=1087$ (\SItab{tab:model_params}). $\mathcal{S}_0$ is omitted, since without floods $\lambda$ has no effect.}
    \label{fig:foreclosure_time_lambda}
\end{figure}

\SIfig{fig:distress_time} reports the distribution of accumulated distress months among foreclosed households, complementing the mortgage age at foreclosure in Figure~\ref{fig:foreclosure_time}. \SIfig{fig:foreclosure_time_lambda} separates the mortgage age at foreclosure by the damage share $\lambda$, which Figure~\ref{fig:foreclosure_time} pools.

At the calibrated parametrisation, financial foreclosures follow a median of $27$ distress months, $92\%$ of them strategic triggers resolved through forbearance, while strategic foreclosures follow a median of $6$ distress months. The histories preceding financial foreclosure therefore should not be interpreted as uninterrupted financial hardship.

Flooding changes the mix of channels rather than the channels themselves. From $\mathcal{S}_0$ to $\mathcal{S}_4$ the strategic share rises from zero to $60\%$, lowering the pooled median age at foreclosure from $62$ to $37$ months and the pooled median distress count from $27$ to $13$. The within-channel medians barely move: financial foreclosures stay between $62$ and $64$ months of age and $27$ and $28$ distress months, strategic foreclosures between $14$ and $15$ months of age and $6$ and $7$ distress months (panel (a) of Figure~\ref{fig:foreclosure_time} and first column of \SIfig{fig:distress_time}). The damage share $\lambda$ sets the size of this shift: at $\mathcal{S}_4$ the strategic share is $34\%$, $61\%$ and $72\%$ at $\lambda = 0.25$, $0.5$ and $0.75$, while the median age at financial foreclosure stays between $63$ and $64$ months and that at strategic foreclosure falls from $21$ to $13$ months (\SIfig{fig:foreclosure_time_lambda}). Narrowing the gap $\beta_1-\beta_2$ has the same effect as flooding (\SIfig{fig:distress_time}, second column). As the gap narrows from $0.7$ to $0.45$, the strategic share in $\mathcal{S}_1$ to $\mathcal{S}_4$ rises from $21\%$ to $96\%$, and the pooled median distress count, between $16$ and $19$ months for gaps of $0.5$ and above, falls to $7$. At a gap of $0.45$, outside the damage-required region, strategic foreclosure no longer requires flood damage, and $91\%$ of foreclosures are strategic even in $\mathcal{S}_0$. Within the region, the financial median rises from $19$ to $27$ distress months as the gap narrows from $0.7$ to $0.5$, since these households accumulate more forborne strategic triggers. A lower $\beta_3$, which raises the debt pressure at every debt level, acts in the same direction: from $\beta_3=2$ to $\beta_3=0.5$ the strategic share rises from $45\%$ to $62\%$, with the within-channel medians unchanged (third column).

\clearpage

\end{document}